\documentclass[pra,twocolumn,amsmath,amssymb,superscriptaddress]{revtex4-1}

\usepackage{epsfig,amsmath}
\usepackage{subfigure}
\usepackage{graphicx}
\usepackage{dcolumn}
\usepackage{stmaryrd}
\usepackage{mathrsfs}
\usepackage{pifont}
\usepackage{amsthm}
\usepackage{amssymb}
\usepackage{bm}
\usepackage{latexsym}
\usepackage[colorlinks=true,linkcolor=blue,citecolor=blue]{hyperref}
\usepackage{color}
\usepackage{epstopdf}

\begin{document}

\title{Generation of Bell and GHZ states from a hybrid qubit-photon-magnon system}

\author{Shi-fan Qi}
\affiliation{Department of Physics, Zhejiang University, Hangzhou 310027, Zhejiang, China}

\author{Jun Jing}
\email{jingjun@zju.edu.cn}
\affiliation{Department of Physics, Zhejiang University, Hangzhou 310027, Zhejiang, China}

\date{\today}

\begin{abstract}
We propose a level-resolved protocol in a hybrid architecture for connecting a superconducting qubit and a magnon mode contained within a microwave cavity (resonator) to generate the local and global entangled states, enabling a wide range of applications in quantum communication, quantum metrology, and quantum information processing. Exploiting the high-degree of controllability in such a hybrid qubit-photon-magnon system, we derive effective Hamiltonians at the second- or the third-order resonant points by virtue of the strong counter-rotating interactions between the resonator and the qubit and between the resonator and the magnon. Consequently, we can efficiently generate the Bell states of the photon-magnon and the qubit-magnon subsystems and the Greenberger-Horne-Zeilinger state of the whole hybrid system. We also check the robustness of our protocol against the environmental noise by the Lindblad master equation. Our work makes this hybrid platform of high-degree of controllability a high-fidelity candidate for the realization of the maximally-entangled multiple states.
\end{abstract}

\date{\today}

\maketitle

\section{Introduction}

Hybrid quantum systems have attracted great attentions due to their diversified applications in quantum computing~\cite{quantumcomputing}, quantum communications~\cite{quantumcommunication}, and quantum sensing~\cite{quantumsense}. During the past decade, the hybrid quantum systems consisting of the collective spin excitations in the ferromagnetic crystals have been applied to novel quantum technologies~\cite{magnon,magnon2,magnon4,magnon5,magnon6,magnoncavity} by virtue of the long coherent-time of the magnet-spin ensemble. The strong dipole-transition for efficiently coupling to the microwave photons and phonons allows the construction of the hybrid magnon-photon, magnon-phonon, and magnon-photon-phonon systems in both theoretical proposals~\cite{yigcavity1,yigcavity2,mppentang} and experimental demonstrations~\cite{yigcavity3,yigcavity4,yigcavity5,yigcavity6}. In more recent experiments~\cite{magnonqubit,magnonqubit2}, a hybrid qubit-photon-magnon system has been realized, opening a new avenue to studying the intermediate transitions by the interaction Hamiltonian of different constituents. It is expected to have many interesting applications, including providing a platform to generate hybrid entangled states.

The entangled states~\cite{quantumentanglement} are essential ingredients for extracting quantum advantage in the quantum communication protocols~\cite{quantumcommun}, such as quantum key distribution~\cite{keydistribute}, quantum secret sharing~\cite{quantumsecretsharing}, and quantum secure direct communication~\cite{quantumsecurecommun}. Many quantum platforms~\cite{trapion,ghzstate,superconduct2,superconduct3,Schrodingerstate} and protocols for preparing and measuring the entangled states have therefore been intensively pursued for a long time~\cite{GHZ,ghzstate,GHZstate2,bellstate3,multiphoton,photonic,noon1,noon2} and are still under active investigation. The simplest and the most popular maximally-entangled-states are Bell states involving only two discrete systems~\cite{quantumentanglement,bellstate}, classified to two groups, i.e., $|\psi_\pm\rangle=(|01\rangle\pm|10\rangle)/\sqrt{2}$ and $|\varphi_\pm\rangle=(|00\rangle\pm|11\rangle)/\sqrt{2}$. Here $|0\rangle$ and $|1\rangle$ are respectively the ground and the excited states of a two-level system. The Bell states are fundamental and important in both quantum cryptography~\cite{cryptography,cryptography2} and quantum teleportation~\cite{teleportation}. In a recent experiment~\cite{yigcavity3}, the single-excitation Bell states $|\psi_\pm\rangle$ of the photon-magnon system have been steadily generated under the parity-time-symmetry broken condition. However, how to generate the double-excitation Bell states $|\varphi_\pm\rangle$ in a hybrid photon-magnon system is still an open question. More generally, the maximally entangled states for the $N$-partite discrete quantum systems, such as the Greenberger-Horne-Zeilinger (GHZ) states~\cite{GHZ,ghzstate}, i.e., $(|0\rangle^{\otimes N}+|1\rangle^{\otimes N})/\sqrt{2}$ and the Werner (W) states~\cite{wstate,wernerstate}, i.e., $(|100\cdots0\rangle+|010\cdots0\rangle+\cdots+|000\cdots1\rangle)/\sqrt{N}$, are also of great practical importance. Note that the multiple GHZ states are natural extensions of the double-excitation Bell states $|\varphi_\pm\rangle$ and cannot be converted to the W-states with nonzero probability~\cite{quantumentanglement}.

This work focuses on generating the local Bell states $|\varphi_\pm\rangle$ and the global GHZ states in a tripartite hybrid system consisting of a superconducting qubit and a magnon mode embedded in a microwave cavity (see the diagram in Fig.~\ref{diagram}), taking advantage of the tunability of the transition frequencies of qubit~\cite{cqed} and magnon~\cite{magnon} and their respective strong or ultrastrong couplings to the cavity mode (resonator). Due to the counter-rotating terms with no conservation of the excitation-number in the Rabi interactions~\cite{cq,ultrastrong,ultrastrong2}, it is possible to design an effective Hamiltonian as well as the generation protocol for connecting states in the whole Hilbert space nearby the desired multi-excitation resonant points. The effective Hamiltonian for the interested indirect transitions can be extracted using the high-order Fermi's Golden rule~\cite{fermigolden,secondorder,noon3,ae,ae1}. It states that if the shortest path connecting $|i\rangle$ and $|f\rangle$ (two eigenstates of the free Hamiltonian) is an $n$th-order process, then the effective coupling strength (in the leading order) between them reads,
\begin{equation}\label{Fermi}
g_{\rm eff}=\sum_{m_1 m_2...m_{n-1}}\frac{V_{fm_{n-1}}...V_{m_2m_1}V_{m_1i}}{(E_i-E_{m_{n-1}})...(E_i-E_{m_1})},
\end{equation}
where $V_{m_2m_1}\equiv\langle m_2|V|m_1\rangle$ is the transition matrix element of the interaction Hamiltonian and $E_i$ is the eigenvalue of the $i$th eigenstate of the free Hamiltonian (with no degeneracy).

The rest part of the work is organized as follows. In Sec.~\ref{model}, we introduce the total Hamiltonian of the hybrid qubit-photon-magnon system. In Sec.~\ref{bell}, we show that at a desired point, where the transition frequency of the qubit is nearly resonant with the frequency sum of the cavity and magnon modes, an effective Hamiltonian can be constructed to describe a photon and a magnon simultaneously excited by annihilating an atomic excitation. This ``three-wave-mixing'' Hamiltonian is applied to generate the double-excitation Bell states of the photon-magnon subsystem in Sec.~\ref{Bellstep}. And then in Sec.~\ref{Bellnum}, the fidelity of generation is numerically estimated with experimental parameters. In Sec.~\ref{GHZ}, the working point of our protocol moves to the near-resonant point for the transition frequency of the qubit and the detuning of the photon and magnon modes, where one can construct an effective Hamiltonian to simultaneously excite a qubit and a magnon by one photon. Then we show that the GHZ state of the whole hybrid system can be concisely generated by a three-step scheme in Sec.~\ref{GHZstep}, whose fidelity under dissipation is evaluated in Sec.~\ref{GHZnum}. The protocol in Sec.~\ref{GHZstep} is slightly modified to generate the Bell state of the qubit-magnon subsystem in Sec.~\ref{discussion}. We finally conclude this work in Sec.~\ref{conclusion}. The derivation details for the effective Hamiltonians in Secs.~\ref{bell} and \ref{GHZ} can be found respectively in Appendices~\ref{appa} and \ref{appb}.

\section{Model Hamiltonian of hybrid system}\label{model}

\begin{figure}[htbp]
\centering
\includegraphics[width=0.45\textwidth]{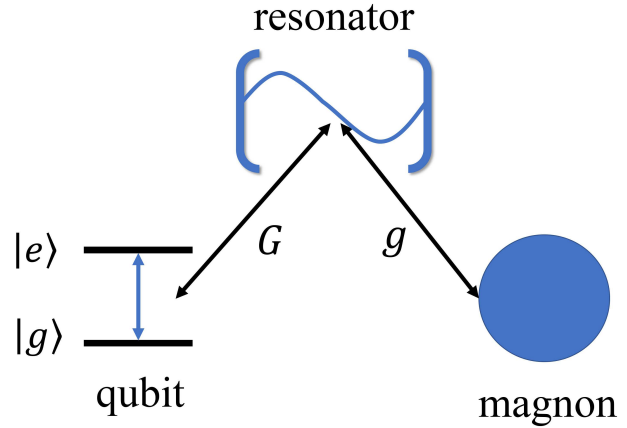}
\caption{A tripartite hybrid system: a single-mode cavity (resonator) is coupled to a superconducting qubit and a YIG sphere (in the Kittel mode, a uniform magnetostatic mode) with the coupling strengths $G$ and $g$, respectively. }\label{diagram}
\end{figure}

The hybrid quantum model we considered in this work consists of a superconducting qubit, a microwave cavity/resonator and a ferromagnetic crystal in the Kittle mode, as shown in Fig.~\ref{diagram}. The interaction between the qubit and the resonator is described by a general Rabi model, and simultaneously the magnetostatic mode (magnon) is coupled to the resonator via the magnetic dipole interaction. Thus the Hamiltonian for the whole system~\cite{magnonqubit,magnonqubit2,ultrastrong2,secondorder} $(\hbar\equiv1)$ can be written as
\begin{equation}\label{Hamiltonian}
\begin{aligned}
&H=H_0+V,\\
&H_0=\omega_aa^\dag a+\omega_mm^\dag m+\omega_q\sigma_+\sigma_-,\\
&V=g(a+a^\dag)(m+m^\dag)+G(a+a^\dag)(\sigma_x \cos\theta+\sigma_z\sin\theta).
\end{aligned}
\end{equation}
Here $a (a^\dag)$ and $m (m^\dag)$ are the annihilation (creation) operators of the photon and magnon modes, respectively. $\omega_a$ and $\omega_m$ are their respective eigen-frequencies. $g$ is the coupling strength between the magnon and the photon mode. The two levels of the qubit are labeled by $|g\rangle$ and $|e\rangle$, indicating the ground and the excited states, respectively. The Pauli operators for the qubit are then written as $\sigma_+\equiv|e\rangle\langle g|$, $\sigma_-\equiv|g\rangle\langle e|$, $\sigma_x\equiv |e\rangle\langle g|+|g\rangle\langle e|$, and $\sigma_z\equiv|e\rangle\langle e|-|g\rangle\langle g|$. We do not apply the rotating-wave approximation to the interaction Hamiltonian $V$. The angle $\theta$ parameterizes the amount of the longitudinal and the transversal couplings between the qubit and the resonator, which is adjustable and independent of the coupling strength $G$. Arbitrary mixture of the longitudinal and the transversal couplings has been realized in the circuit-QED experiments~\cite{cq,ultrastrong}, where the coupling strengths can reach the ultrastrong coupling regime.

Our approach works in the dispersive regime of the hybrid model in Eq.~(\ref{Hamiltonian}), where the coupling strengths are much smaller than the transition frequencies of the three subsystems and their detunings, i.e., $g,G\ll\omega_a$, $\omega_m$, $\omega_q$, $|\omega_a-\omega_m|$, $|\omega_a-\omega_q|$, $|\omega_m-\omega_q|$. However, it is crucial to know precisely their respective ranges of validity. We would apply the high-order Fermi's Golden rule in Eq.~(\ref{Fermi}) based on the standard perturbation theory to obtain the effective Hamiltonians at the near-resonant points in charge of the desired Rabi oscillations, through which we can generate the Bell states of the photon-magnon subsystem and the GHZ states of the whole qubit-photon-magnon system. For each effective Hamiltonian, a pair of the effective coupling strength $g_{\rm eff}$ and the energy shift $\delta$ in the leading order of $g$ and $G$ can be analytically derived. In Appendices~\ref{appa} and \ref{appb}, we will benchmark the ranges of validity of the coupling strengths $g,G$ by comparing the analytical results of $g_{\rm eff}$ and $\delta$ by the effective Hamiltonian and their numerical simulation over the whole Hilbert space.

\section{Generating Bell state of photon-magnon subsystem}

\subsection{Effective Hamiltonian}\label{bell}

In this section, we propose a protocol to generate the double-excitation Bell state of photon-magnon subsystem. The basic mechanism in this protocol is similar to the three-wave mixing schemes~\cite{threewave1,threewave2} in the nonlinear quantum optics. At a special point in the parametric space, the de-excitation of the superconducting qubit gives rise to a magnon-photon pair. State transfer occurs by the Rabi oscillation  $|e00\rangle\equiv|e\rangle_q|0\rangle_a|0\rangle_m\leftrightarrow|g11\rangle$, where $|e00\rangle$ and $|g11\rangle$ are two eigenstates of the free Hamiltonian $H_0$ in Eq.~(\ref{Hamiltonian}).

\begin{figure}[htbp]
\centering
\includegraphics[width=0.45\textwidth]{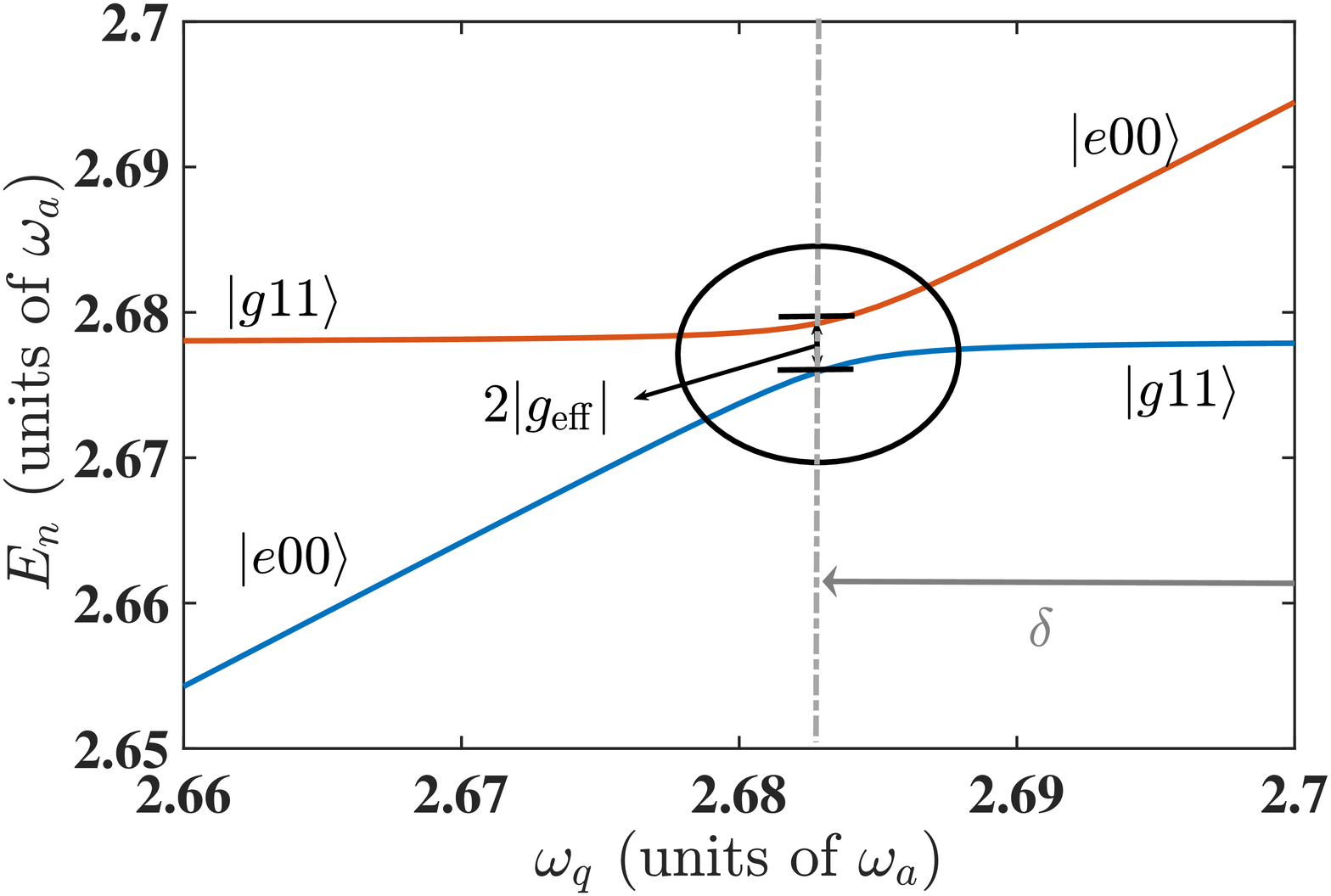}
\caption{Energy levels and the avoided level crossing for the states $|e00\rangle$ and $|g11\rangle$, normalized by $\omega_a$ and plotted as functions of the qubit transition frequency $\omega_q$. The avoided level crossings of these two eigenstates are distinguished by the dark circle. Here the parameters are fixed as $\omega_m=1.7\omega_a$, $\theta=\pi/4$, and $g=G=0.1\omega_a$. }\label{e1p1m1eigen}
\end{figure}

As shown in Fig.~\ref{e1p1m1eigen}, the interaction Hamiltonian $V$ in Eq.~(\ref{Hamiltonian}) shifts the point of the avoided level crossing for $|e00\rangle$ and $|g11\rangle$ from the exact double-resonant point by $\delta$, which satisfies $\omega_q=\omega_a+\omega_m+\delta$. In Fig.~\ref{e1p1m1eigen}, the eigenvalues $E_n$ of both states are obtained by a standard numerical diagonalization over the full Hamiltonian in a truncated Hilbert space. A sufficiently large number of energy eigenstates have been used to ensure that the result is not significantly affected by the truncation. The avoided level crossing is distinguished with a black circle, demonstrating the nonlinear resonance between the states $|e00\rangle$ and $|g11\rangle$ with an effective transition rate $g_{\rm eff}$. This phenomenon can be well described by the effective Hamiltonian up to the leading-order process involving all the coupling strengths, provided that $g,G\ll\omega_a$, $\omega_m$, $|\omega_a-\omega_m|$. The derivation details to obtain the effective Hamiltonian are provided in Appendix~\ref{appa}. In the subspace spanned by $\{|e00\rangle, |g11\rangle\}$, we have
\begin{equation}\label{Heff1}
H_{\rm eff}=g_{\rm eff}\left(|e00\rangle\langle g11|+|g11\rangle\langle e00|\right),
\end{equation}
at the resonant point $\omega_q=\omega_a+\omega_m+\delta$. We find that the magnitudes of the energy shift
\begin{equation}\label{delta1}
\delta=-2G^2\cos^2\theta\left(\frac{1}{\omega_m}+\frac{1}{2\omega_a+\omega_m}\right)-\frac{2g^2}{\omega_a+\omega_m},
\end{equation}
and the coupling strength $g_{\rm eff}$
\begin{equation}\label{geff1}
g_{\rm eff}=\frac{2G^2g\sin(2\theta)}{\omega_m(\omega_a-\omega_m)},
\end{equation}
are in the second and third orders of the coupling strengths $g$ and $G$, respectively. $g_{\rm eff}$ achieves the maximum value when $\theta=\pi/4$. Then we stick to this sweet spot in the following.

\begin{figure}[http]
\centering
\includegraphics[width=0.45\linewidth]{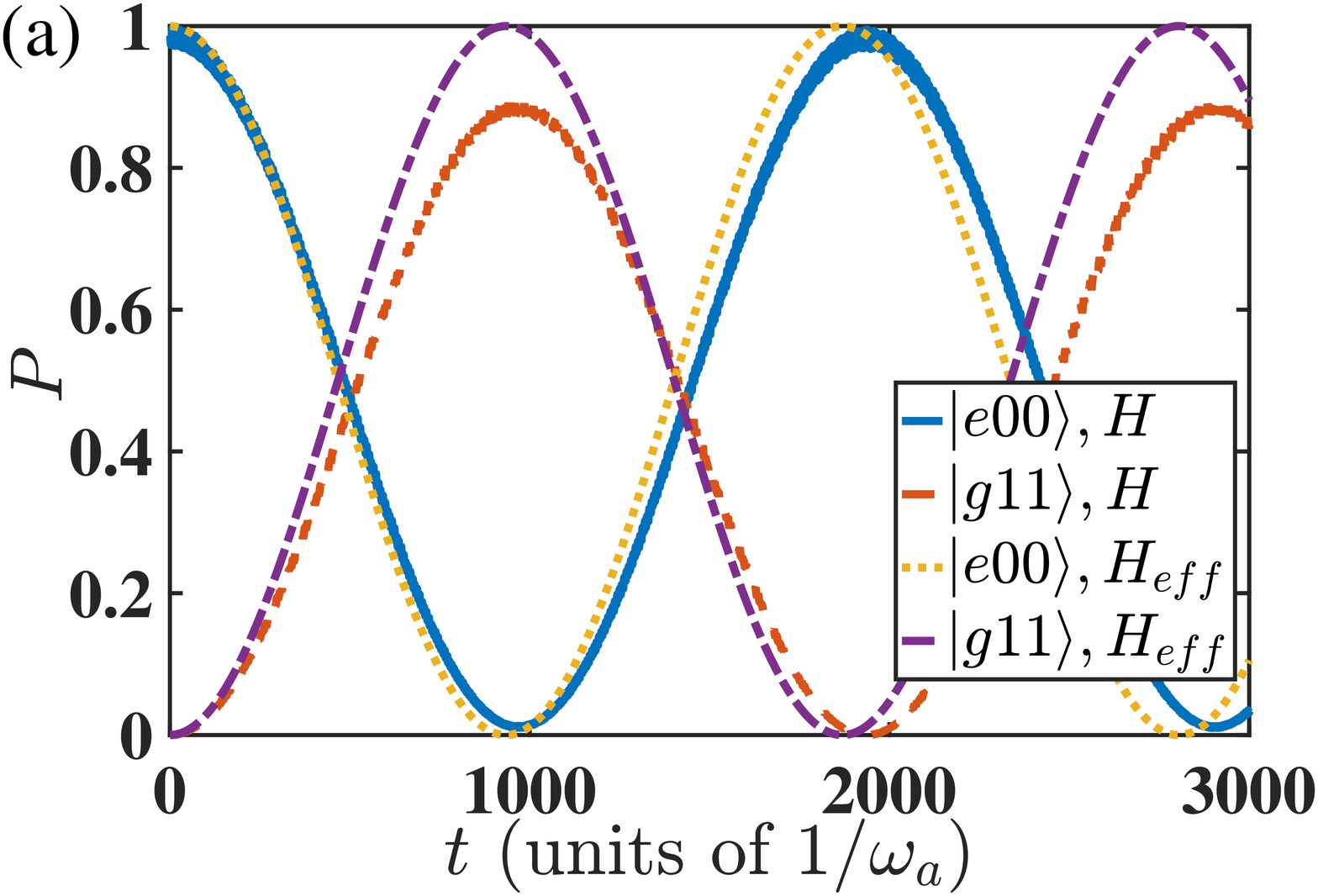}
\includegraphics[width=0.45\linewidth]{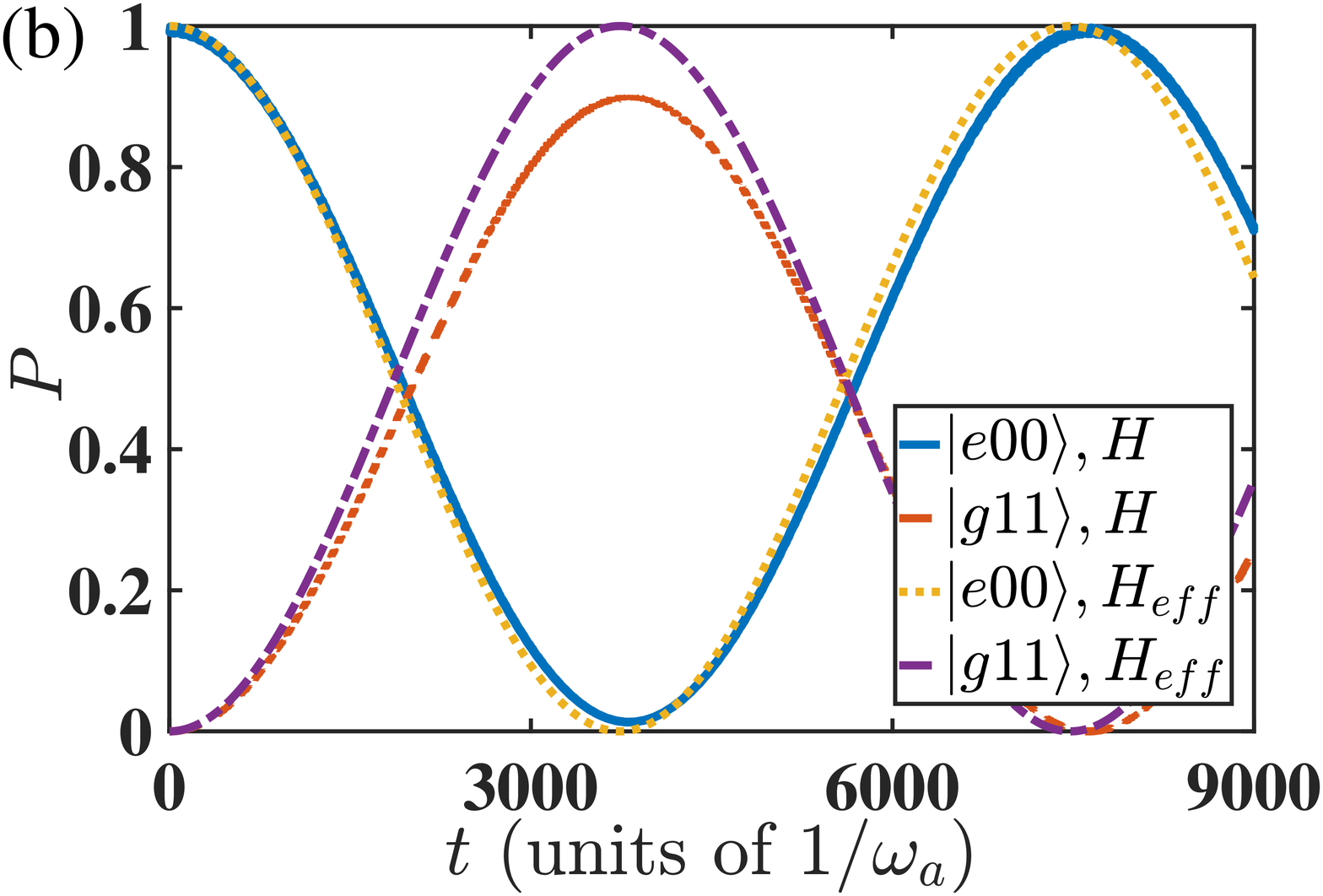}
\includegraphics[width=0.45\linewidth]{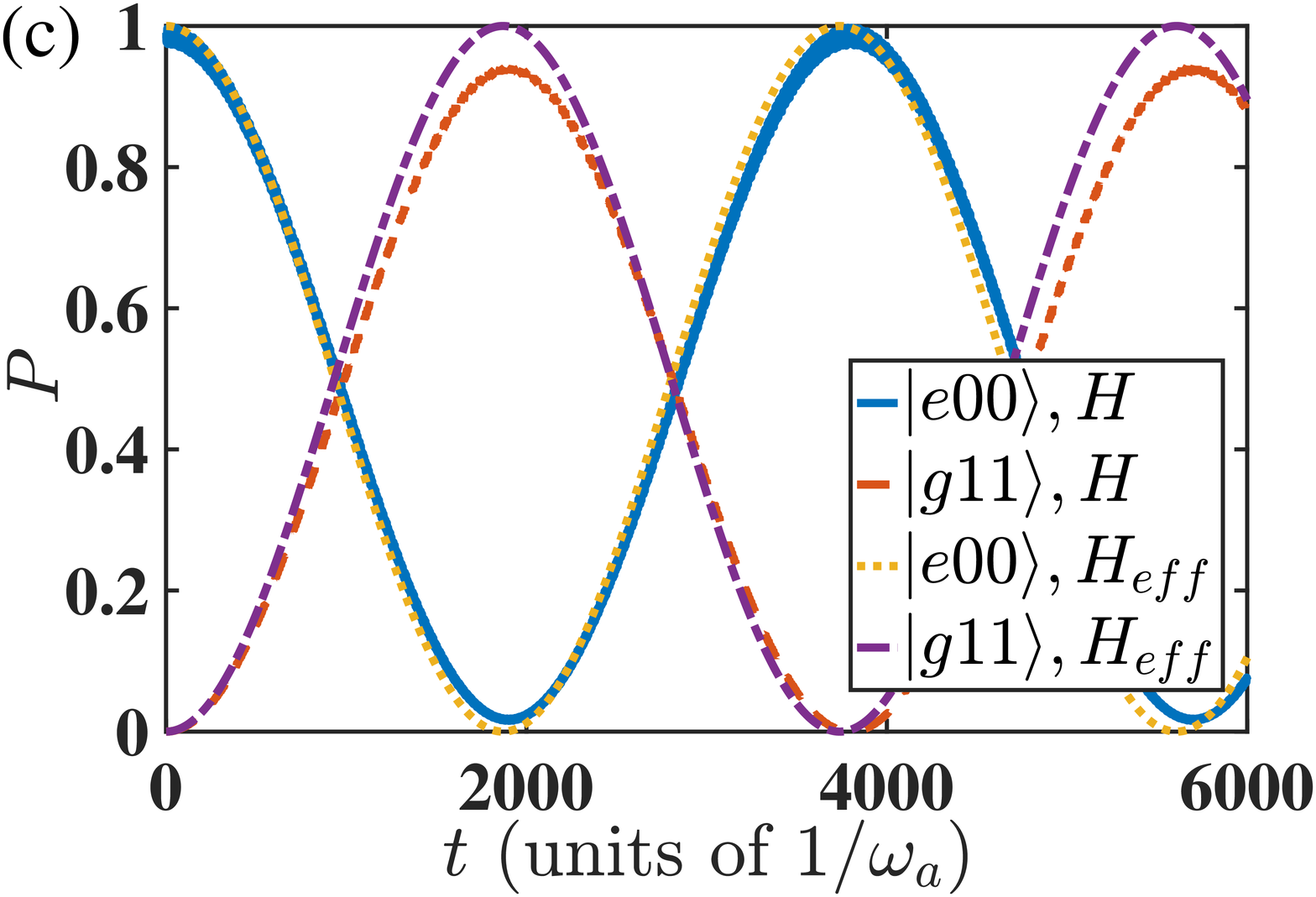}
\includegraphics[width=0.45\linewidth]{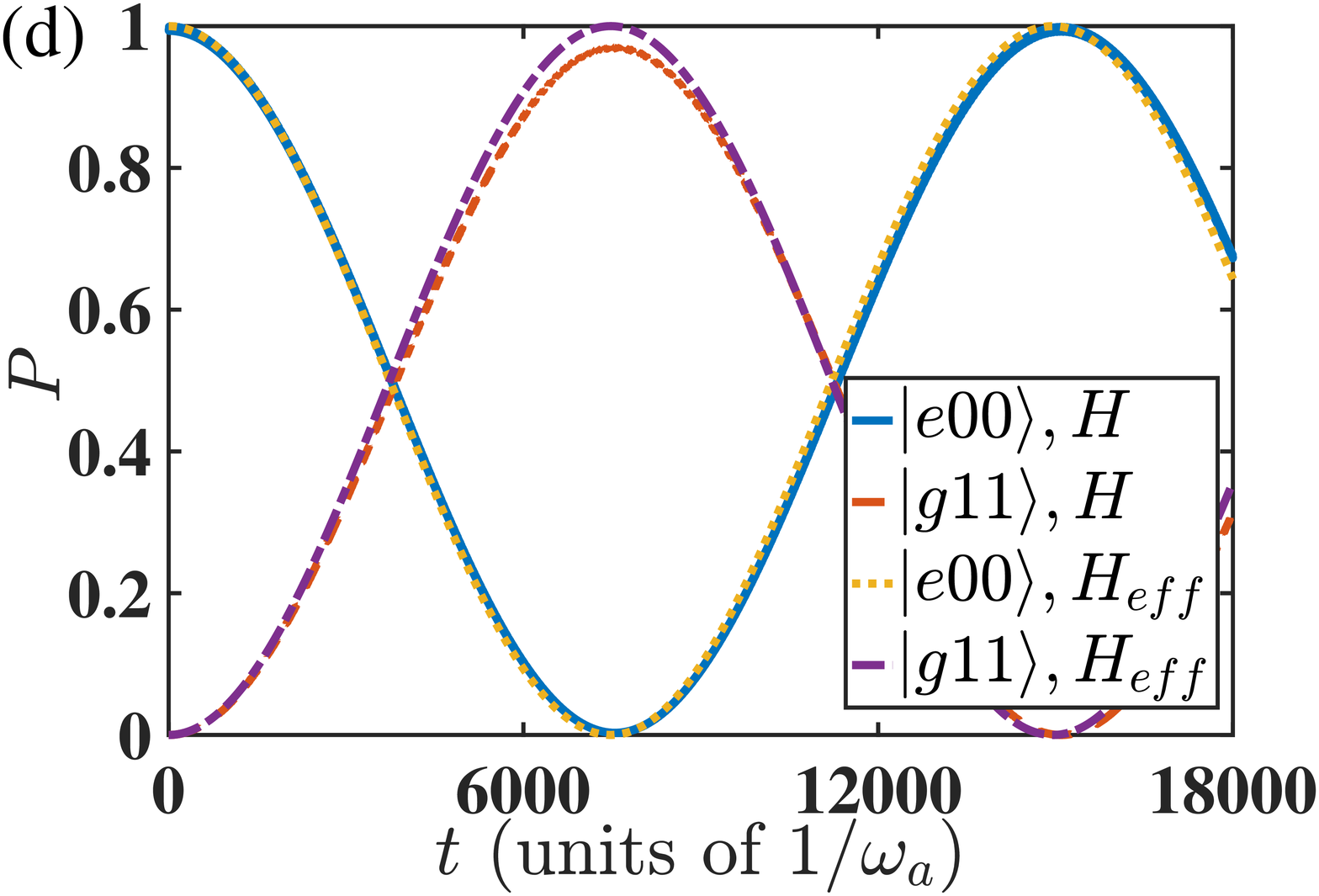}
\caption{Time evolution of the state population of the initial state $|e00\rangle$ and the target state $|g11\rangle$ under various coupling strengths to show the difference between the effective Hamiltonian and the full Hamiltonian at the avoided-level-crossing point shown in Fig.~\ref{e1p1m1eigen}. (a) $g=G=0.1\omega_a$, (b) $g=0.1\omega_a, G=0.05\omega_a$, (c) $g=0.05\omega_a, G=0.1\omega_a$, and (d) $g=G=0.05\omega_a$. }\label{e1p1m1ut}
\end{figure}

Under the effective Hamiltonian in Eq.~(\ref{Heff1}), a completed Rabi oscillation between the states $|e00\rangle$ and $|g11\rangle$ could be accurately established with a period of $\tau=\pi/|g_{\rm eff}|$ as demonstrated by the yellow dotted lines and the purple dash-dotted lines, respectively, in Fig.~\ref{e1p1m1ut}. We also plot the time evolutions under the full Hamiltonian $H$ in Eq.~(\ref{Hamiltonian}), where the blue solid lines and the red dashed lines represent the state-populations of $|e00\rangle$ and $|g11\rangle$, respectively. Note all the Rabi oscillations are ideal by the effective Hamiltonian, while the high-order effect emerges in the numerical evaluations by the full Hamiltonian.

Figures~\ref{e1p1m1ut}(a), (b), (c), and (d) vary with decreasing coupling strengths of $g$ and $G$ while fixing the other parameters. In the first period of the Rabi oscillation, one can observe that the maximum population of the state $|g11\rangle$ is about $P_{\rm max}=0.88$, $0.91$, $0.95$, and $0.98$, respectively under these four situations (see the red-dashed lines). The practical period of the Rabi oscillation between $|e00\rangle$ and $|g11\rangle$ becomes longer with smaller coupling strengths of $g$ and $G$ and it is slightly greater than that determined by the effective Hamiltonian. Similar to the maximum population, it is found that the generation time of the target state $\tau/2$ approaches also the ideal result by the effective Hamiltonian with decreasing coupling strengths. The relative errors of the period are about $2.7\%$, $4.3\%$, $0.4\%$, and $0.3\%$ for Figs.~\ref{e1p1m1ut}(a), (b), (c), and (d), respectively. Roughly, the reduction of $g$ has more impacts than that of $G$. In general, we can estimate from the four sub-figures that the analytical results from $H_{\rm eff}$ become gradually close to the numerical results from $H$ by decreasing the original coupling strengths $g$ and $G$. It is interesting to find that the range of validity of the effective Hamiltonian has approached the ultrastrong regime with $g=G=0.05\omega_a$.

\subsection{Generation protocol for Bell state}\label{Bellstep}

With the effective Hamiltonian in Eq.~(\ref{Heff1}), one can generate the Bell state of the photon-magnon subsystem by the following two-step protocol.

Step-$1$: The qubit frequency is adjusted to be far-off-resonant with both the resonator and the magnon modes (Note the latter two modes has already been set to be off-resonant). The whole hybrid system is prepared in the ground state of the free Hamiltonian, i.e., $|g\rangle_q|0\rangle_a|0\rangle_m=|g\rangle|00\rangle$. Then we perform a single-qubit gate operation $U$ on the qubit~\cite{onegate} and the operator reads,
\begin{equation}\label{U}
U=e^{i\frac{\pi}{4}\vec{\sigma}\cdot\vec{n}}=\begin{bmatrix}
1 & ie^{-i\phi}\\
ie^{i\phi} & 1
\end{bmatrix},
\end{equation}
where $\vec{\sigma}\equiv(\sigma_x, \sigma_y, \sigma_z)$ and $\vec{n}=(\cos\phi, \sin\phi, 0)$. It rotates the qubit into a superposed state
\begin{equation}\label{psi0s}
\frac{1}{\sqrt{2}}\left(|g\rangle+ie^{i\phi}|e\rangle\right),
\end{equation}
where the phase $\phi$ is tunable as desired and determines the final local phase of the double-excitation Bell states. For example, when $\phi=0,\pi$, the generated Bell state is $(|00\rangle\pm|11\rangle)/\sqrt{2}$.

Step-$2$: Thus the state of the whole system is now
\begin{equation}\label{psit0}
|\varphi(0)\rangle=\frac{1}{\sqrt{2}}\left(|g00\rangle+ie^{i\phi}|e00\rangle\right).
\end{equation}
Then we tune the qubit frequency to be nearly-resonant with the sum of the frequencies of the photon and magnon modes in an adiabatic way. As we shown in Eq.~(\ref{Heff1}), it will create an effective transition rate $g_{\rm eff}$ between the states $|e00\rangle$ and $|g11\rangle$, and the state $|g00\rangle$ is unaffected. The system state then evolves with time as
\begin{equation}\label{psit}
\begin{aligned}
|\varphi(t)\rangle&=\frac{1}{\sqrt{2}}\big[|g00\rangle+ie^{i\phi}\cos(g_{\rm eff}t)|e00\rangle\\
&+e^{i\phi}\sin(g_{\rm eff}t)|g11\rangle\big].
\end{aligned}
\end{equation}
After a time $T=\pi/(2|g_{\rm eff}|)$ (one half of the Rabi oscillation), the state evolves to
\begin{equation}\label{psiTT}
\begin{aligned}
|\varphi(T)\rangle&=\frac{1}{\sqrt{2}}\left(|g00\rangle+e^{i\phi}|g11\rangle\right)\\
&=|g\rangle\otimes\frac{1}{\sqrt{2}}\left(|00\rangle+e^{i\phi}|11\rangle\right).
\end{aligned}
\end{equation}
Then we tune the qubit-frequency faraway from the preceding point of the avoided level crossing and the Bell state $(|00\rangle+e^{i\phi}|11\rangle)/\sqrt{2}$ of the photon-magnon subsystem can be survival due to the large detuning between the resonator and the magnon mode.

\subsection{The fidelity of Bell state under dissipation}\label{Bellnum}

The fidelity of the generated state can be studied using the master equation approach by taking into account the dissipations from all parts of the hybrid system. By applying the standard Markovian approximation to the individual external environments (assumed to be at the vacuum states), we arrive at the Lindblad master equation for the density operator $\rho(t)$ of the hybrid system consisting of the qubit, the resonator and the magnon mode~\cite{bellstate},
\begin{equation}\label{lindblad}
\begin{aligned}
\dot{\rho}(t)&=-i[H_{\rm diag}, \rho]\\
&+\kappa_a\mathcal{L}[X_a]\rho(t)+\kappa_m\mathcal{L}[X_m]\rho(t)+\gamma\mathcal{L}[S_-]\rho(t).
\end{aligned}
\end{equation}
Here $H_{\rm diag}$ indicates that the full Hamiltonian $H$ is now expressed by its eigenvectors $|E_n\rangle$'s. $\kappa_a$, $\kappa_m$, and $\gamma$ are the dissipation rates for the resonator, the magnon, and the qubit, respectively. And the superoperator $\mathcal{L}[O]$ is defined as
\begin{equation}\label{L}
\mathcal{L}[O]\rho\equiv\frac{1}{2}\left(2O\rho O^\dag-O^\dag O\rho-\rho O^\dag O\right).
\end{equation}
Here $O=X_a, X_m, S_-$ are the dressed lowering operators, defined respectively in terms of their bare counterparts $o=a, m, \sigma_-$ as
\begin{equation}\label{O}
O\equiv\sum_{E_n>E_m}\langle E_m|(o+o^{\dag})|E_n\rangle|E_m\rangle\langle E_n|.
\end{equation}
To simplify the robustness estimation of our protocol but with no loss of generality, we assume all of the decoherence rates to be in the same order of magnitude $\kappa_a=\kappa_m=\gamma=\kappa$. It is consistent with the relative decoherence rates obtained in recent experiments~\cite{magnonqubit,magnonqubit2,ultrastrong}, i.e., $\kappa_a/\omega_a\sim 10^{-6}-10^{-5}$, $\kappa_m/\omega_m\sim 10^{-6}-10^{-5}$, and $\gamma/\omega_q\sim 10^{-6}-10^{-5}$.

\begin{figure}[htbp]
\centering
\includegraphics[width=0.4\textwidth]{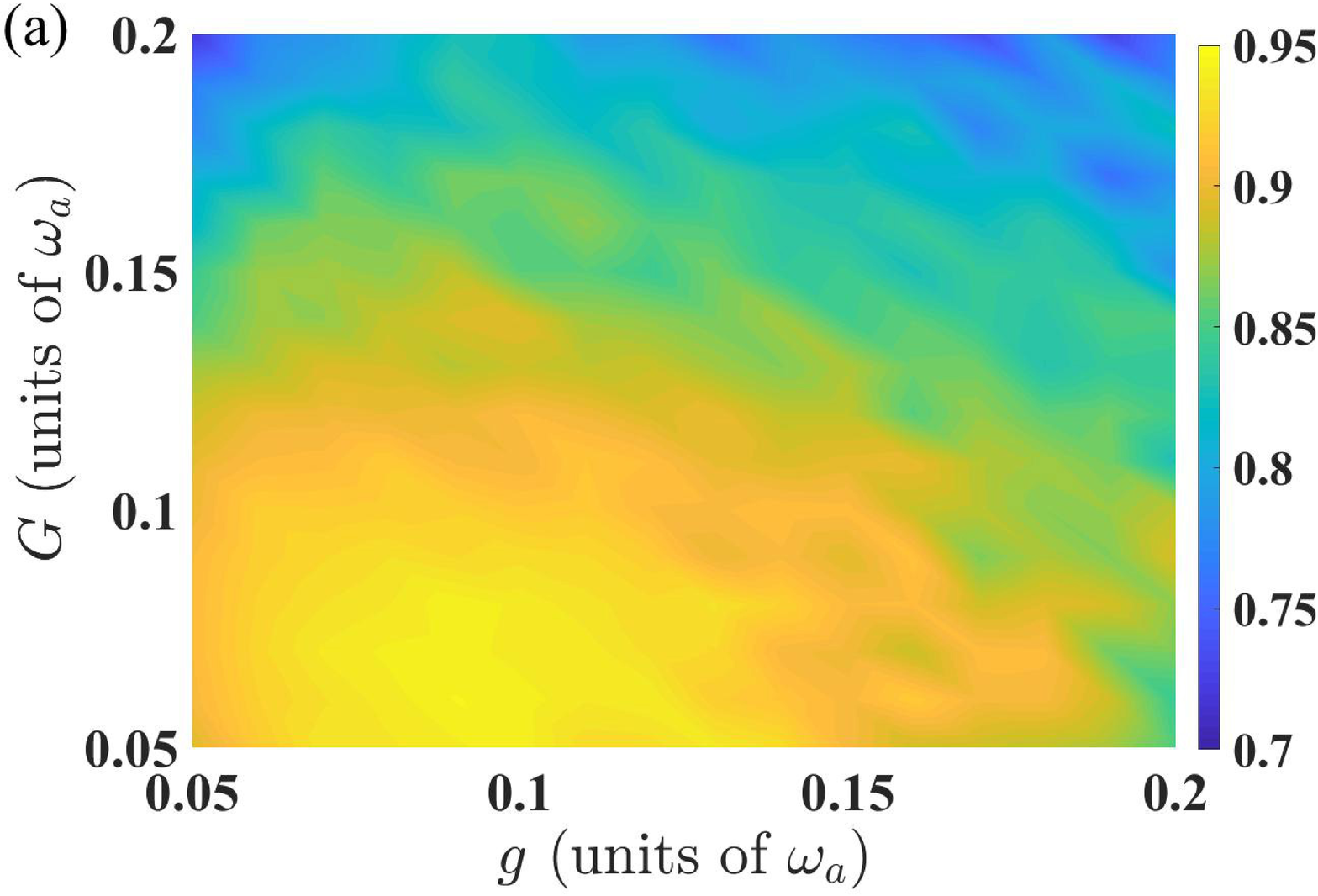}
\includegraphics[width=0.4\textwidth]{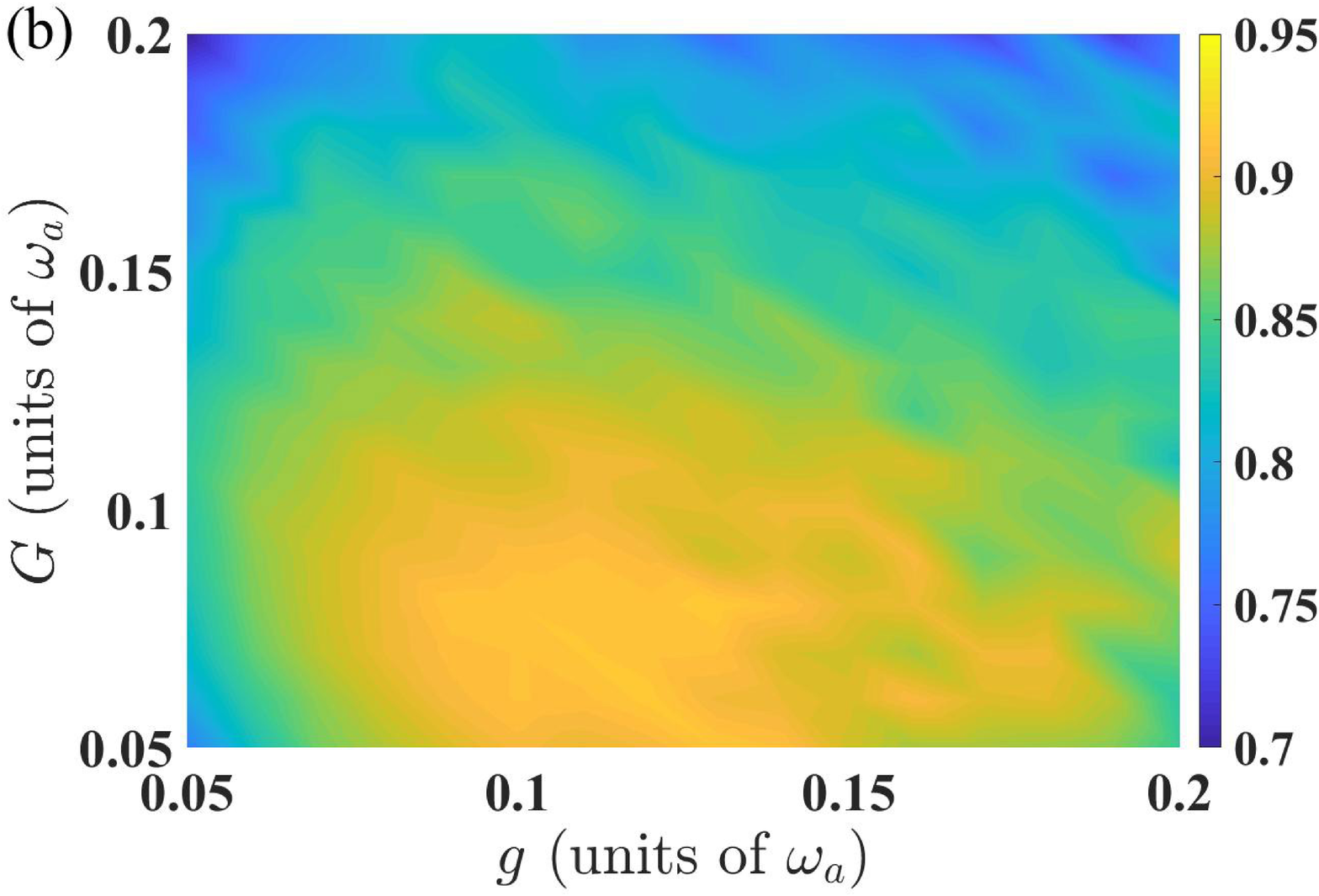}
\caption{The final fidelity $F(T)$ in the parametric space of the coupling strengths $g$ and $G$ under (a) $\kappa_a=\kappa_m=\gamma=0$ and (b) $\kappa_a=\kappa_m=\gamma=10^{-5}\omega_a$. The other parameters are fixed as $\omega_m=1.7\omega_a$ and $\theta=\pi/4$.}\label{e1p1m1master}
\end{figure}

The state-generation fidelity $F$ is defined as $F(t)=\langle\varphi(t)|\rho|\varphi(t)\rangle$, where $|\varphi(T)\rangle$ is the target state. Here the phase $\phi$ is set as zero, then $|\varphi(T)\rangle=(|00\rangle+|11\rangle)/\sqrt{2}$.

In Fig.~\ref{e1p1m1master}(a) and (b), we demonstrate and compare the generation fidelities $F(T)$ of the Bell state with no decoherence (under the full Hamiltonian rather than the effective Hamiltonian) and that with dissipation (under the Lindblad master equation). To be consistent with the dynamics in Fig.~\ref{e1p1m1ut}, it is shown in Fig.~\ref{e1p1m1master}(a) that the generation fidelity is enhanced when the coupling strengths $g$ and $G$ are reduced. One can see that the fidelity approaches $0.95$ when both the coupling strengths $g$ and $G$ are reduced to about $0.05\omega_a$. This observation supports again the validity of our effective Hamiltonian. In contrast, a smaller coupling strength does not always give rise to a higher fidelity. As shown in Fig.~\ref{e1p1m1master}(b), it is found that the generation fidelity is about $0.77$ under the coupling strengths $g=G=0.05\omega_a$, and about $0.90$ when $g=G=0.1\omega_a$. A compromise in terms of the coupling strength in fidelity is expected due to the fact that the period of the desired Rabi oscillation is inversely proportional to the coupling strengths and the dissipation becomes more destructive under a longer evolution time. A complete scanning over the parametric space demonstrates a remarkable working regime for the generation of the Bell state: $0.07<g/\omega_a<0.13$ and $0.05<G/\omega_a<0.1$, where we have $F>0.9$.

\begin{figure}[htbp]
\centering
\includegraphics[width=0.45\textwidth]{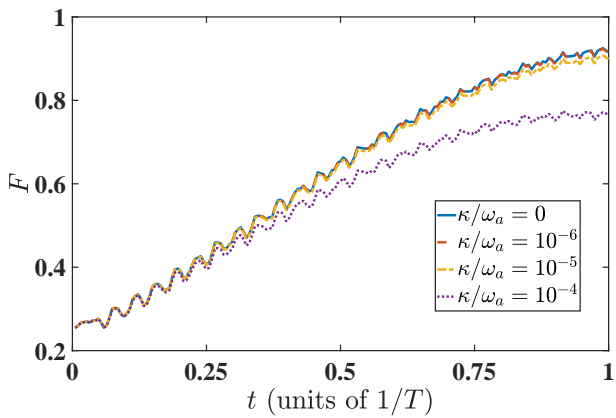}
\caption{Time evolution of the fidelity by the master equation~(\ref{lindblad}) under different dissipation rates. The other parameters are fixed as $\omega_m=1.7\omega_a$, $G=g=0.1\omega_a$, and $\theta=\pi/4$.}\label{e1p1m1lindblad}
\end{figure}

We then pick up a pair of $g$ and $G$ to plot the fidelity dynamics under different dissipation rates in Fig.~\ref{e1p1m1lindblad}. One can observe that the protocol works well for $\kappa\le10^{-5}\omega_a$, producing the desired Bell state with a fidelity over $0.90$, close to $0.93$ in the situation with no decoherence. The fidelity $F$ can be maintained above $0.78$ even when $\kappa$ is enhanced to $10^{-4}\omega_a$. It means that our protocol is robust even when all the decoherence channels of the hybrid system are simultaneously switched on.

\section{Generating GHZ state of qubit-photon-magnon system}

\subsection{Effective Hamiltonian}\label{GHZ}

This section is devoted to generating the GHZ state of the whole hybrid system, which shares the same key step or basic mechanism with the protocol for generating the Bell state in Sec.~\ref{bell}. At the desired point of the avoided level crossing, one can prepare the ``excited'' state $|e11\rangle\equiv|e\rangle_q|1\rangle_a|1\rangle_m$ from $|g20\rangle$, where the annihilation of one photon excites the qubit and the magnon mode simultaneously.

\begin{figure}[htbp]
\centering
\includegraphics[width=0.45\textwidth]{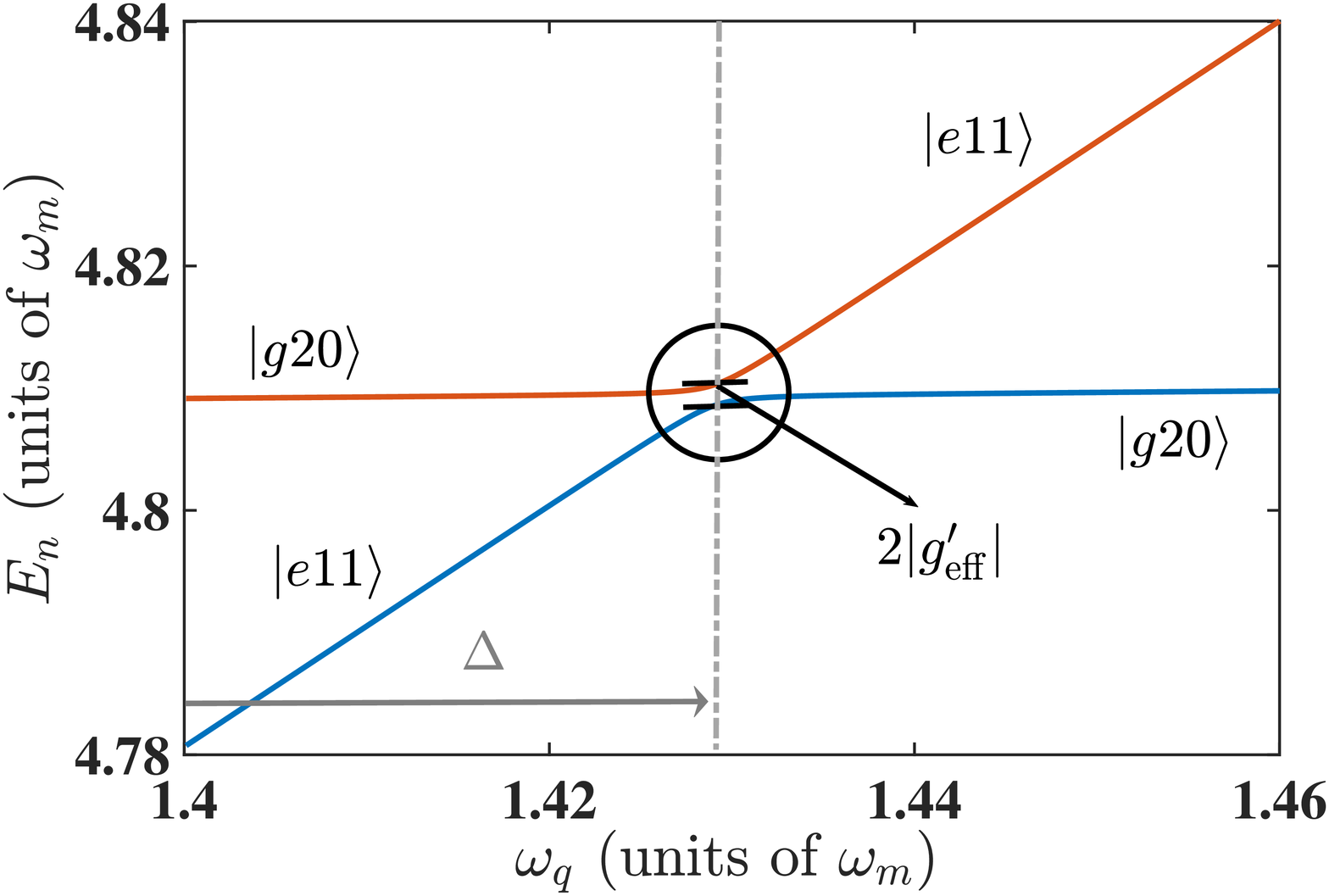}
\caption{Energy levels and the avoided level crossing for $|g20\rangle\leftrightarrow|e11\rangle$, normalized by $\omega_m$ and plotted as functions of the qubit transition frequency $\omega_q$. Here the parameters are fixed as $\omega_a=2.4\omega_m$, $g=G=0.1\omega_m$, and $\theta=\pi/4$.}\label{GHZeigen}
\end{figure}

In Fig.~\ref{GHZeigen}, the avoided level crossing for $|g20\rangle$ and $|e11\rangle$ is distinguished in the dark circle, demonstrating an effective transition rate $|g'_{\rm eff}|$. The energy shift $\Delta$ between the avoided-level-crossing point and the exact double-resonant point arises from the interaction Hamiltonian $V$ and then defined by $\omega_q=\omega_a-\omega_m+\Delta$. According to Appendix~\ref{appb}, the effective Hamiltonian reads,
\begin{equation}\label{Heff2}
H'_{\rm eff}=g'_{\rm eff}\left(|e11\rangle\langle g20|+|g20\rangle\langle e11|\right),
\end{equation}
with the effective coupling strength
\begin{equation}\label{geffghz}
g'_{\rm eff}=-\frac{\sqrt{2}G^2g\sin(2\theta)(\omega_a+3\omega_m)}{\omega_a\omega_m(\omega_a+\omega_m)}.
\end{equation}
And the energy shift reads
\begin{equation}\label{deltaghz}
\Delta=4G^2\cos^2\theta\left(\frac{1}{\omega_m}-\frac{1}{2\omega_a-\omega_m}\right)+\frac{2g^2}{\omega_a-\omega_m}.
\end{equation}

\begin{figure}[htbp]
\centering
\includegraphics[width=0.45\linewidth]{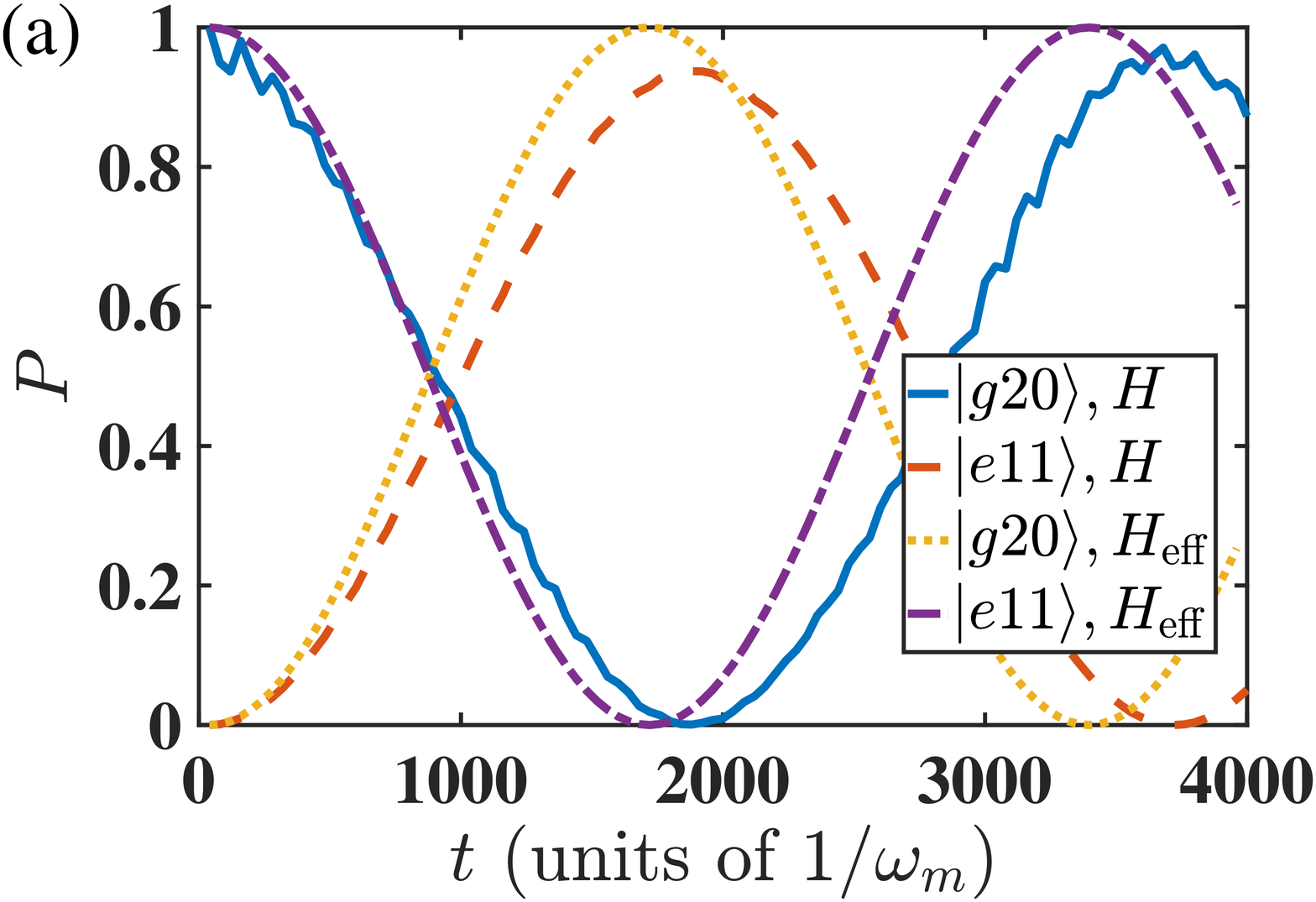}
\includegraphics[width=0.45\linewidth]{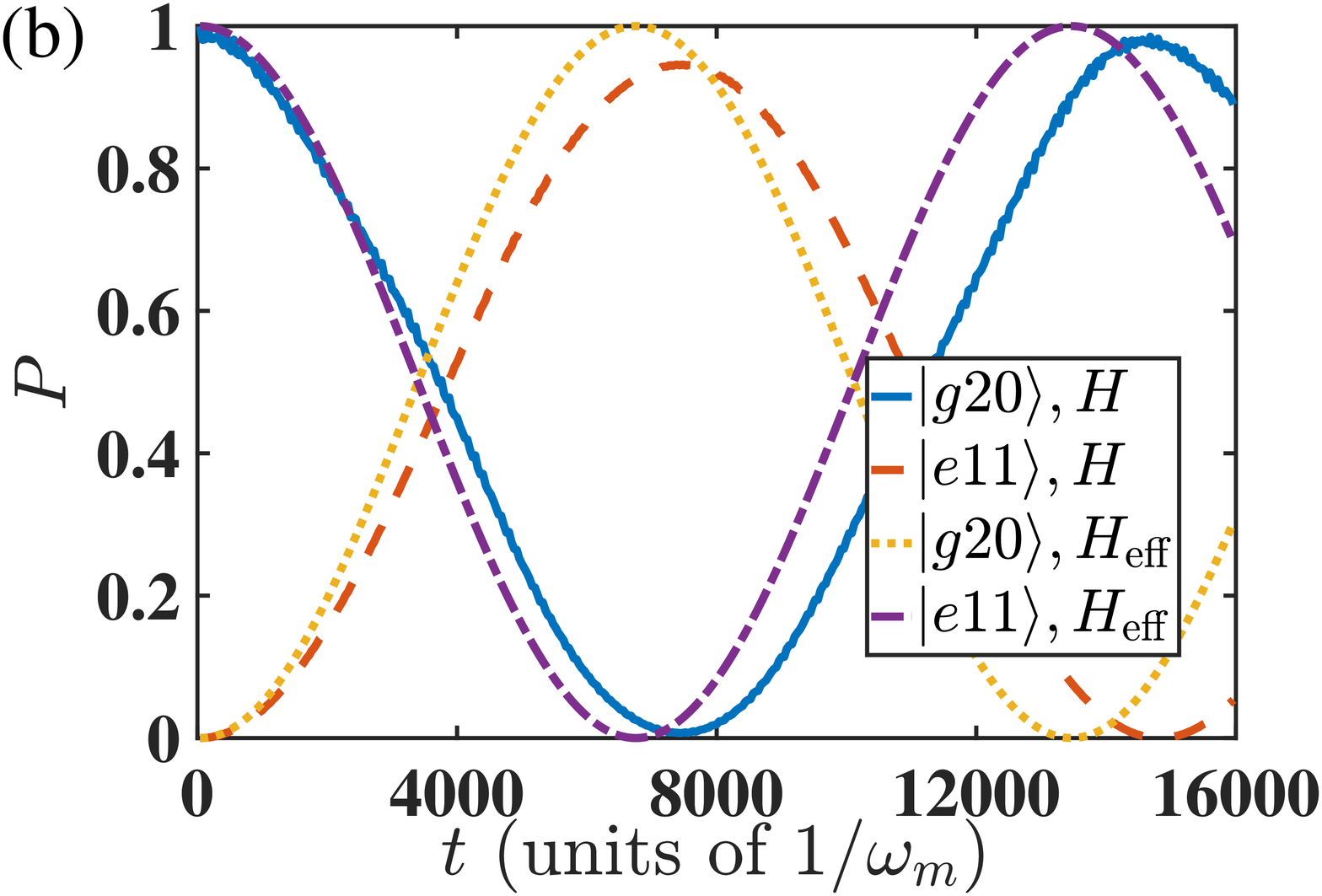}
\includegraphics[width=0.45\linewidth]{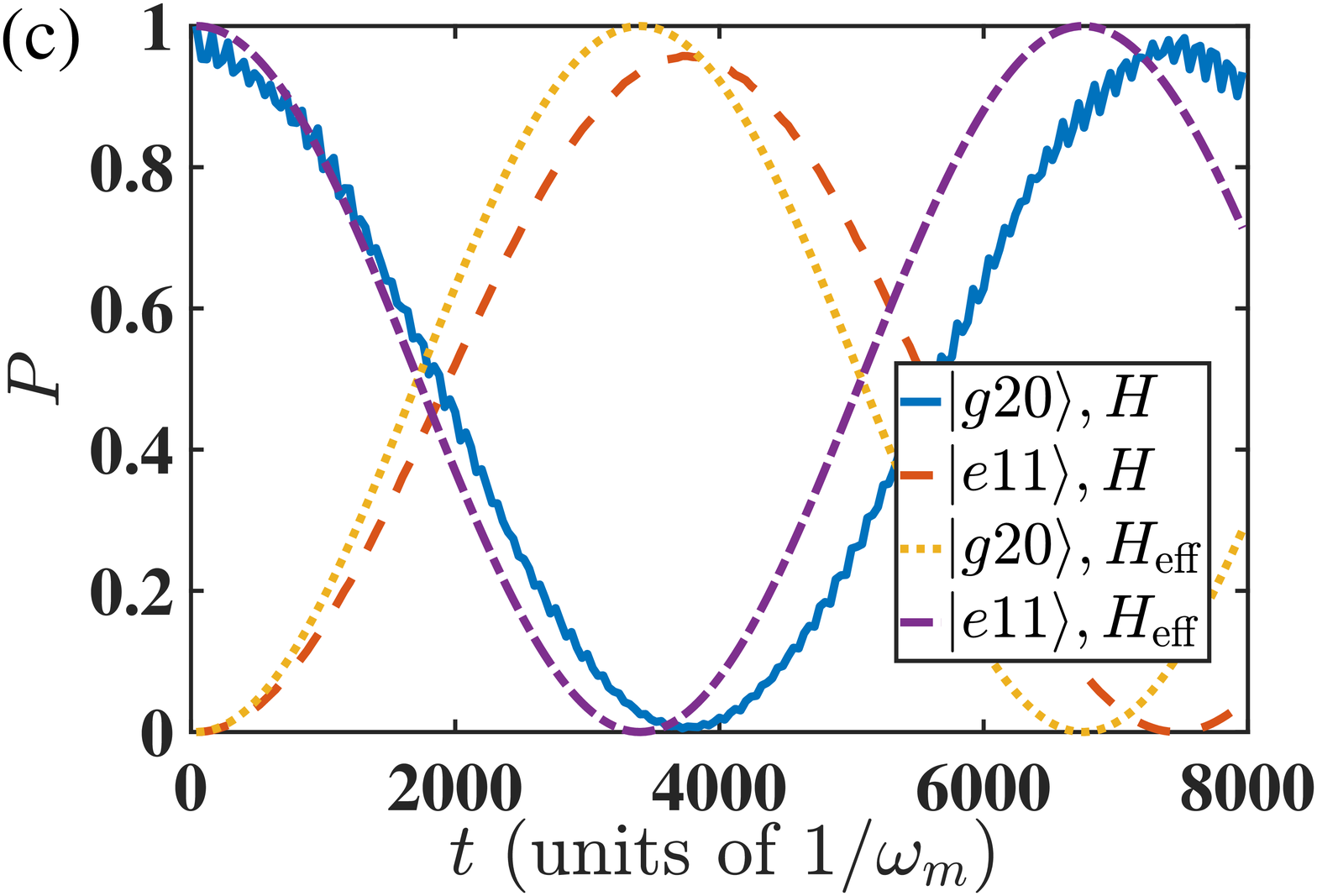}
\includegraphics[width=0.45\linewidth]{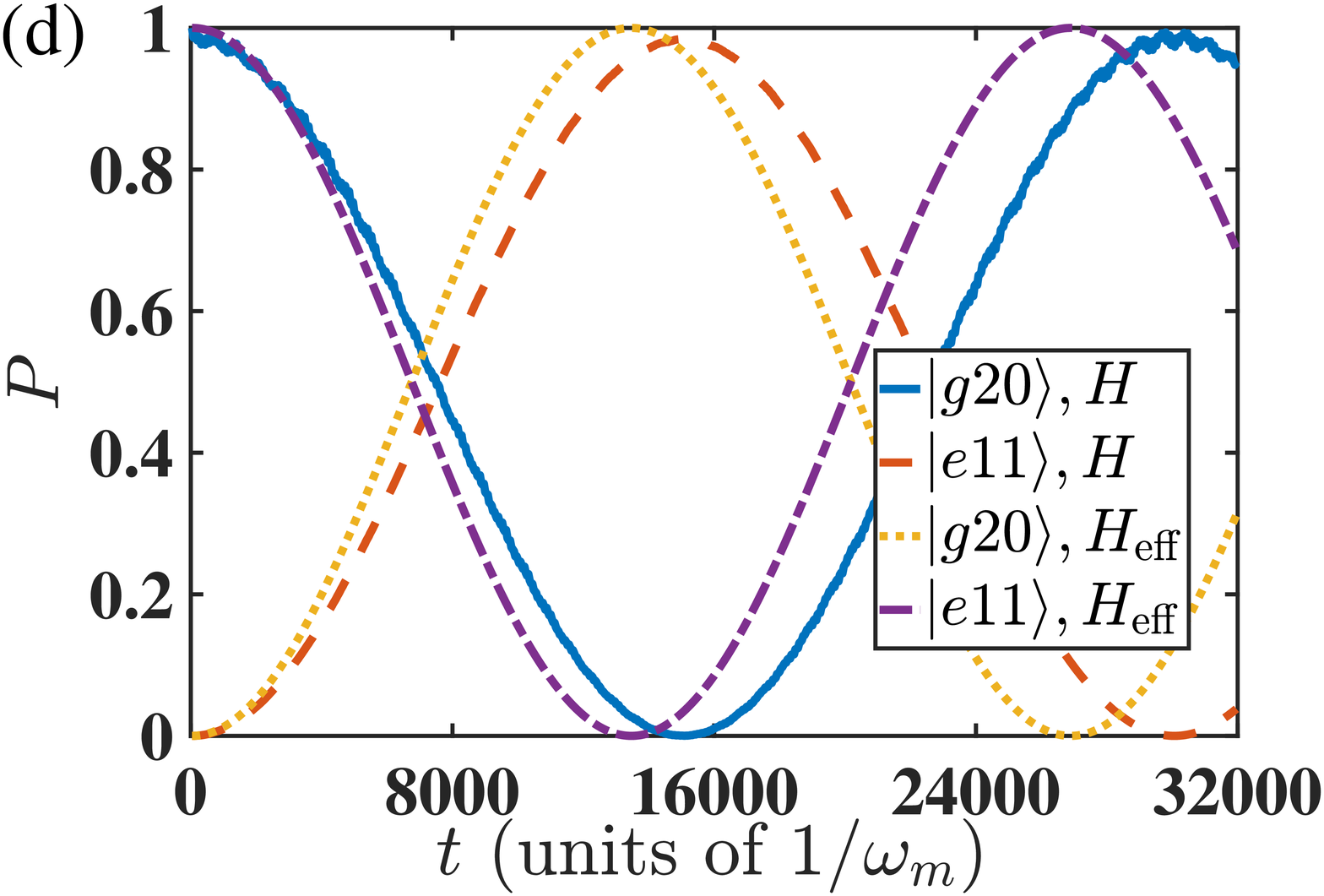}
\caption{Time evolution of the state population of the initial state $|g20\rangle$ and the target state $|e11\rangle$ under various coupling strengths to show the difference between the effective Hamiltonian and the full Hamiltonian at the avoided-level-crossing point shown in Fig.~\ref{GHZeigen}. (a) $g=G=0.1\omega_m$, (b) $g=0.1\omega_m, G=0.05\omega_m$, (c) $g=0.05\omega_m, G=0.1\omega_m$, and (d) $g=G=0.05\omega_m$. }\label{p1e1m1ut}
\end{figure}

Under the effective Hamiltonian in Eq.~(\ref{Heff2}), a completed Rabi oscillation between states $|g20\rangle$ and $|e11\rangle$ could be accurately established with a period of $\tau=\pi/|g'_{\rm eff}|$ as demonstrated by the yellow dotted lines and the purple dash-dotted lines, respectively in Fig.~\ref{p1e1m1ut}. To estimate the range of validity of  the effective Hamiltonian, we also present the time evolutions under the full Hamiltonian $H$ in Eq.~(\ref{Hamiltonian}), where the blue solid lines and the red dashed lines indicate the states populations for $|g20\rangle$ and $|e11\rangle$, respectively. In Figs.~\ref{p1e1m1ut}(a), (b), (c), and (d), one can observe that the maximum population of the state $|e11\rangle$ in the first period of Rabi oscillation is gradually enhanced with the decreasing coupling strengths of $g$ and $G$ while fixing the other parameters, i.e., $P_{\rm max}=0.94$, $0.95$, $0.96$, and $0.98$, respectively. While the relative errors of the period between the effective Hamiltonian and the total Hamiltonian have no clear dependence on the coupling strengths. It indicates that our estimation over the effective coupling rate $|g'_{\rm eff}|$ for the transition $|e11\rangle\leftrightarrow|g20\rangle$ in Fig.~\ref{GHZeigen} is not as accurate as that for the transition $|e00\rangle\leftrightarrow|g11\rangle$ in Fig.~\ref{e1p1m1eigen}.

\subsection{Generation protocol for GHZ state}\label{GHZstep}

With the effective Hamiltonian in Eq.~(\ref{Heff2}), one can generate the GHZ state for the whole qubit-photon-magnon system by the following three-step protocol.

Step-$1$: The transition frequency of the superconducting qubit is tuned to be far-off-resonant with those for both photon and magnon modes. And the whole system is initially at the ground state of the free Hamiltonian, i.e., in the state $|g\rangle_q|0\rangle_a|0\rangle_m=|g\rangle|00\rangle$. Then we rotate the qubit state into a superposed state
\begin{equation}\label{psi0sGHZ}
\frac{1}{\sqrt{2}}\left(|g\rangle-e^{i\phi}|e\rangle\right),
\end{equation}
with a single-qubit gate
\begin{equation}\label{UGHZ}
U=e^{i\frac{\pi}{4}\vec{\sigma}\cdot\vec{n}}=\begin{bmatrix}
1 & -e^{-i\phi}\\
-e^{i\phi} & 1
\end{bmatrix},
\end{equation}
where $\vec{n}=(\sin\phi, \cos\phi, 0)$. The phase $\phi$ is tunable as desired and determines the final local phase in the GHZ states $(|g00\rangle+e^{i\phi}|e11\rangle)/\sqrt{2}$.

Step-$2$: The state of the whole system is thus written as
\begin{equation}\label{psi0s}
|\varphi'(0)\rangle=\frac{1}{\sqrt{2}}\left(|g00\rangle-e^{i\phi}|e00\rangle\right).
\end{equation}
Then we tune the qubit frequency adiabatically into the nearly-two-photon resonance with the resonator mode, namely, $\omega_q=2\omega_a+\tilde{\delta}$. Note the magnon mode is far off-resonant from them. In this case, the full Hamiltonian of the system in Eq.~(\ref{Hamiltonian}) turns out to describe a two-photon Jaynes-Cummings model~\cite{bellstate},
\begin{equation}\label{Hefftwophoton}
\tilde{H}_{\rm eff}=\tilde{g}_{\rm eff}\left(|e00\rangle\langle g20|+|g20\rangle\langle e00|\right),
\end{equation}
where $\tilde{g}_{\rm eff}=-\sqrt{2}G^2\sin(2\theta)/\omega_a$ and $\tilde{\delta}=-2G^2/\omega_a-2g^2/(\omega_a+\omega_m)+2g^2/(\omega_a-\omega_m)$ could also be directly obtained by the high-order Fermi's Golden rule given in Eq.~(\ref{Fermi}) or the standard perturbation method in Eq.~(\ref{secondp}). The state $|g00\rangle$ is not influenced by $\tilde{H}_{\rm eff}$, and the state $|e00\rangle$ evolves with time as
\begin{equation}\label{psitau}
\begin{aligned}
|\varphi'(t)\rangle&=\frac{1}{\sqrt{2}}\big[|g00\rangle-e^{i\phi}\cos(\tilde{g}_{\rm eff}t)|e00\rangle\\
&+ie^{i\phi}\sin(\tilde{g}_{\rm eff}t)|g20\rangle\big].
\end{aligned}
\end{equation}
After a time $T'=\pi/|2\tilde{g}_{\rm eff}|$, the state becomes
\begin{equation}\label{psitaus}
|\varphi'(T')\rangle=\frac{1}{\sqrt{2}}\left(|g00\rangle+ie^{i\phi}|g20\rangle\right).
\end{equation}

Step-$3$: We then tune the qubit frequency into the near-resonance with the detuning between the photon and magnon modes, i.e., the avoided-level-crossing point shown in Fig.~\ref{GHZeigen}: $\omega_q=\omega_a-\omega_m+\Delta$. Then driven by Eq.~(\ref{Heff2}), the state of the hybrid system evolves as
\begin{equation}\label{psit}
\begin{aligned}
|\varphi'(T'+t)\rangle&=\frac{1}{\sqrt{2}}\big[|g00\rangle+ie^{i\phi}\cos(g'_{\rm eff}t)|g20\rangle\\
&+e^{i\phi}\sin(g'_{\rm eff}t)|e11\rangle\big].
\end{aligned}
\end{equation}
After a time $T=\pi/(2|g'_{\rm eff}|)$, it turns out to be
\begin{equation}\label{psiTT}
|\varphi'(T'+T)\rangle=\frac{1}{\sqrt{2}}\left(|g00\rangle+e^{i\phi}|e11\rangle\right).
\end{equation}
Then we tune the qubit faraway from the resonant point and the GHZ state $(|g00\rangle+e^{i\phi}|e11\rangle)/\sqrt{2}$ can be maintained for the transition frequencies of the three subsystems are now far-off-detuning from each other.

\subsection{The fidelity of GHZ state under dissipation}\label{GHZnum}

The generation fidelity of the GHZ state can be also studied using the standard Lindblad master equation~(\ref{lindblad}). The fidelity is defined as $F(t)=\langle\varphi(T'+t)|\rho|\varphi(T'+t)\rangle$ with $|\varphi(T'+T)\rangle=(|g00\rangle+|e11\rangle)/\sqrt{2}$ in Eq.~(\ref{psiTT}) and $\phi=0$.

\begin{figure}[htbp]
\centering
\includegraphics[width=0.4\textwidth]{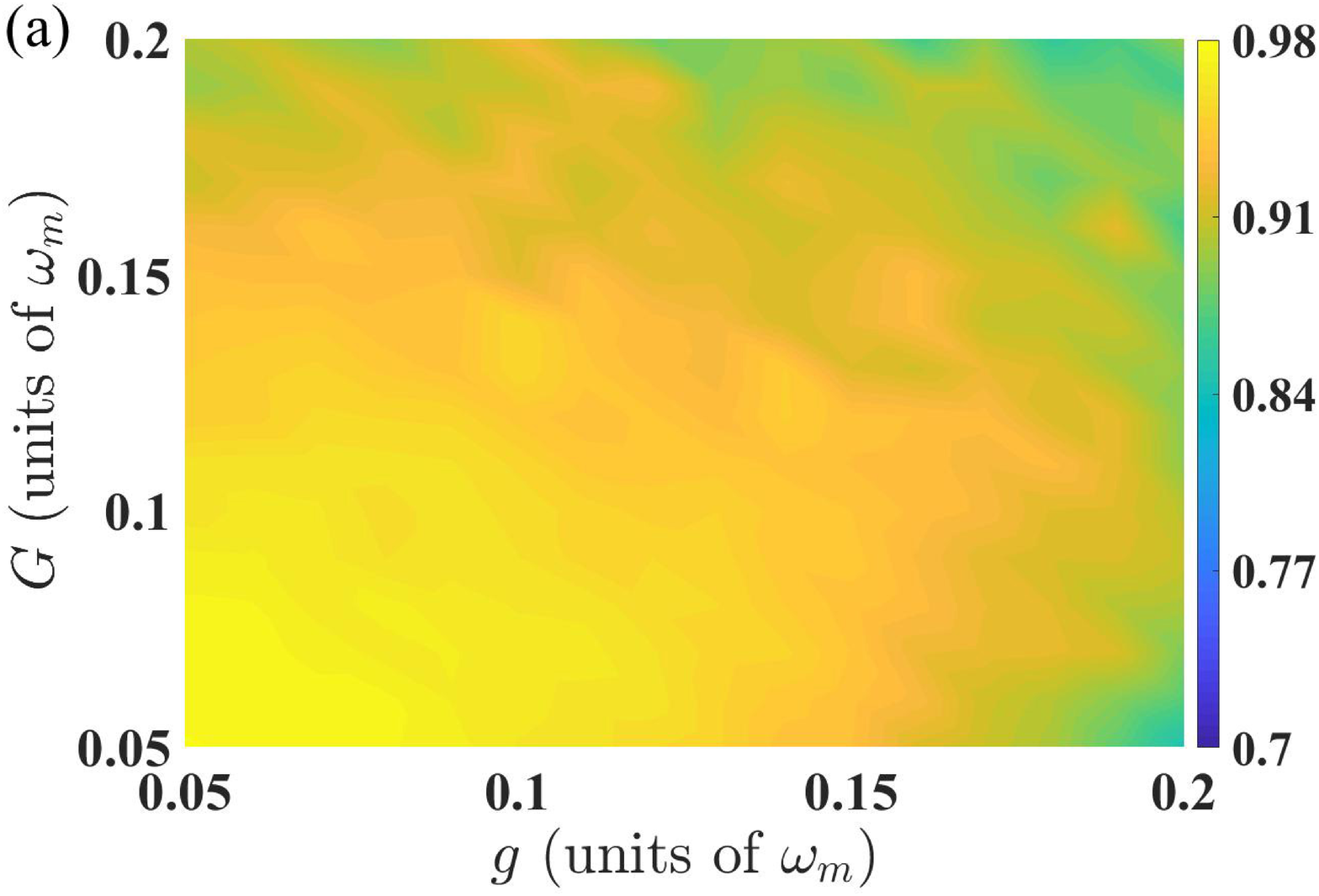}
\includegraphics[width=0.4\textwidth]{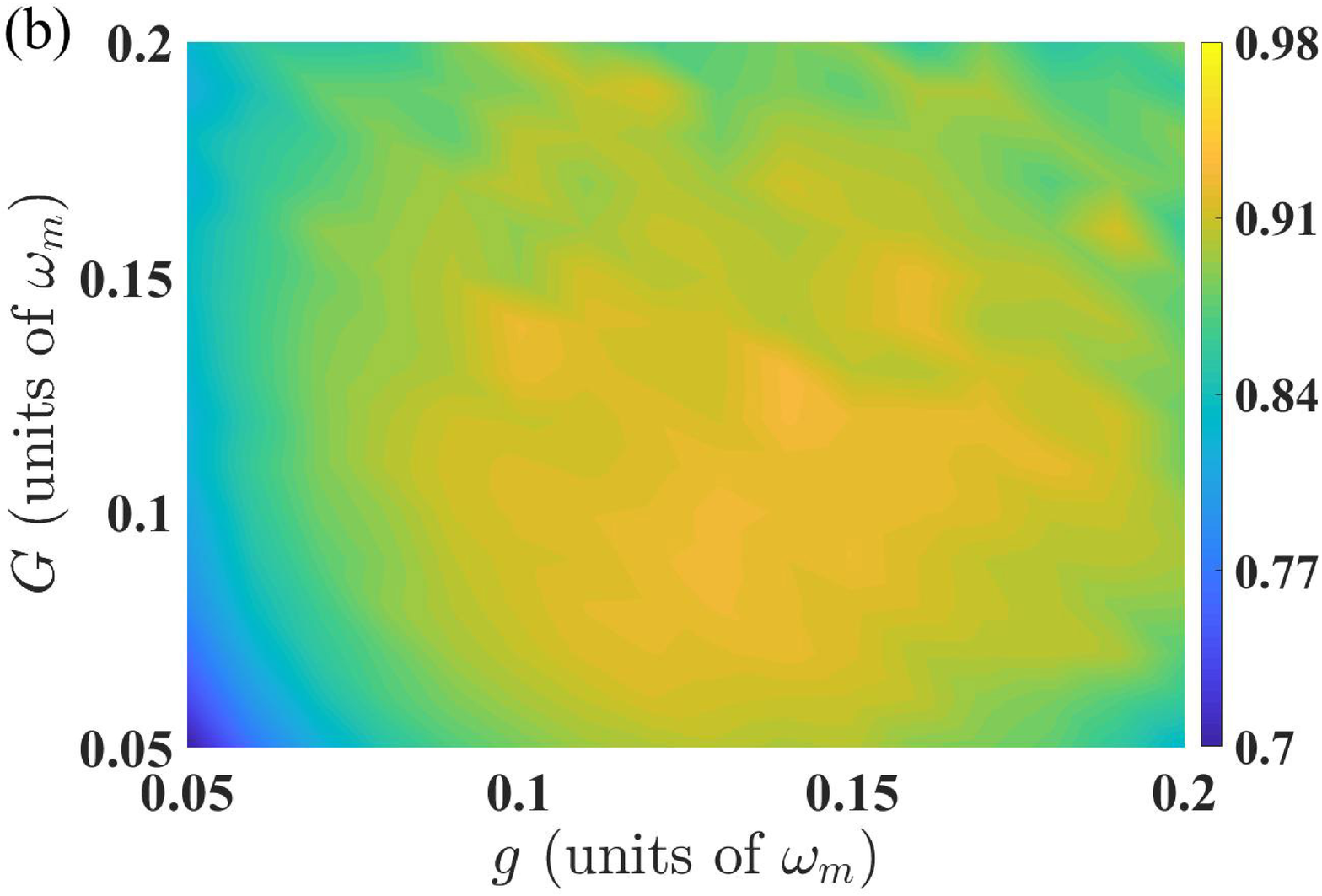}
\caption{The final fidelity $F(T)$ in the parametric space of the coupling strengths $g$ and $G$ under (a) $\kappa_a=\kappa_m=\gamma=0$ and (b) $\kappa_a=\kappa_m=\gamma=10^{-5}\omega_m$. The other parameters are fixed as $\omega_a=2.4\omega_m$ and $\theta=\pi/4$.}\label{fidelity}
\end{figure}

In Fig.~\ref{fidelity}(a) and (b), we demonstrate and compare the final fidelity $F(T)$ of the GHZ state with no decoherence and that under the simultaneous dissipations from qubit, resonator, and magnon. In Fig.~\ref{fidelity}(a), one can observe that a high-fidelity of state-generation can be maintained when the coupling strengths $g$ and $G$ are reduced, being consistent with the Rabi oscillations in Fig.~\ref{p1e1m1ut}. The fidelity is greater than $0.95$ for the coupling strengths $g, G<0.2\omega_m$ and is even close to $0.98$ when $g=0.05\omega_m$ and $G=0.05\omega_m$. This justifies our effective Hamiltonian in Eq.~(\ref{Heff2}). In the presence of the dissipation, however, the dependence of the fidelity on the coupling strengths in Fig.~\ref{fidelity}(b) is not monotonic as shown in Fig.~\ref{fidelity}(a). For example, when $g=G=0.05\omega_m$, the generation fidelity is about $0.70$; in contrast, when $g=G=0.1\omega_m$, it is over $0.95$. During the long-time evolution induced by small coupling strengths, the effect from decoherence on the fidelity will destroy the fidelity of the final GHZ state. When $0.1\le g/\omega_m\le0.17$ and $0.1\le G/\omega_m\le0.17$, we have an optimized regime with $F>0.9$.

\begin{figure}[htbp]
\centering
\includegraphics[width=0.45\textwidth]{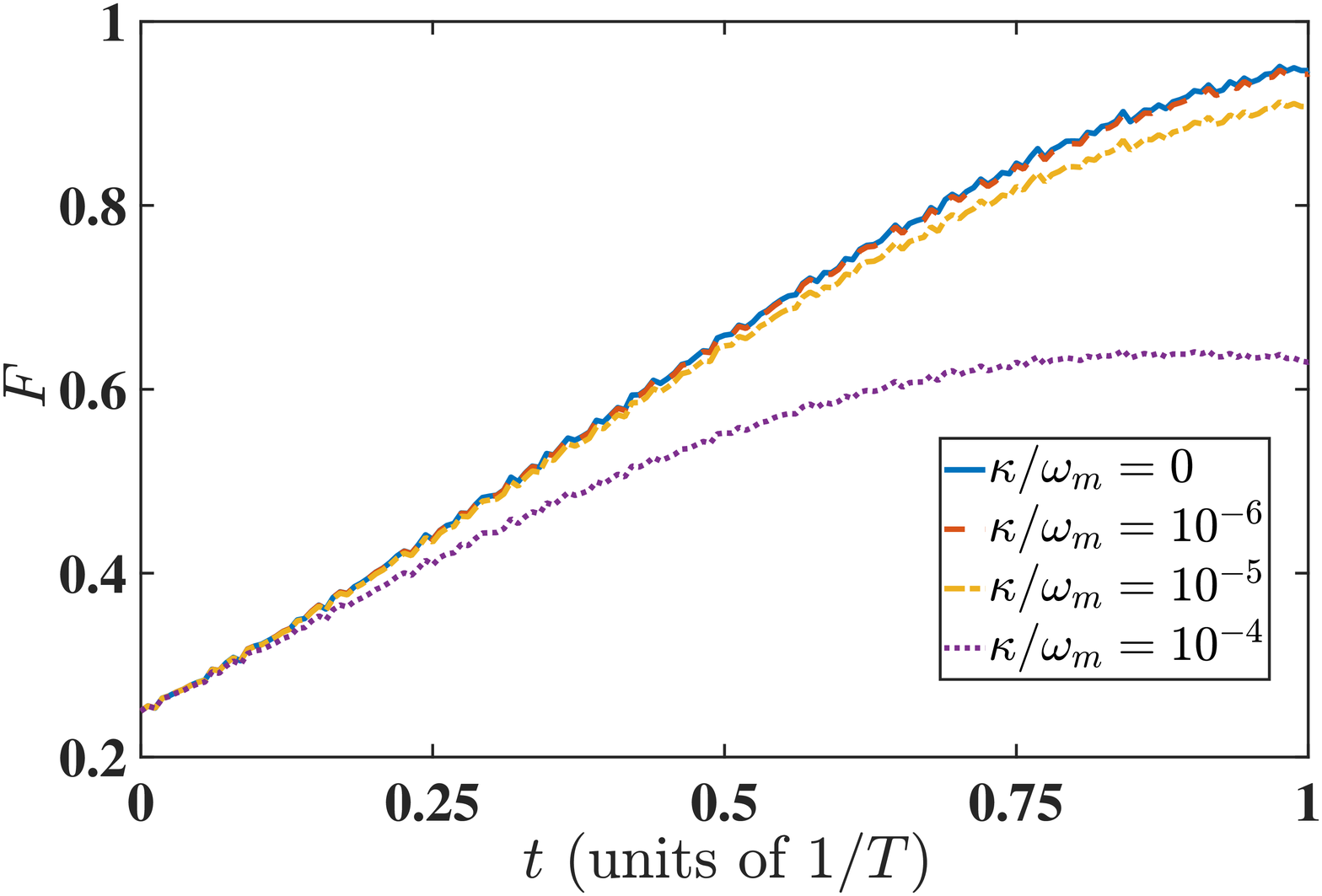}
\caption{Time evolution of the fidelity of the hybrid GHZ state under the master equation~(\ref{lindblad}) under different decoherence rates. Here the parameters are fixed as $\omega_a=2.4\omega_m$, $G=g=0.1\omega_m$ and $\theta=\pi/4$.}\label{GHZlindblad}
\end{figure}

In Fig.~\ref{GHZlindblad}, we present the dynamics of the generation fidelity with a fixed pair of $g$ and $G$. One can observe that the protocol for the hybrid system works well for $\kappa\le10^{-5}\omega_m$, producing a global GHZ state with a fidelity over $0.95$, close to $0.98$ in the ideal situation with no decoherence. It decreases to $0.62$ when $\kappa$ is enhanced to $10^{-4}\omega_m$. The global GHZ state is therefore less robust than the local Bell state.

\section{Discussion}\label{discussion}

After a minor modification, the protocol for GHZ state in Sec.~\ref{GHZ} can be immediately applied to generate the double-excitation Bell state of the qubit-magnon subsystem. It is still of a three-step scheme. Step-$1$ remains invariant, i.e., the whole system again starts from Eq.~(\ref{psi0sGHZ}):
\begin{equation}
|\varphi'(0)\rangle=\frac{1}{\sqrt{2}}\left(|g\rangle_q-e^{i\phi}|e\rangle_q\right)|0\rangle_a|0\rangle_m.
\end{equation}

Step-$2$: We now tune the qubit frequency to be single-photon resonant with the resonator mode instead of the double-photon resonance as in Sec.~\ref{GHZstep}. Around this point, the ground state $|g00\rangle$ holds and the state $|e00\rangle$ undergoes a Rabi oscillation with an effective transition rate $G$. Then after an evolution time $T'=\pi/(2G)$, the system state evolves into
\begin{equation}\label{psitaus}
|\varphi'(T')\rangle=\frac{1}{\sqrt{2}}\left(|g00\rangle+ie^{i\phi}|g10\rangle\right).
\end{equation}

Step-$3$: We then again tune the qubit frequency to satisfy $\omega_q=\omega_a-\omega_m+\Delta$, which is the avoided-level-crossing point for the states $|g10\rangle$ and $|e01\rangle$. The effective Hamiltonian becomes $g'_{\rm eff}(|g10\rangle\rangle e01|+|e01\rangle\rangle g10|)$. Although formally it seems almost the same as Eq.~(\ref{Heff2}), here $g'_{\rm eff}$ is yet not the same as Eq.~(\ref{geffghz}) since the transition paths as well as the transition rates connecting $|g10\rangle$ and $|e01\rangle$ are not the same as those connecting $|g20\rangle$ and $|e11\rangle$. In the current case, it is found that
\begin{equation}\label{deltabell2}
g'_{\rm eff}=-\frac{2G^2g\sin(2\theta)}{\omega_m(\omega_a+\omega_m)}.
\end{equation}
And after a time $T=\pi/(2|g'_{\rm eff}|)$, the state in Eq.~(\ref{psitaus}) becomes
\begin{equation}\label{psibell2}
|\varphi'(T'+T)\rangle=\frac{1}{\sqrt{2}}\left(|g00\rangle+e^{i\phi}|e01\rangle\right),
\end{equation}
which is the desired Bell state for the qubit-magnon subsystem, since the middle state for the resonator is now separable. At this moment, one can tune the qubit frequency faraway from the resonant point and then the Bell state $(|g0\rangle+e^{i\phi}|e1\rangle)/\sqrt{2}$ can be maintained.

\section{Conclusion}\label{conclusion}

The protocols for generating local and global entangled states we proposed can be performed in a hybrid setup consisting of a single superconducting qubit, a microwave resonator, and a YIG sphere (magnon)~\cite{magnonqubit,magnonqubit2}. The resonator is simultaneously strongly coupled with the magnon via the magnetic dipole interaction, and with the qubit via a general Rabi interaction. In recent experiments~\cite{magnonqubit,magnonqubit2,ultrastrong}, the coupling strength between photon and qubit $G/2\pi\approx120$ MHz, the coupling strength between photon and magnon $g/2\pi\approx20$ MHz, and the transition frequencies of photon mode, magnon mode and qubit are almost in the same order of GHz. Thus the generation time $T$ of our protocols is nearly about $0.1\sim10\mu s$. Note our target entangled state is of a ``discrete-variable'' type rather than a ``continuous-variable'' one in Ref.~\cite{mppentang}. Our study is of interested in pursuit of the entangled states with the counter-rotating interaction and of importance to control the quantum state in a level-resolved hybrid system.

In conclusion, we have presented a concise protocol for the deterministic generation of local Bell state of the photon-magnon or the qubit-magnon subsystems, and global GHZ state of the whole qubit-photon-magnon system. Our protocol relies on the effective Hamiltonian at the avoided-level-crossing points, which reserves the effects of the counter-rotating interactions and the leading-order contributions of the state transitions. By properly tuning the transition frequency of the superconducting qubit, various scenarios of ``three-wave-mixing'' relevant to either double-excitation Bell state or GHZ state are constructed. Moreover, the generation fidelities of these entangled states are numerically estimated with the standard Lindblad master equation and our protocol is found to be robust against the external dissipation noises.

\section*{Acknowledgments}

We acknowledge grant support from the National Science Foundation of China (Grants No. 11974311 and No. U1801661), and the Zhejiang Provincial Natural Science Foundation of China under Grant No. LD18A040001.

\appendix

\section{Effective Hamiltonian for generating Bell state of photon-magnon system}\label{appa}

\begin{figure}[htbp]
\centering
\includegraphics[width=0.35\textwidth]{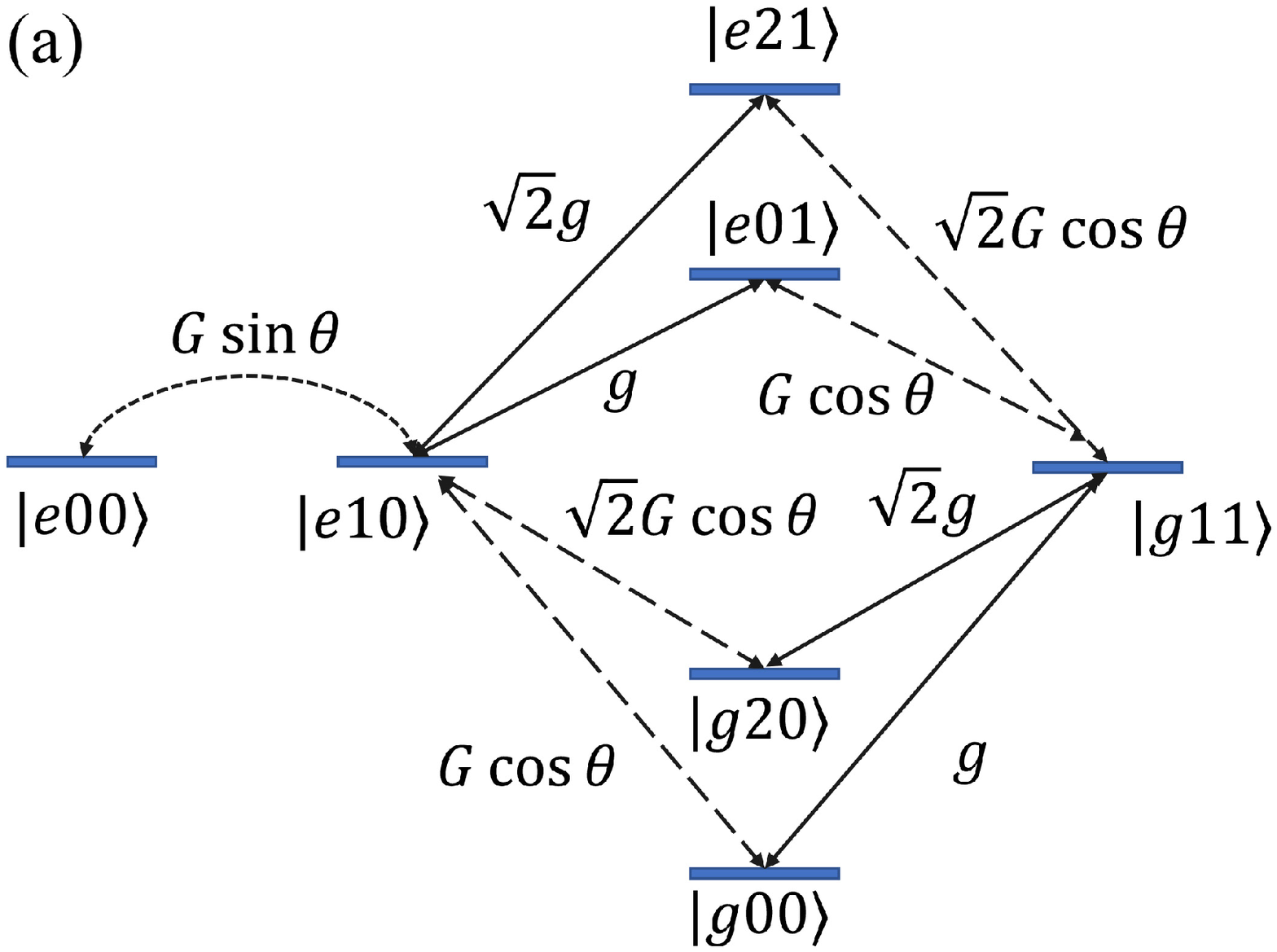}
\includegraphics[width=0.35\textwidth]{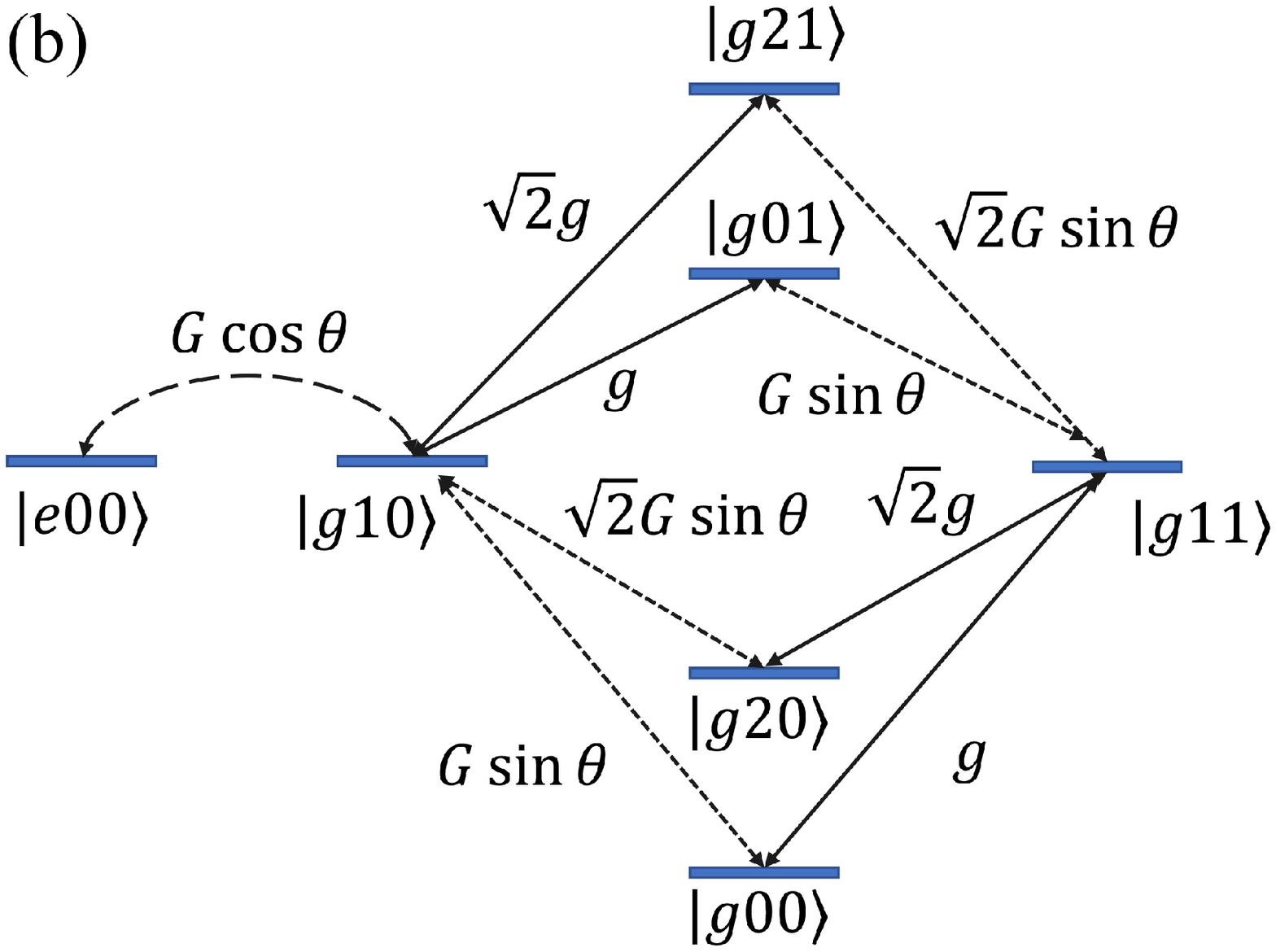}
\includegraphics[width=0.35\textwidth]{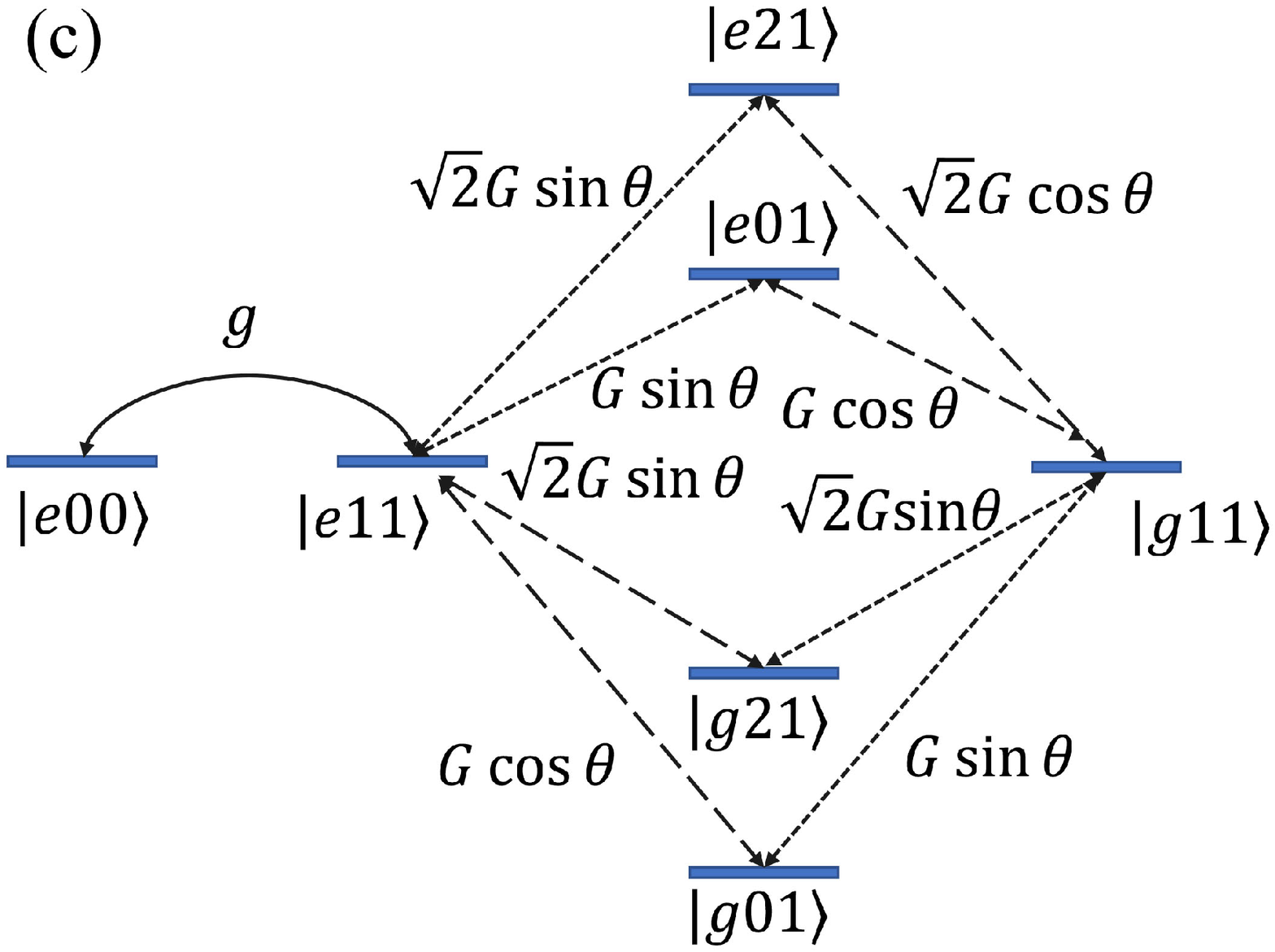}
\caption{All $12$ third-order (leading-order) paths connecting the states $|e00\rangle$ and $|g11\rangle$. (a), (b), and (c) are distinguished by the first intermediate state. Solid lines represent the coupling between photon and magnon, and dashed (dotted) lines mark the coupling between photon and qubit via the transition induced by $\sigma_x$ ($\sigma_z$) in the general Rabi model. }\label{e1p1m1path}
\end{figure}

The interaction Hamiltonian $V$ in Eq.~(\ref{Hamiltonian}) including the photon-magnon coupling and the general Rabi interaction between qubit and resonator can be regarded as a perturbation provided that $g,G\ll\omega_a$, $\omega_m$, $|\omega_a-\omega_m|$, while the results are found to have a broader range of validity in terms of coupling strength. The existence of $V$ gives rise to nonzero shifts of the eigenstructure of the unperturbed Hamiltonian $H_0$ in Eq.~(\ref{Hamiltonian}). To the second order of $g,G$, the shift of the $i$th eigenstate $|E_i\rangle$ is given by
\begin{equation}\label{secondp}
\epsilon=\sum_{n\neq i}\frac{V_{in}V_{ni}}{E_i-E_n},
\end{equation}
where $V_{ni}\equiv\langle E_n|V|E_i\rangle$ and $E_n$ is the $n$th eigenenergy. Our protocol to generating the Bell state of the photon-magnon subsystem is based on the ``three-wave mixing'' of  $|e00\rangle\equiv|e\rangle_q|0\rangle_a|0\rangle_m\leftrightarrow|g11\rangle$, which consists of $12$ third-order paths as the leading-order contribution as plotted in Fig.~\ref{e1p1m1path}. Due to the high-order Fermi's Golden rule in Eq.~(\ref{Fermi}) or the standard perturbation theory~\cite{noon3}, the third-order effective coupling strength between any eigenstates $|i\rangle$ and $|j\rangle$ of the unperturbed Hamiltonian $H_0$ is given by
\begin{equation}\label{thirdp}
g_{\rm eff}=\sum_{n,m\neq i,j}\frac{V_{jn}V_{nm}V_{mi}}{(E_i-E_n)(E_i-E_m)}.
\end{equation}

Consequently, the effective Hamiltonian in the interested subspace spanned by $\{|e00\rangle, |g11\rangle\}$ can be expressed as
\begin{equation}\label{Heff1s}
\begin{aligned}
H_{\rm eff}&=(\omega_q+\epsilon_1)|e00\rangle\langle e00|+(\omega_a+\omega_m+\epsilon_2)|g11\rangle\langle g11|\\
&+g_{\rm eff}\left(|e00\rangle\langle g11|+|g11\rangle\langle e00|\right),
\end{aligned}
\end{equation}
where $\epsilon_1$ and $\epsilon_2$ are the energy shifts of the states $|e00\rangle$ and $|g11\rangle$, respectively, due to the interaction Hamiltonian $V$, and $g_{\rm eff}$ is the effective coupling strength (transition rate). These three coefficients are to be determined by summarizing all the leading-order contributions from the paths connecting the initial state $|e00\rangle$ and the target state $|g11\rangle$ in Eqs.~(\ref{secondp}) and (\ref{thirdp}).

We first consider the energy shift $\epsilon_1$ of state $|e00\rangle$. Summarizing all paths from $|e00\rangle$ to itself through one intermediate state, i.e., $|e00\rangle\to|e10\rangle\to|e00\rangle$ in Fig.~\ref{e1p1m1path}(a), $|e00\rangle\to|g10\rangle\to|e00\rangle$ in Fig.~\ref{e1p1m1path}(b), and $|e00\rangle\to|e11\rangle\to|e00\rangle$ in Fig.~\ref{e1p1m1path}(c), we can obtain the second-order energy correction (shift) $\epsilon_1$ for the state $|e00\rangle$  according to Eq.~(\ref{secondp}):
\begin{equation}\label{shift1}
\epsilon_1=\frac{G^2\sin^2\theta}{-\omega_a}+\frac{G^2\cos^2\theta}{\omega_q-\omega_a}-\frac{g^2}{\omega_a+\omega_m}.
\end{equation}
Similarly, we have the energy shift
\begin{equation}\label{shift2}
\epsilon_2=\frac{G^2\sin^2\theta}{-\omega_a}+\frac{G^2\cos^2\theta}{\omega_a-\omega_q}
-\frac{2G^2\cos^2\theta}{\omega_q+\omega_a}-\frac{3g^2}{\omega_a+\omega_m},
\end{equation}
for the state $|g11\rangle$. Note a completed Rabi oscillation between $|e00\rangle$ and $|g11\rangle$ demands an exact resonant condition in Eq.~(\ref{Heff1s}), i.e., the diagonal terms in the first line of $H_{\rm eff}$ becomes the identity operator in the subspace. We thus have $\omega_q+\epsilon_1=\omega_a+\omega_m+\epsilon_2$ and then
\begin{equation*}
\begin{aligned}
\delta&\equiv\omega_q-\omega_a-\omega_m=\epsilon_2-\epsilon_1\\
&=-2G^2\cos^2\theta\left(\frac{1}{\omega_q-\omega_a}+\frac{1}{\omega_a+\omega_q}\right)-\frac{2g^2}{\omega_a+\omega_m}\\
&=-2G^2\cos^2\theta\left(\frac{1}{\omega_m}+\frac{1}{2\omega_a+\omega_m}\right)-\frac{2g^2}{\omega_a+\omega_m}\\
&-2G^2\cos^2\theta\left[\frac{1}{\omega_m^2}+\frac{1}{(2\omega_a+\omega_m)^2}\right]\delta+\mathcal{O}(\delta^2)\\
&=A-B\delta+\mathcal{O}(\delta^2),
\end{aligned}
\end{equation*}
where
\begin{eqnarray*}
A&\equiv&-2G^2\cos^2\theta\left(\frac{1}{\omega_m}-\frac{1}{2\omega_a+\omega_m}\right)-\frac{2g^2}{\omega_a+\omega_m}, \\
B&\equiv&2G^2\cos^2\theta\left[\frac{1}{\omega_m^2}+\frac{1}{(2\omega_a+\omega_m)^2}\right],
\end{eqnarray*}
and $\mathcal{O}(\delta^2)$ represents all the higher orders of $\delta$ than the first order in Taylor expansion. Then $\delta$ is consistently solved as $\delta=A/(1+B)$ up to the second-order correction. Note $B\approx\mathcal{O}(G^2/\omega^2_a)$, so that up to the second-order of coupling strengths $g$ and $G$, we have
\begin{equation}\label{delta1}
\delta=-2G^2\cos^2\theta\left(\frac{1}{\omega_m}+\frac{1}{2\omega_a+\omega_m}\right)-\frac{2g^2}{\omega_a+\omega_m}.
\end{equation}

\begin{figure}[htbp]
\centering
\includegraphics[width=0.45\linewidth]{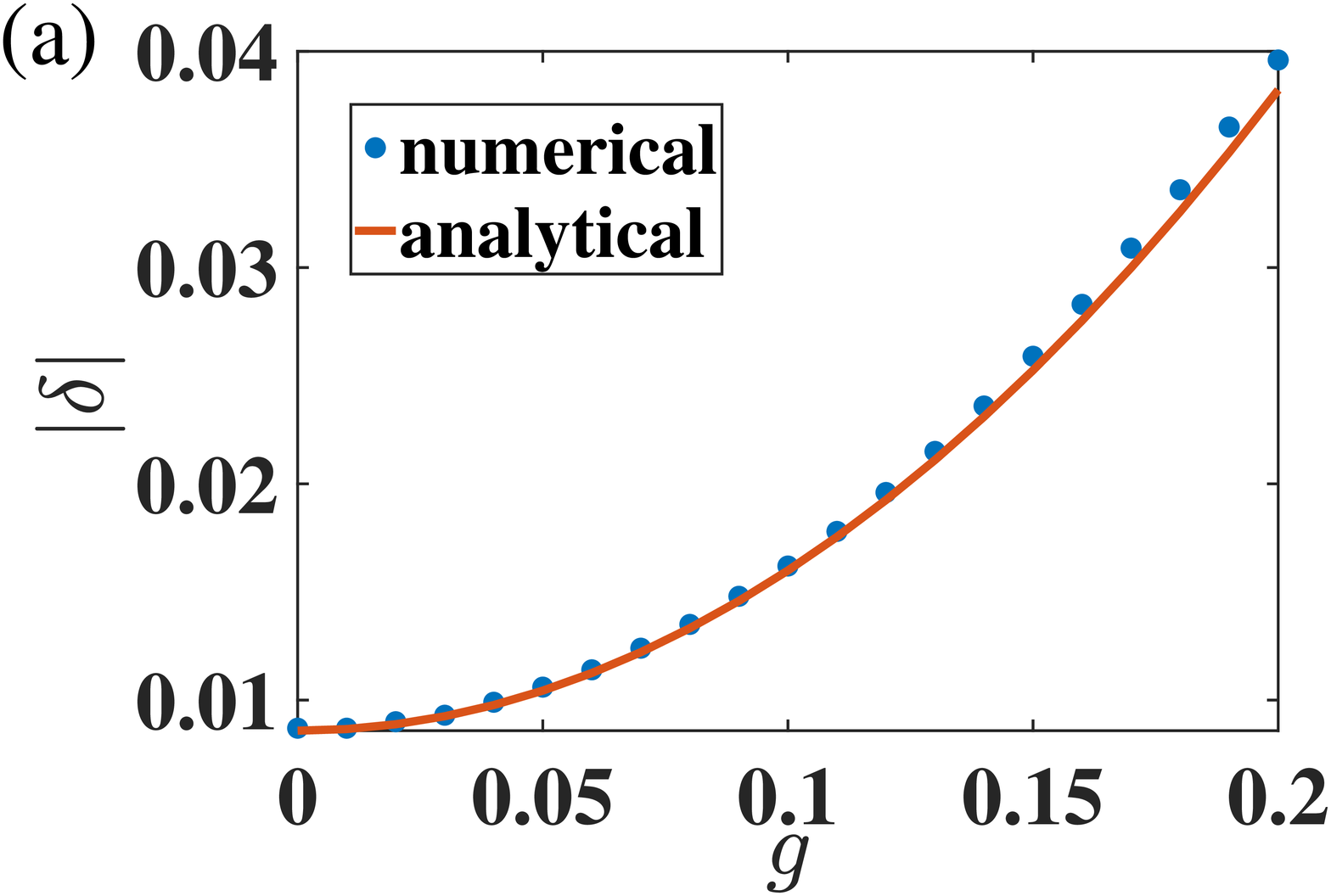}
\includegraphics[width=0.45\linewidth]{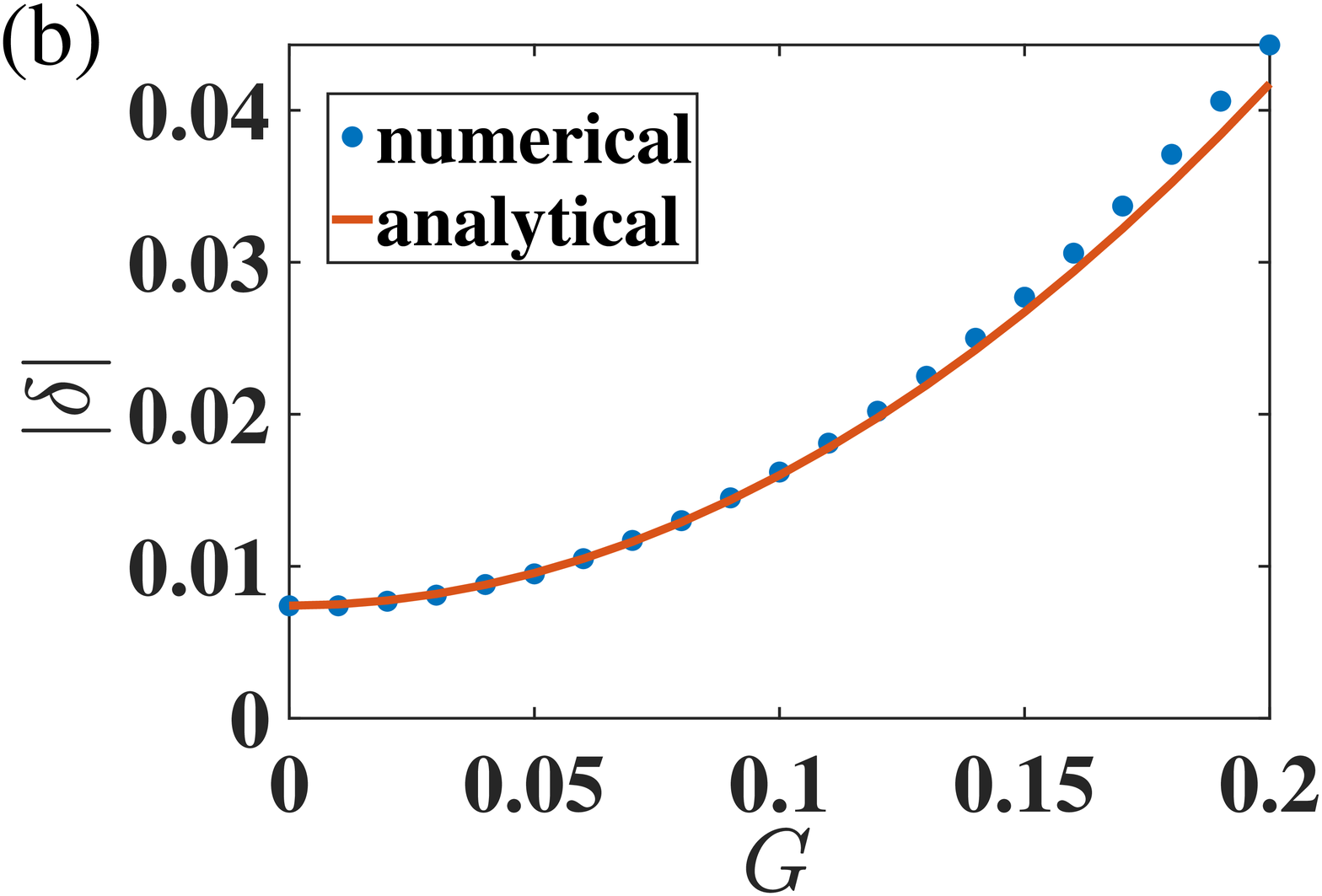}
\includegraphics[width=0.45\linewidth]{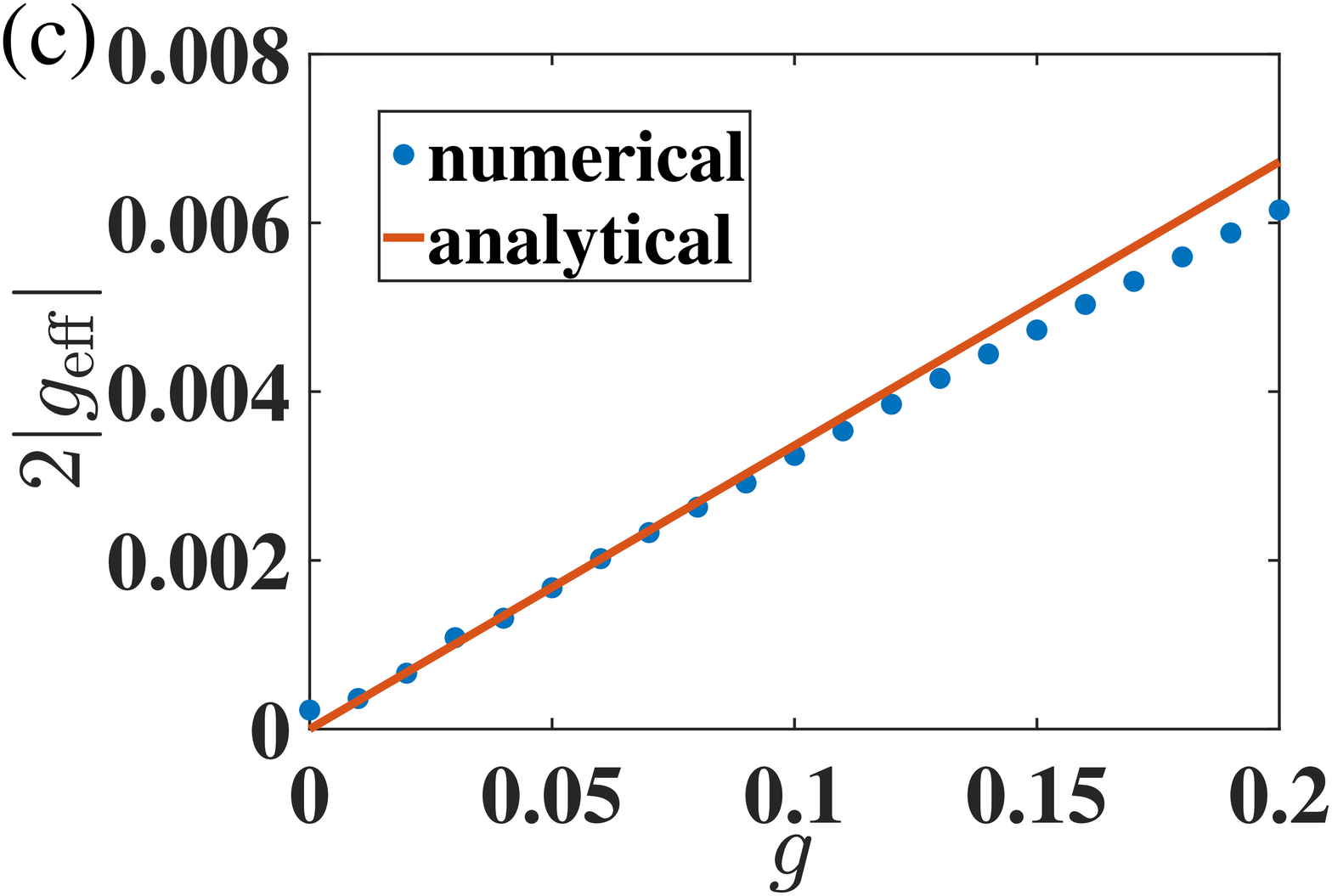}
\includegraphics[width=0.45\linewidth]{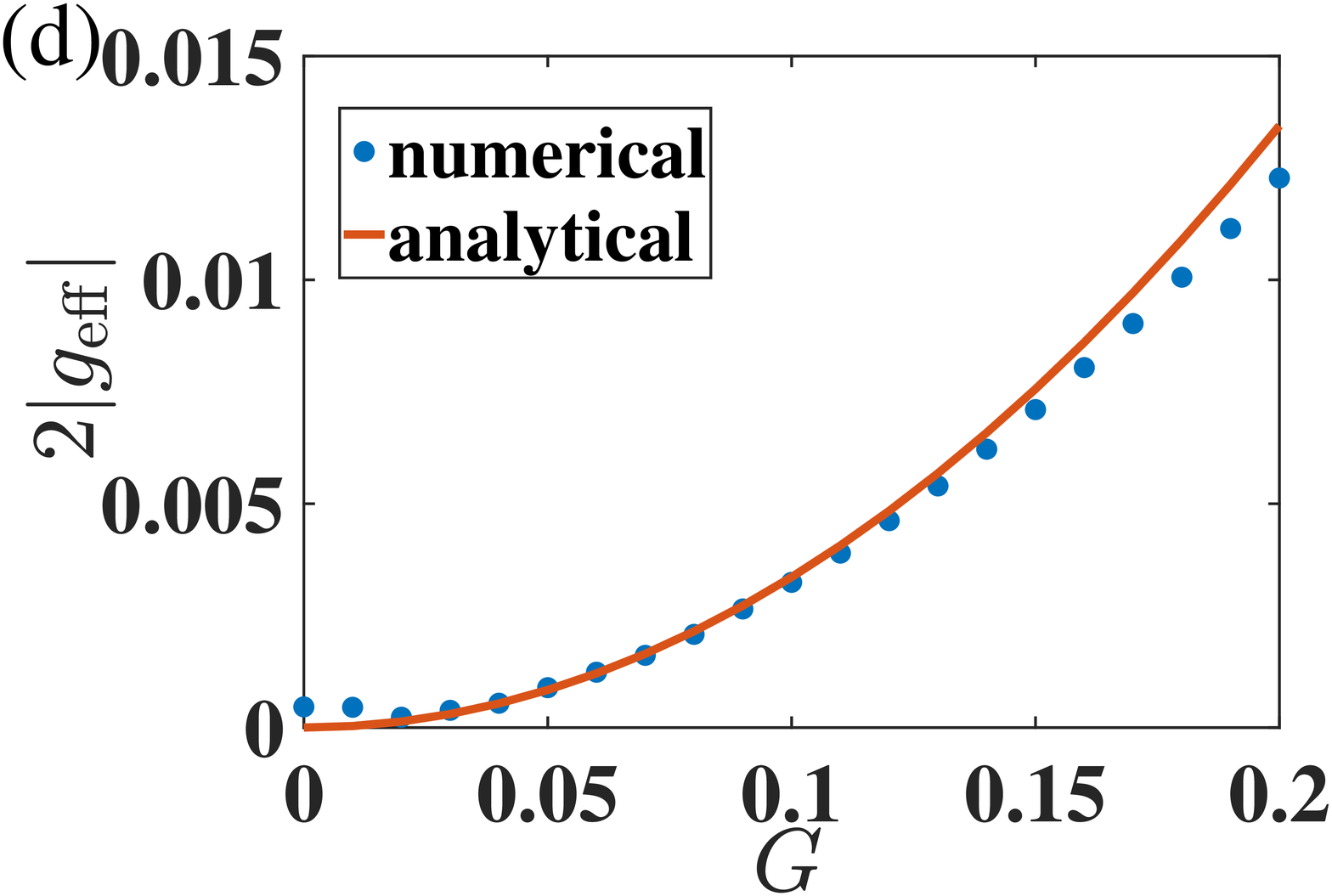}
\caption{(a) and (b), Comparison between the numerically evaluated energy shift $|\delta|$ (in units of $\omega_a$) from the full Hamiltonian (blue dots) and the corresponding analytical results in Eq.~(\ref{delta1}) from the third-order perturbation theory (orange solid line), respectively. (c) and (d), Comparison between the numerically evaluated effective coupling strength $2|g_{\rm{eff}}|$ (in units of $\omega_a$) (blue dots) and the corresponding analytical results in Eq.~(\ref{geff1}) from the third-order perturbation theory (orange solid line), respectively. Here $\omega_m=1.7\omega_a$ and $\theta=\pi/4$ and in (a) and (c) $G=0.1\omega_a$ and in (b) and (d) $g=0.1\omega_a$.}\label{e1p1m1geff}
\end{figure}

Next we consider the leading-order contributions to the effective coupling strength $g_{\rm eff}$ from all the $12$ three-order paths in Fig.~\ref{e1p1m1path} connecting $|e00\rangle$ and $|g11\rangle$, e.g., $|e00\rangle\to|e10\rangle\to|g00\rangle\to|g11\rangle$. The transition rate for each path up to the third order of the coupling strengths $g$ and $G$ can be obtained by virtue of Eq.~(\ref{thirdp}). In Fig.~\ref{e1p1m1path}(a), we have
\begin{eqnarray*}
g_1&=&\frac{G^2g\sin(2\theta)}{\omega_a(2\omega_a+\omega_m)},\\
g_2&=&\frac{G^2g\sin(2\theta)}{2\omega_a\omega_m},\\
g_3&=&\frac{G^2g\sin(2\theta)}{\omega_a(2\omega_a-\omega_q)}\approx\frac{G^2g\sin(2\theta)}{\omega_a(\omega_a-\omega_m)},\\
g_4&=&-\frac{G^2g\sin(2\theta)}{2\omega_a\omega_q}\approx-\frac{G^2g\sin(2\theta)}{2\omega_a(\omega_a+\omega_m)}.
\end{eqnarray*}
In Fig.~\ref{e1p1m1path}(b), we have
\begin{eqnarray*}
g_5&=&-\frac{G^2g\sin(2\theta)}{(\omega_q-\omega_a)(\omega_q-2\omega_a-\omega_m)}\approx\frac{G^2g\sin(2\theta)}{\omega_a\omega_m},\\
g_6&=&-\frac{G^2g\sin(2\theta)}{(\omega_q-\omega_a)(\omega_q-\omega_m)}\approx-\frac{G^2g\sin(2\theta)}{2\omega_a\omega_m},\\
g_7&=&-\frac{G^2g\sin(2\theta)}{(\omega_q-\omega_a)(\omega_q-2\omega_a-\omega_m)}\approx\frac{G^2g\sin(2\theta)}{\omega_m(\omega_a-\omega_m)},\\
g_8&=&-\frac{G^2g\sin(2\theta)}{2(\omega_q-\omega_a)\omega_q}\approx-\frac{G^2g\sin(2\theta)}{2\omega_m(\omega_a+\omega_m)}.
\end{eqnarray*}
And in Fig.~\ref{e1p1m1path}(c), we have
\begin{eqnarray*}
g_9&=&\frac{G^2g\sin(2\theta)}{(\omega_a+\omega_m)(2\omega_a+\omega_m)},\\
g_{10}&=&\frac{G^2g\sin(2\theta)}{2\omega_m(\omega_a+\omega_m)},\\
g_{11}&=&\frac{G^2g\sin(2\theta)}{(\omega_a+\omega_m)(\omega_q-2\omega_a-\omega_m)}\approx-\frac{G^2g\sin(2\theta)}{\omega_a(\omega_a+\omega_m)},\\
g_{12}&=&\frac{G^2g\sin(2\theta)}{2(\omega_a+\omega_m)(\omega_q-\omega_m)}\approx\frac{G^2g\sin(2\theta)}{2\omega_a(\omega_a+\omega_m)}.
\end{eqnarray*}
The total effective coupling strength therefore reads
\begin{equation}\label{geff1}
g_{\rm eff}=\sum_{k=1}^{12}g_k=\frac{2G^2g\sin(2\theta)}{\omega_m(\omega_a-\omega_m)}.
\end{equation}
Eventually, the effective Hamiltonian in Eq.~(\ref{Heff1s}) becomes
\begin{equation}\label{Heff1app}
H_{\rm eff}=g_{\rm eff}\left(|e00\rangle\langle g11|+|g11\rangle\langle e00|\right).
\end{equation}

Both $\delta$ and $g_{\rm eff}$ can also by numerically evaluated in the whole Hilbert space of the full Hamiltonian. They can be shown around the avoided-level-crossing points in Fig.~\ref{e1p1m1eigen}. To demonstrate the ranges of validity of Eqs.~(\ref{delta1}) and (\ref{geff1}), the analytical and numerical results for their magnitudes are directly compared in Fig.~\ref{e1p1m1geff} as functions of the normalized coupling strengths $g/\omega_a$ and $G/\omega_a$, respectively. In Figs.~\ref{e1p1m1geff}(a) and (b), one can observe that the energy shifts do match with their numerical results for the normalized coupling strengths $g,G\le0.15\omega_a$. In Figs.~\ref{e1p1m1geff}(c) and (d), the effective coupling strengths are expected to provide good description for the normalized coupling strengths $g\le0.1\omega_a$ and $G\le0.12\omega_a$. For an even larger $g$ and $G$, higher-order contributions have to be considered to capture the whole effect from the interaction Hamiltonian $V$ modifying the eigenstates of the bare system. Note Eqs.~(\ref{delta1}) and (\ref{geff1}) provide up to the second-order and the third-order expressions for $\delta$ and $g_{\rm eff}$, respectively.

\section{Effective Hamiltonian for generating GHZ state of the whole hybrid system}\label{appb}

This Appendix contributes to generating the GHZ state of the whole hybrid system by virtue of an effective Hamiltonian that yields the Rabi oscillation between the desired states $|e11\rangle$ and $|g20\rangle$. Similar to Appendix~\ref{appa}, we also need to find out all the paths connecting these two states by the full Hamiltonian~(\ref{Hamiltonian}) in the leading-order. And then we can determine both the energy shifts and the effective transition rate for them. In contrast to the ``three-wave mixing'' applied in the generation of the Bell state of the photon-magnon subsystem, here the procedure occurs around the near-resonant point $\omega_a\approx\omega_q+\omega_m$.

The effective Hamiltonian in the subspace spanned by $\{|e11\rangle,|g20\rangle\}$ can be also expressed by
\begin{equation}\label{Heff2s}
\begin{aligned}
H'_{\rm eff}&=(\omega_q+\epsilon'_1)|e11\rangle\langle e11|+(\omega_a-\omega_m+\epsilon'_2)|g20\rangle\langle g20|\\
&+g'_{\rm eff}\left(|e11\rangle\langle g20|+|g20\rangle\langle e11|\right),
\end{aligned}
\end{equation}
where $\epsilon'_1$ and $\epsilon'_2$ are the individual energy shifts induced by the second-order transitions for the states $|e11\rangle$ and $|g20\rangle$ to themselves, respectively, such as $|e11\rangle\to|e21\rangle\to|e11\rangle$, and $g'_{\rm eff}$ is the effective coupling strength from $|e11\rangle$ to $|g20\rangle$ in the leading order.

Summarizing all paths from $|e11\rangle$ to $|e11\rangle$ through an intermediate state, one can obtain the second-order energy correction (shift) $\epsilon'_1$ according to Eq.~(\ref{secondp}):
\begin{equation}\label{shift21}
\epsilon'_1=-\frac{G^2\sin^2\theta}{\omega_a}-\frac{G^2\cos^2\theta}{2\omega_a-\omega_m}
-\frac{2G^2\cos^2\theta}{\omega_m}-\frac{3g^2}{\omega_a+\omega_m}.
\end{equation}
In the same way, we have the energy shift $\epsilon'_2$
\begin{equation}\label{shift22}
\begin{aligned}
\epsilon'_2&=-\frac{G^2\sin^2\theta}{\omega_a}+\frac{2G^2\cos^2\theta}{\omega_m}
-\frac{3G^2\cos^2\theta}{2\omega_a-\omega_m}\\
&+\frac{2g^2}{\omega_a-\omega_m}-\frac{3g^2}{\omega_a+\omega_m}
\end{aligned}
\end{equation}
for the state $|g20\rangle$. These two shifts are required to fill the gap between $\omega_a$ and $\omega_q+\omega_m$ to facilitate a completed Rabi oscillation between $|e11\rangle$ and $|g20\rangle$. Thus up to the second-order of the coupling strengths $g$ and $G$, we have
\begin{equation}\label{delta2s}
\begin{aligned}
\Delta&\equiv\omega_q-\omega_a+\omega_m=\epsilon'_2-\epsilon'_1\\
&\approx4G^2\cos^2\theta\left(\frac{1}{\omega_m}-\frac{1}{2\omega_a-\omega_m}\right)+\frac{2g^2}{\omega_a-\omega_m}.
\end{aligned}
\end{equation}

\begin{figure}[htbp]
\centering
\includegraphics[width=0.45\linewidth]{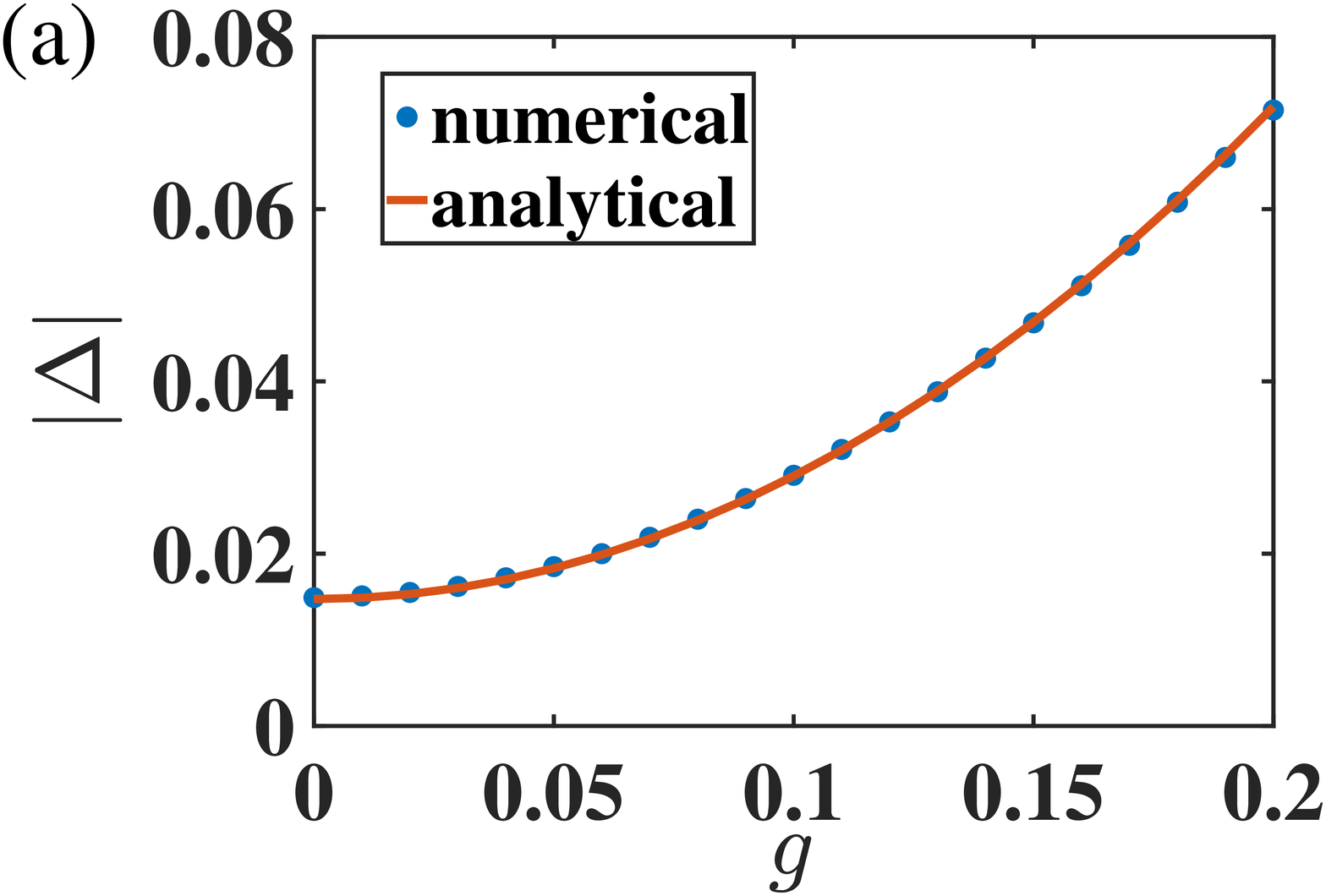}
\includegraphics[width=0.45\linewidth]{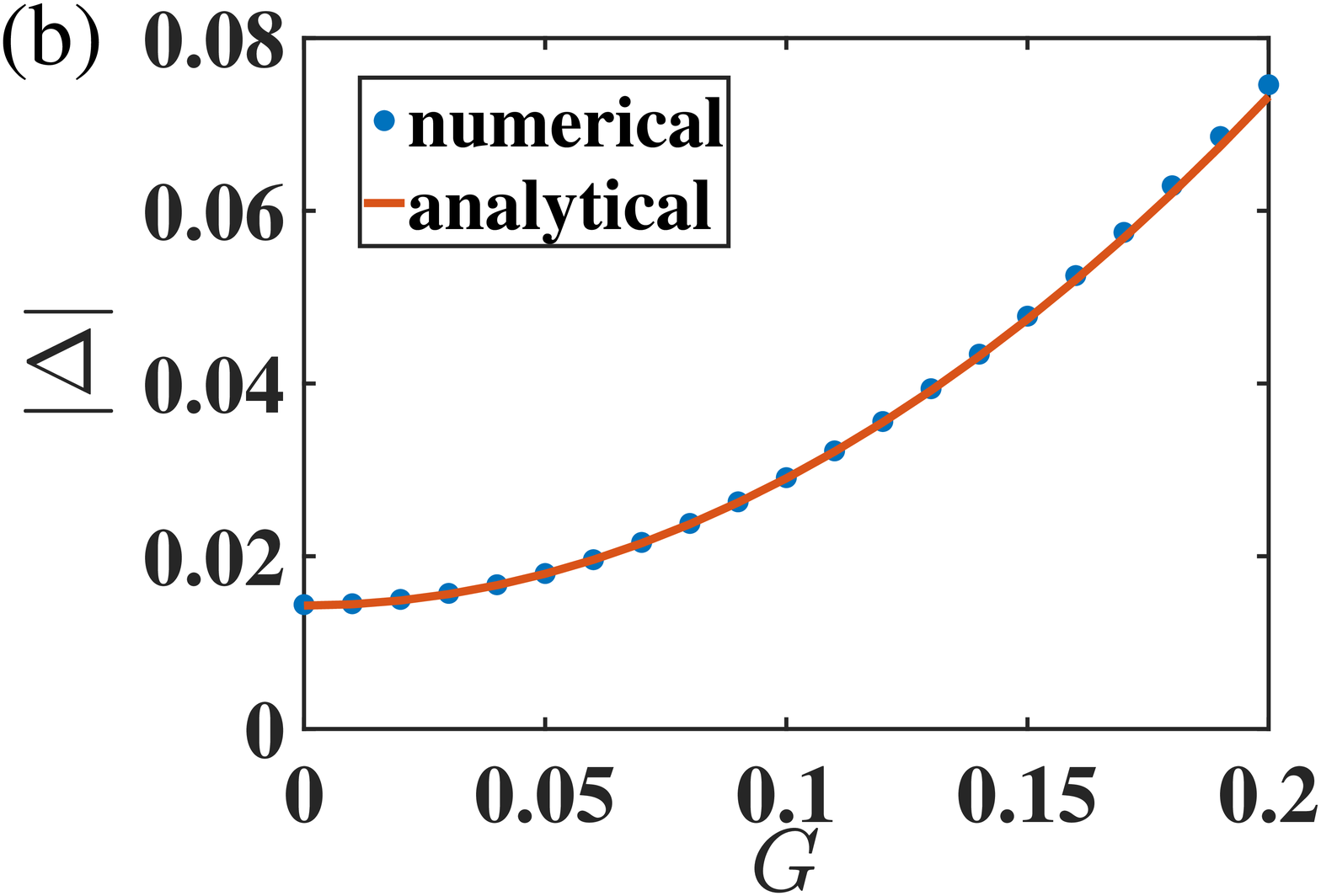}
\includegraphics[width=0.45\linewidth]{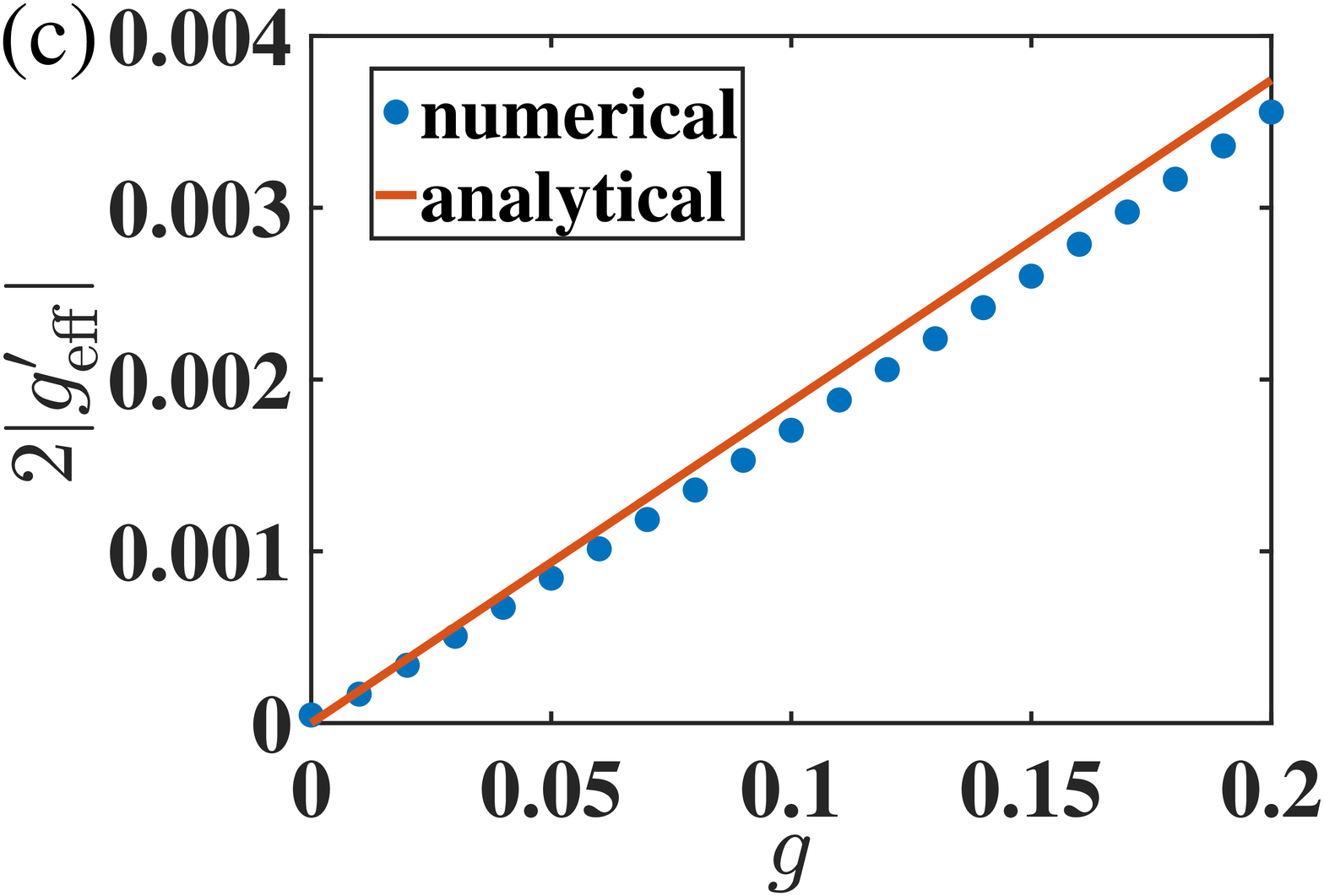}
\includegraphics[width=0.45\linewidth]{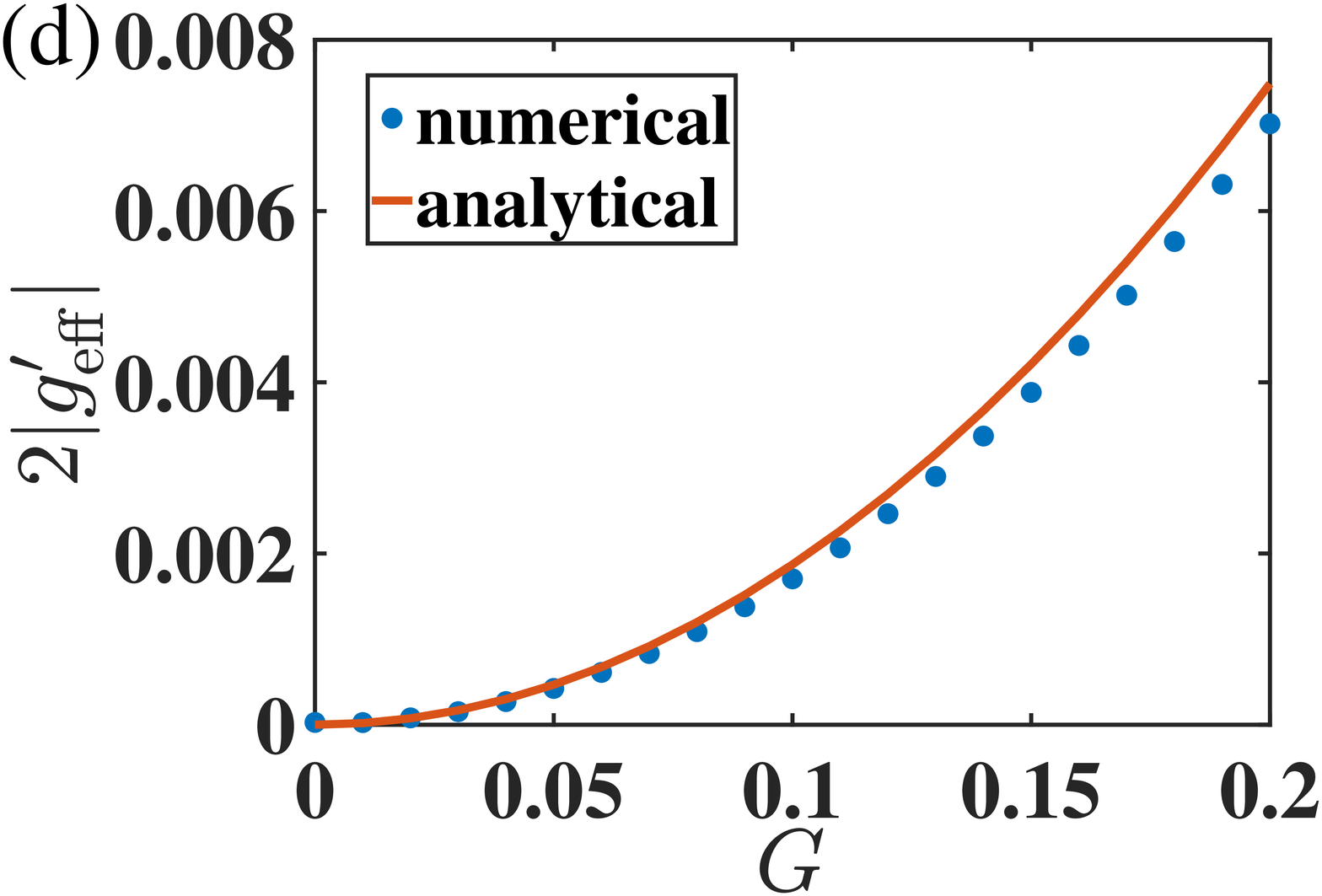}
\caption{(a) and (b), Comparison between the numerically evaluated energy shift $|\Delta|$ (in units of $\omega_m$) from the full Hamiltonian (blue dots) and the corresponding analytical results in Eq.~(\ref{delta2s}) from the third-order perturbation theory (orange solid line), respectively. (c) and (d), Comparison between the numerically evaluated effective coupling strength $2|g'_{\rm{eff}}|$ (in units of $\omega_m$) (blue dots) and the corresponding analytical results in Eq.~(\ref{geff2}) from the third-order perturbation theory (orange solid line), respectively. Here $\omega_a=2.4\omega_m$ and $\theta=\pi/4$ and in (a) and (c) $G=0.1\omega_m$ and in (b) and (d) $g=0.1\omega_m$.}\label{GHZgeff}
\end{figure}

Next we consider all the leading-order contributions to the effective coupling strength connecting $|g20\rangle$ and $|e11\rangle$, e.g., $|g20\rangle\to|g01\rangle\to|e00\rangle\to|e11\rangle$. By virtue of Eq.~(\ref{thirdp}) and collecting all $18$ paths, one can get the effective coupling strength
\begin{equation}\label{geff2}
g'_{\rm eff}=-\frac{\sqrt{2}G^2g\sin(2\theta)(\omega_a+3\omega_m)}{\omega_a\omega_m(\omega_a+\omega_m)}
\end{equation}
up to the third order of the coupling strengths $g$ and $G$. Thus the effective Hamiltonian in Eq.~(\ref{Heff2s}) can be eventually written as
\begin{equation}\label{Heff2app}
H'_{\rm eff}=g'_{\rm eff}\left(|e11\rangle\langle g20|+|g20\rangle\langle e11|\right).
\end{equation}

Both $\Delta$ and $g'_{\rm eff}$, as shown around the avoided-level-crossing point in Fig.~\ref{GHZeigen}, can be justified by comparing the preceding analytical results and the numerical simulation over the whole Hilbert space. In Fig.~\ref{GHZgeff}, their magnitudes are plotted as functions of the normalized coupling strengths $g/\omega_m$ or $G/\omega_m$. In Figs.~\ref{GHZgeff}(a) and (b), it is demonstrated that the energy shifts in Eq.~(\ref{delta2s}) describe well the numerical results at least for the normalized coupling strengths $g,G\le 0.2\omega_m$, which are definitely of the ultrastrong coupling regime. While in Figs.~\ref{GHZgeff}(c) and (d), one can see that the effective coupling strengths yield perfect results for the normalized interaction strengths $g/\omega_m\le0.1$ and $G/\omega_m \le 0.15$. For an even larger $g$ and $G$, higher-order contribution has to be considered to capture the whole effect from the interaction Hamiltonian modifying the eigenstructure of the bare system.

\bibliographystyle{apsrevlong}
\bibliography{reference}

\begin{thebibliography}{53}%
\makeatletter
\providecommand \@ifxundefined [1]{%
 \@ifx{#1\undefined}
}%
\providecommand \@ifnum [1]{%
 \ifnum #1\expandafter \@firstoftwo
 \else \expandafter \@secondoftwo
 \fi
}%
\providecommand \@ifx [1]{%
 \ifx #1\expandafter \@firstoftwo
 \else \expandafter \@secondoftwo
 \fi
}%
\providecommand \natexlab [1]{#1}%
\providecommand \enquote  [1]{``#1''}%
\providecommand \bibnamefont  [1]{#1}%
\providecommand \bibfnamefont [1]{#1}%
\providecommand \citenamefont [1]{#1}%
\providecommand \href@noop [0]{\@secondoftwo}%
\providecommand \href [0]{\begingroup \@sanitize@url \@href}%
\providecommand \@href[1]{\@@startlink{#1}\@@href}%
\providecommand \@@href[1]{\endgroup#1\@@endlink}%
\providecommand \@sanitize@url [0]{\catcode `\\12\catcode `\$12\catcode
  `\&12\catcode `\#12\catcode `\^12\catcode `\_12\catcode `\%12\relax}%
\providecommand \@@startlink[1]{}%
\providecommand \@@endlink[0]{}%
\providecommand \url  [0]{\begingroup\@sanitize@url \@url }%
\providecommand \@url [1]{\endgroup\@href {#1}{\urlprefix }}%
\providecommand \urlprefix  [0]{URL }%
\providecommand \Eprint [0]{\href }%
\providecommand \doibase [0]{http://dx.doi.org/}%
\providecommand \selectlanguage [0]{\@gobble}%
\providecommand \bibinfo  [0]{\@secondoftwo}%
\providecommand \bibfield  [0]{\@secondoftwo}%
\providecommand \translation [1]{[#1]}%
\providecommand \BibitemOpen [0]{}%
\providecommand \bibitemStop [0]{}%
\providecommand \bibitemNoStop [0]{.\EOS\space}%
\providecommand \EOS [0]{\spacefactor3000\relax}%
\providecommand \BibitemShut  [1]{\csname bibitem#1\endcsname}%
\let\auto@bib@innerbib\@empty
\bibitem [{\citenamefont {Ladd}\ \emph {et~al.}(2010)\citenamefont {Ladd},
  \citenamefont {Jelezko}, \citenamefont {Laflamme}, \citenamefont {Nakamura},
  \citenamefont {Monroe},\ and\ \citenamefont {O'Brien}}]{quantumcomputing}%
  \BibitemOpen
  \bibfield  {author} {\bibinfo {author} {\bibfnamefont {T.~D.}\ \bibnamefont
  {Ladd}}, \bibinfo {author} {\bibfnamefont {F.}~\bibnamefont {Jelezko}},
  \bibinfo {author} {\bibfnamefont {R.}~\bibnamefont {Laflamme}}, \bibinfo
  {author} {\bibfnamefont {Y.}~\bibnamefont {Nakamura}}, \bibinfo {author}
  {\bibfnamefont {C.}~\bibnamefont {Monroe}}, \ and\ \bibinfo {author}
  {\bibfnamefont {J.~L.}\ \bibnamefont {O'Brien}},\ }\bibfield  {title} {\emph
  {\bibinfo {title} {Quantum computers},\ }}\href {\doibase
  https://doi.org/10.1038/nature08812} {\bibfield  {journal} {\bibinfo
  {journal} {Nature (London)}\ }\textbf {\bibinfo {volume} {464}},\ \bibinfo
  {pages} {45} (\bibinfo {year} {2010})}\BibitemShut {NoStop}%
\bibitem [{\citenamefont {Reiserer}\ and\ \citenamefont
  {Rempe}(2015)}]{quantumcommunication}%
  \BibitemOpen
  \bibfield  {author} {\bibinfo {author} {\bibfnamefont {A.}~\bibnamefont
  {Reiserer}}\ and\ \bibinfo {author} {\bibfnamefont {G.}~\bibnamefont
  {Rempe}},\ }\bibfield  {title} {\emph {\bibinfo {title} {Cavity-based quantum
  networks with single atoms and optical photons},\ }}\href {\doibase
  10.1103/RevModPhys.87.1379} {\bibfield  {journal} {\bibinfo  {journal} {Rev.
  Mod. Phys.}\ }\textbf {\bibinfo {volume} {87}},\ \bibinfo {pages} {1379}
  (\bibinfo {year} {2015})}\BibitemShut {NoStop}%
\bibitem [{\citenamefont {Degen}\ \emph {et~al.}(2017)\citenamefont {Degen},
  \citenamefont {Reinhard},\ and\ \citenamefont {Cappellaro}}]{quantumsense}%
  \BibitemOpen
  \bibfield  {author} {\bibinfo {author} {\bibfnamefont {C.~L.}\ \bibnamefont
  {Degen}}, \bibinfo {author} {\bibfnamefont {F.}~\bibnamefont {Reinhard}}, \
  and\ \bibinfo {author} {\bibfnamefont {P.}~\bibnamefont {Cappellaro}},\
  }\bibfield  {title} {\emph {\bibinfo {title} {Quantum sensing},\ }}\href
  {\doibase 10.1103/RevModPhys.89.035002} {\bibfield  {journal} {\bibinfo
  {journal} {Rev. Mod. Phys.}\ }\textbf {\bibinfo {volume} {89}},\ \bibinfo
  {pages} {035002} (\bibinfo {year} {2017})}\BibitemShut {NoStop}%
\bibitem [{\citenamefont {Lachance-Quirion}\ \emph {et~al.}(2019)\citenamefont
  {Lachance-Quirion}, \citenamefont {Tabuchi}, \citenamefont {Gloppe},
  \citenamefont {Usami},\ and\ \citenamefont {Nakamura}}]{magnon}%
  \BibitemOpen
  \bibfield  {author} {\bibinfo {author} {\bibfnamefont {D.}~\bibnamefont
  {Lachance-Quirion}}, \bibinfo {author} {\bibfnamefont {Y.}~\bibnamefont
  {Tabuchi}}, \bibinfo {author} {\bibfnamefont {A.}~\bibnamefont {Gloppe}},
  \bibinfo {author} {\bibfnamefont {K.}~\bibnamefont {Usami}}, \ and\ \bibinfo
  {author} {\bibfnamefont {Y.}~\bibnamefont {Nakamura}},\ }\bibfield  {title}
  {\emph {\bibinfo {title} {Hybrid quantum systems based on magnonics},\
  }}\href {\doibase 10.7567/1882-0786/ab248d} {\bibfield  {journal} {\bibinfo
  {journal} {Appl. Phys. Express}\ }\textbf {\bibinfo {volume} {12}},\ \bibinfo
  {pages} {070101} (\bibinfo {year} {2019})}\BibitemShut {NoStop}%
\bibitem [{\citenamefont {Li}\ \emph {et~al.}(2020)\citenamefont {Li},
  \citenamefont {Zhang}, \citenamefont {Tyberkevych}, \citenamefont {Kwok},\
  and\ \citenamefont {Novosad}}]{magnon2}%
  \BibitemOpen
  \bibfield  {author} {\bibinfo {author} {\bibfnamefont {Y.}~\bibnamefont
  {Li}}, \bibinfo {author} {\bibfnamefont {W.}~\bibnamefont {Zhang}}, \bibinfo
  {author} {\bibfnamefont {V.}~\bibnamefont {Tyberkevych}}, \bibinfo {author}
  {\bibfnamefont {W.~K.}\ \bibnamefont {Kwok}}, \ and\ \bibinfo {author}
  {\bibfnamefont {V.}~\bibnamefont {Novosad}},\ }\bibfield  {title} {\emph
  {\bibinfo {title} {Hybrid magnonics: physics, circuits, and applications for
  coherent information processing},\ }}\href {\doibase
  https://doi.org/10.1063/5.0020277} {\bibfield  {journal} {\bibinfo  {journal}
  {J. Appl. Phys.}\ }\textbf {\bibinfo {volume} {128}},\ \bibinfo {pages}
  {130902} (\bibinfo {year} {2020})}\BibitemShut {NoStop}%
\bibitem [{\citenamefont {Potts}\ and\ \citenamefont {Davis}(2020)}]{magnon4}%
  \BibitemOpen
  \bibfield  {author} {\bibinfo {author} {\bibfnamefont {C.~A.}\ \bibnamefont
  {Potts}}\ and\ \bibinfo {author} {\bibfnamefont {J.~P.}\ \bibnamefont
  {Davis}},\ }\bibfield  {title} {\emph {\bibinfo {title} {Strong magnon-photon
  coupling within a tunable cryogenic microwave cavity},\ }}\href {\doibase
  https://doi.org/10.1063/5.0015660} {\bibfield  {journal} {\bibinfo  {journal}
  {Appl. Phys. Lett.}\ }\textbf {\bibinfo {volume} {116}},\ \bibinfo {pages}
  {263503} (\bibinfo {year} {2020})}\BibitemShut {NoStop}%
\bibitem [{\citenamefont {Li}\ and\ \citenamefont {Zhu}(2019)}]{magnon5}%
  \BibitemOpen
  \bibfield  {author} {\bibinfo {author} {\bibfnamefont {J.}~\bibnamefont
  {Li}}\ and\ \bibinfo {author} {\bibfnamefont {S.-Y.}\ \bibnamefont {Zhu}},\
  }\bibfield  {title} {\emph {\bibinfo {title} {Entangling two magnon modes via
  magnetostrictive interaction},\ }}\href {\doibase
  https://doi.org/10.1088/1367-2630/ab3508} {\bibfield  {journal} {\bibinfo
  {journal} {New. J. Phys.}\ }\textbf {\bibinfo {volume} {21}},\ \bibinfo
  {pages} {085001} (\bibinfo {year} {2019})}\BibitemShut {NoStop}%
\bibitem [{\citenamefont {Wang}\ \emph {et~al.}(2018)\citenamefont {Wang},
  \citenamefont {Zhang}, \citenamefont {Zhang}, \citenamefont {Li},
  \citenamefont {Hu},\ and\ \citenamefont {You}}]{magnon6}%
  \BibitemOpen
  \bibfield  {author} {\bibinfo {author} {\bibfnamefont {Y.-P.}\ \bibnamefont
  {Wang}}, \bibinfo {author} {\bibfnamefont {G.-Q.}\ \bibnamefont {Zhang}},
  \bibinfo {author} {\bibfnamefont {D.}~\bibnamefont {Zhang}}, \bibinfo
  {author} {\bibfnamefont {T.-F.}\ \bibnamefont {Li}}, \bibinfo {author}
  {\bibfnamefont {C.-M.}\ \bibnamefont {Hu}}, \ and\ \bibinfo {author}
  {\bibfnamefont {J.~Q.}\ \bibnamefont {You}},\ }\bibfield  {title} {\emph
  {\bibinfo {title} {Bistability of cavity magnon polaritons},\ }}\href
  {\doibase 10.1103/PhysRevLett.120.057202} {\bibfield  {journal} {\bibinfo
  {journal} {Phys. Rev. Lett.}\ }\textbf {\bibinfo {volume} {120}},\ \bibinfo
  {pages} {057202} (\bibinfo {year} {2018})}\BibitemShut {NoStop}%
\bibitem [{\citenamefont {Zhang}\ \emph {et~al.}(2016)\citenamefont {Zhang},
  \citenamefont {Zou}, \citenamefont {Jiang},\ and\ \citenamefont
  {Tang}}]{magnoncavity}%
  \BibitemOpen
  \bibfield  {author} {\bibinfo {author} {\bibfnamefont {X.}~\bibnamefont
  {Zhang}}, \bibinfo {author} {\bibfnamefont {C.-L.}\ \bibnamefont {Zou}},
  \bibinfo {author} {\bibfnamefont {L.}~\bibnamefont {Jiang}}, \ and\ \bibinfo
  {author} {\bibfnamefont {H.}~\bibnamefont {Tang}},\ }\bibfield  {title}
  {\emph {\bibinfo {title} {Cavity magnonmechanics},\ }}\href
  {https://www.science.org/doi/10.1126/sciadv.1501286} {\bibfield  {journal}
  {\bibinfo  {journal} {Sci. Adv.}\ }\textbf {\bibinfo {volume} {2}},\ \bibinfo
  {pages} {e1501286} (\bibinfo {year} {2016})}\BibitemShut {NoStop}%
\bibitem [{\citenamefont {Soykal}\ and\ \citenamefont
  {Flatt\'e}(2010{\natexlab{a}})}]{yigcavity1}%
  \BibitemOpen
  \bibfield  {author} {\bibinfo {author} {\bibfnamefont {O.~O.}\ \bibnamefont
  {Soykal}}\ and\ \bibinfo {author} {\bibfnamefont {M.~E.}\ \bibnamefont
  {Flatt\'e}},\ }\bibfield  {title} {\emph {\bibinfo {title} {Strong field
  interactions between a nanomagnet and a photonic cavity},\ }}\href {\doibase
  10.1103/PhysRevLett.104.077202} {\bibfield  {journal} {\bibinfo  {journal}
  {Phys. Rev. Lett.}\ }\textbf {\bibinfo {volume} {104}},\ \bibinfo {pages}
  {077202} (\bibinfo {year} {2010}{\natexlab{a}})}\BibitemShut {NoStop}%
\bibitem [{\citenamefont {Soykal}\ and\ \citenamefont
  {Flatt\'e}(2010{\natexlab{b}})}]{yigcavity2}%
  \BibitemOpen
  \bibfield  {author} {\bibinfo {author} {\bibfnamefont {O.~O.}\ \bibnamefont
  {Soykal}}\ and\ \bibinfo {author} {\bibfnamefont {M.~E.}\ \bibnamefont
  {Flatt\'e}},\ }\bibfield  {title} {\emph {\bibinfo {title} {Size dependence
  of strong coupling between nanomagnets and photonic cavities},\ }}\href
  {\doibase 10.1103/PhysRevB.82.104413} {\bibfield  {journal} {\bibinfo
  {journal} {Phys. Rev. B}\ }\textbf {\bibinfo {volume} {82}},\ \bibinfo
  {pages} {104413} (\bibinfo {year} {2010}{\natexlab{b}})}\BibitemShut
  {NoStop}%
\bibitem [{\citenamefont {Li}\ \emph {et~al.}(2018)\citenamefont {Li},
  \citenamefont {Zhu},\ and\ \citenamefont {Agarwal}}]{mppentang}%
  \BibitemOpen
  \bibfield  {author} {\bibinfo {author} {\bibfnamefont {J.}~\bibnamefont
  {Li}}, \bibinfo {author} {\bibfnamefont {S.-Y.}\ \bibnamefont {Zhu}}, \ and\
  \bibinfo {author} {\bibfnamefont {G.~S.}\ \bibnamefont {Agarwal}},\
  }\bibfield  {title} {\emph {\bibinfo {title} {Magnon-photon-phonon
  entanglement in cavity magnomechanics},\ }}\href {\doibase
  10.1103/PhysRevLett.121.203601} {\bibfield  {journal} {\bibinfo  {journal}
  {Phys. Rev. Lett.}\ }\textbf {\bibinfo {volume} {121}},\ \bibinfo {pages}
  {203601} (\bibinfo {year} {2018})}\BibitemShut {NoStop}%
\bibitem [{\citenamefont {Yuan}\ \emph {et~al.}(2020)\citenamefont {Yuan},
  \citenamefont {Yan}, \citenamefont {Zheng}, \citenamefont {He}, \citenamefont
  {Xia},\ and\ \citenamefont {Yung}}]{yigcavity3}%
  \BibitemOpen
  \bibfield  {author} {\bibinfo {author} {\bibfnamefont {H.~Y.}\ \bibnamefont
  {Yuan}}, \bibinfo {author} {\bibfnamefont {P.}~\bibnamefont {Yan}}, \bibinfo
  {author} {\bibfnamefont {S.}~\bibnamefont {Zheng}}, \bibinfo {author}
  {\bibfnamefont {Q.~Y.}\ \bibnamefont {He}}, \bibinfo {author} {\bibfnamefont
  {K.}~\bibnamefont {Xia}}, \ and\ \bibinfo {author} {\bibfnamefont {M.-H.}\
  \bibnamefont {Yung}},\ }\bibfield  {title} {\emph {\bibinfo {title} {Steady
  bell state generation via magnon-photon coupling},\ }}\href {\doibase
  10.1103/PhysRevLett.124.053602} {\bibfield  {journal} {\bibinfo  {journal}
  {Phys. Rev. Lett.}\ }\textbf {\bibinfo {volume} {124}},\ \bibinfo {pages}
  {053602} (\bibinfo {year} {2020})}\BibitemShut {NoStop}%
\bibitem [{\citenamefont {Tabuchi}\ \emph {et~al.}(2014)\citenamefont
  {Tabuchi}, \citenamefont {Ishino}, \citenamefont {Ishikawa}, \citenamefont
  {Yamazaki}, \citenamefont {Usami},\ and\ \citenamefont
  {Nakamura}}]{yigcavity4}%
  \BibitemOpen
  \bibfield  {author} {\bibinfo {author} {\bibfnamefont {Y.}~\bibnamefont
  {Tabuchi}}, \bibinfo {author} {\bibfnamefont {S.}~\bibnamefont {Ishino}},
  \bibinfo {author} {\bibfnamefont {T.}~\bibnamefont {Ishikawa}}, \bibinfo
  {author} {\bibfnamefont {R.}~\bibnamefont {Yamazaki}}, \bibinfo {author}
  {\bibfnamefont {K.}~\bibnamefont {Usami}}, \ and\ \bibinfo {author}
  {\bibfnamefont {Y.}~\bibnamefont {Nakamura}},\ }\bibfield  {title} {\emph
  {\bibinfo {title} {Hybridizing ferromagnetic magnons and microwave photons in
  the quantum limit},\ }}\href {\doibase 10.1103/PhysRevLett.113.083603}
  {\bibfield  {journal} {\bibinfo  {journal} {Phys. Rev. Lett.}\ }\textbf
  {\bibinfo {volume} {113}},\ \bibinfo {pages} {083603} (\bibinfo {year}
  {2014})}\BibitemShut {NoStop}%
\bibitem [{\citenamefont {Zhang}\ \emph {et~al.}(2014)\citenamefont {Zhang},
  \citenamefont {Zou}, \citenamefont {Jiang},\ and\ \citenamefont
  {Tang}}]{yigcavity5}%
  \BibitemOpen
  \bibfield  {author} {\bibinfo {author} {\bibfnamefont {X.}~\bibnamefont
  {Zhang}}, \bibinfo {author} {\bibfnamefont {C.-L.}\ \bibnamefont {Zou}},
  \bibinfo {author} {\bibfnamefont {L.}~\bibnamefont {Jiang}}, \ and\ \bibinfo
  {author} {\bibfnamefont {H.~X.}\ \bibnamefont {Tang}},\ }\bibfield  {title}
  {\emph {\bibinfo {title} {Strongly coupled magnons and cavity microwave
  photons},\ }}\href {\doibase 10.1103/PhysRevLett.113.156401} {\bibfield
  {journal} {\bibinfo  {journal} {Phys. Rev. Lett.}\ }\textbf {\bibinfo
  {volume} {113}},\ \bibinfo {pages} {156401} (\bibinfo {year}
  {2014})}\BibitemShut {NoStop}%
\bibitem [{\citenamefont {Goryachev}\ \emph {et~al.}(2014)\citenamefont
  {Goryachev}, \citenamefont {Farr}, \citenamefont {Creedon}, \citenamefont
  {Fan}, \citenamefont {Kostylev},\ and\ \citenamefont {Tobar}}]{yigcavity6}%
  \BibitemOpen
  \bibfield  {author} {\bibinfo {author} {\bibfnamefont {M.}~\bibnamefont
  {Goryachev}}, \bibinfo {author} {\bibfnamefont {W.~G.}\ \bibnamefont {Farr}},
  \bibinfo {author} {\bibfnamefont {D.~L.}\ \bibnamefont {Creedon}}, \bibinfo
  {author} {\bibfnamefont {Y.}~\bibnamefont {Fan}}, \bibinfo {author}
  {\bibfnamefont {M.}~\bibnamefont {Kostylev}}, \ and\ \bibinfo {author}
  {\bibfnamefont {M.~E.}\ \bibnamefont {Tobar}},\ }\bibfield  {title} {\emph
  {\bibinfo {title} {High-cooperativity cavity qed with magnons at microwave
  frequencies},\ }}\href {\doibase 10.1103/PhysRevApplied.2.054002} {\bibfield
  {journal} {\bibinfo  {journal} {Phys. Rev. Applied}\ }\textbf {\bibinfo
  {volume} {2}},\ \bibinfo {pages} {054002} (\bibinfo {year}
  {2014})}\BibitemShut {NoStop}%
\bibitem [{\citenamefont {Tabuchi}\ \emph {et~al.}(2015)\citenamefont
  {Tabuchi}, \citenamefont {Ishino}, \citenamefont {Noguchi}, \citenamefont
  {Ishikawa}, \citenamefont {Yamazaki}, \citenamefont {Usami},\ and\
  \citenamefont {Nakamura}}]{magnonqubit}%
  \BibitemOpen
  \bibfield  {author} {\bibinfo {author} {\bibfnamefont {Y.}~\bibnamefont
  {Tabuchi}}, \bibinfo {author} {\bibfnamefont {S.}~\bibnamefont {Ishino}},
  \bibinfo {author} {\bibfnamefont {A.}~\bibnamefont {Noguchi}}, \bibinfo
  {author} {\bibfnamefont {T.}~\bibnamefont {Ishikawa}}, \bibinfo {author}
  {\bibfnamefont {R.}~\bibnamefont {Yamazaki}}, \bibinfo {author}
  {\bibfnamefont {K.}~\bibnamefont {Usami}}, \ and\ \bibinfo {author}
  {\bibfnamefont {Y.}~\bibnamefont {Nakamura}},\ }\bibfield  {title} {\emph
  {\bibinfo {title} {Coherent coupling between a ferromagnetic magnon and a
  superconducting qubit},\ }}\href
  {https://www.science.org/lookup/doi/10.1126/science.aaa3693} {\bibfield
  {journal} {\bibinfo  {journal} {Science}\ }\textbf {\bibinfo {volume}
  {349}},\ \bibinfo {pages} {405} (\bibinfo {year} {2015})}\BibitemShut
  {NoStop}%
\bibitem [{\citenamefont {Lachance-Quirion}\ \emph {et~al.}(2020)\citenamefont
  {Lachance-Quirion}, \citenamefont {Piotr~Wolski}, \citenamefont {Tabuchi},
  \citenamefont {Kono}, \citenamefont {Usami},\ and\ \citenamefont
  {Nakamura}}]{magnonqubit2}%
  \BibitemOpen
  \bibfield  {author} {\bibinfo {author} {\bibfnamefont {D.}~\bibnamefont
  {Lachance-Quirion}}, \bibinfo {author} {\bibfnamefont {S.}~\bibnamefont
  {Piotr~Wolski}}, \bibinfo {author} {\bibfnamefont {Y.}~\bibnamefont
  {Tabuchi}}, \bibinfo {author} {\bibfnamefont {S.}~\bibnamefont {Kono}},
  \bibinfo {author} {\bibfnamefont {K.}~\bibnamefont {Usami}}, \ and\ \bibinfo
  {author} {\bibfnamefont {Y.}~\bibnamefont {Nakamura}},\ }\bibfield  {title}
  {\emph {\bibinfo {title} {Entanglement-based single-shot detection of a
  single magnon with a superconducting qubit},\ }}\href
  {https://www.science.org/lookup/doi/10.1126/science.aaz9236} {\bibfield
  {journal} {\bibinfo  {journal} {Science}\ }\textbf {\bibinfo {volume}
  {367}},\ \bibinfo {pages} {425} (\bibinfo {year} {2020})}\BibitemShut
  {NoStop}%
\bibitem [{\citenamefont {Horodecki}\ \emph {et~al.}(2009)\citenamefont
  {Horodecki}, \citenamefont {Horodecki}, \citenamefont {Horodecki},\ and\
  \citenamefont {Horodecki}}]{quantumentanglement}%
  \BibitemOpen
  \bibfield  {author} {\bibinfo {author} {\bibfnamefont {R.}~\bibnamefont
  {Horodecki}}, \bibinfo {author} {\bibfnamefont {P.}~\bibnamefont
  {Horodecki}}, \bibinfo {author} {\bibfnamefont {M.}~\bibnamefont
  {Horodecki}}, \ and\ \bibinfo {author} {\bibfnamefont {K.}~\bibnamefont
  {Horodecki}},\ }\bibfield  {title} {\emph {\bibinfo {title} {Quantum
  entanglement},\ }}\href {\doibase 10.1103/RevModPhys.81.865} {\bibfield
  {journal} {\bibinfo  {journal} {Rev. Mod. Phys.}\ }\textbf {\bibinfo {volume}
  {81}},\ \bibinfo {pages} {865} (\bibinfo {year} {2009})}\BibitemShut
  {NoStop}%
\bibitem [{\citenamefont {Gisin}\ and\ \citenamefont
  {Thew}(2007)}]{quantumcommun}%
  \BibitemOpen
  \bibfield  {author} {\bibinfo {author} {\bibfnamefont {N.}~\bibnamefont
  {Gisin}}\ and\ \bibinfo {author} {\bibfnamefont {R.}~\bibnamefont {Thew}},\
  }\bibfield  {title} {\emph {\bibinfo {title} {Quantum communication},\
  }}\href@noop {} {\bibfield  {journal} {\bibinfo  {journal} {Nat. Photon.}\
  }\textbf {\bibinfo {volume} {1}},\ \bibinfo {pages} {165} (\bibinfo {year}
  {2007})}\BibitemShut {NoStop}%
\bibitem [{\citenamefont {Ekert}(1991)}]{keydistribute}%
  \BibitemOpen
  \bibfield  {author} {\bibinfo {author} {\bibfnamefont {A.~K.}\ \bibnamefont
  {Ekert}},\ }\bibfield  {title} {\emph {\bibinfo {title} {Quantum cryptography
  based on bell's theorem},\ }}\href {\doibase 10.1103/PhysRevLett.67.661}
  {\bibfield  {journal} {\bibinfo  {journal} {Phys. Rev. Lett.}\ }\textbf
  {\bibinfo {volume} {67}},\ \bibinfo {pages} {661} (\bibinfo {year}
  {1991})}\BibitemShut {NoStop}%
\bibitem [{\citenamefont {Hillery}\ \emph {et~al.}(1999)\citenamefont
  {Hillery}, \citenamefont {Bu\ifmmode~\check{z}\else \v{z}\fi{}ek},\ and\
  \citenamefont {Berthiaume}}]{quantumsecretsharing}%
  \BibitemOpen
  \bibfield  {author} {\bibinfo {author} {\bibfnamefont {M.}~\bibnamefont
  {Hillery}}, \bibinfo {author} {\bibfnamefont {V.}~\bibnamefont
  {Bu\ifmmode~\check{z}\else \v{z}\fi{}ek}}, \ and\ \bibinfo {author}
  {\bibfnamefont {A.}~\bibnamefont {Berthiaume}},\ }\bibfield  {title} {\emph
  {\bibinfo {title} {Quantum secret sharing},\ }}\href {\doibase
  10.1103/PhysRevA.59.1829} {\bibfield  {journal} {\bibinfo  {journal} {Phys.
  Rev. A}\ }\textbf {\bibinfo {volume} {59}},\ \bibinfo {pages} {1829}
  (\bibinfo {year} {1999})}\BibitemShut {NoStop}%
\bibitem [{\citenamefont {Deng}\ \emph {et~al.}(2003)\citenamefont {Deng},
  \citenamefont {Long},\ and\ \citenamefont {Liu}}]{quantumsecurecommun}%
  \BibitemOpen
  \bibfield  {author} {\bibinfo {author} {\bibfnamefont {F.~G.}\ \bibnamefont
  {Deng}}, \bibinfo {author} {\bibfnamefont {G.~L.}\ \bibnamefont {Long}}, \
  and\ \bibinfo {author} {\bibfnamefont {X.-S.}\ \bibnamefont {Liu}},\
  }\bibfield  {title} {\emph {\bibinfo {title} {Two-step quantum direct
  communication protocol using the einstein-podolsky-rosen pair block},\
  }}\href {\doibase 10.1103/PhysRevA.68.042317} {\bibfield  {journal} {\bibinfo
   {journal} {Phys. Rev. A}\ }\textbf {\bibinfo {volume} {68}},\ \bibinfo
  {pages} {042317} (\bibinfo {year} {2003})}\BibitemShut {NoStop}%
\bibitem [{\citenamefont {M\o{}lmer}\ and\ \citenamefont
  {S\o{}rensen}(1999)}]{trapion}%
  \BibitemOpen
  \bibfield  {author} {\bibinfo {author} {\bibfnamefont {K.}~\bibnamefont
  {M\o{}lmer}}\ and\ \bibinfo {author} {\bibfnamefont {A.}~\bibnamefont
  {S\o{}rensen}},\ }\bibfield  {title} {\emph {\bibinfo {title} {Multiparticle
  entanglement of hot trapped ions},\ }}\href {\doibase
  10.1103/PhysRevLett.82.1835} {\bibfield  {journal} {\bibinfo  {journal}
  {Phys. Rev. Lett.}\ }\textbf {\bibinfo {volume} {82}},\ \bibinfo {pages}
  {1835} (\bibinfo {year} {1999})}\BibitemShut {NoStop}%
\bibitem [{\citenamefont {Wei}\ \emph {et~al.}(2006)\citenamefont {Wei},
  \citenamefont {Liu},\ and\ \citenamefont {Nori}}]{ghzstate}%
  \BibitemOpen
  \bibfield  {author} {\bibinfo {author} {\bibfnamefont {L.~F.}\ \bibnamefont
  {Wei}}, \bibinfo {author} {\bibfnamefont {Y.-x.}\ \bibnamefont {Liu}}, \ and\
  \bibinfo {author} {\bibfnamefont {F.}~\bibnamefont {Nori}},\ }\bibfield
  {title} {\emph {\bibinfo {title} {Generation and control of
  greenberger-horne-zeilinger entanglement in superconducting circuits},\
  }}\href {\doibase 10.1103/PhysRevLett.96.246803} {\bibfield  {journal}
  {\bibinfo  {journal} {Phys. Rev. Lett.}\ }\textbf {\bibinfo {volume} {96}},\
  \bibinfo {pages} {246803} (\bibinfo {year} {2006})}\BibitemShut {NoStop}%
\bibitem [{\citenamefont {Yang}\ \emph {et~al.}(2016)\citenamefont {Yang},
  \citenamefont {Su}, \citenamefont {Zheng},\ and\ \citenamefont
  {Nori}}]{superconduct2}%
  \BibitemOpen
  \bibfield  {author} {\bibinfo {author} {\bibfnamefont {C.-P.}\ \bibnamefont
  {Yang}}, \bibinfo {author} {\bibfnamefont {Q.-P.}\ \bibnamefont {Su}},
  \bibinfo {author} {\bibfnamefont {S.-B.}\ \bibnamefont {Zheng}}, \ and\
  \bibinfo {author} {\bibfnamefont {F.}~\bibnamefont {Nori}},\ }\bibfield
  {title} {\emph {\bibinfo {title} {Entangling superconducting qubits in a
  multi-cavity system},\ }}\href {\doibase 10.1088/1367-2630/18/1/013025}
  {\bibfield  {journal} {\bibinfo  {journal} {New J. Phys.}\ }\textbf {\bibinfo
  {volume} {18}},\ \bibinfo {pages} {013025} (\bibinfo {year}
  {2016})}\BibitemShut {NoStop}%
\bibitem [{\citenamefont {Erhard}\ \emph {et~al.}(2018)\citenamefont {Erhard},
  \citenamefont {Malik}, \citenamefont {Krenn},\ and\ \citenamefont
  {Zeilinger}}]{superconduct3}%
  \BibitemOpen
  \bibfield  {author} {\bibinfo {author} {\bibfnamefont {M.}~\bibnamefont
  {Erhard}}, \bibinfo {author} {\bibfnamefont {M.}~\bibnamefont {Malik}},
  \bibinfo {author} {\bibfnamefont {M.}~\bibnamefont {Krenn}}, \ and\ \bibinfo
  {author} {\bibfnamefont {A.}~\bibnamefont {Zeilinger}},\ }\bibfield  {title}
  {\emph {\bibinfo {title} {Experimental greenberger-horne-zeilinger
  entanglement beyond qubits},\ }}\href {\doibase
  https://doi.org/10.1038/s41566-018-0257-6} {\bibfield  {journal} {\bibinfo
  {journal} {Nat.Photon.}\ }\textbf {\bibinfo {volume} {12}},\ \bibinfo {pages}
  {759} (\bibinfo {year} {2018})}\BibitemShut {NoStop}%
\bibitem [{\citenamefont {Song}\ \emph {et~al.}(2019)\citenamefont {Song},
  \citenamefont {Xu}, \citenamefont {Li}, \citenamefont {Zhang}, \citenamefont
  {Zhang}, \citenamefont {Liu}, \citenamefont {Guo}, \citenamefont {Wang},
  \citenamefont {Ren}, \citenamefont {Hao}, \citenamefont {Feng}, \citenamefont
  {Fan}, \citenamefont {Zheng}, \citenamefont {Wang}, \citenamefont {Wang},\
  and\ \citenamefont {Zhu}}]{Schrodingerstate}%
  \BibitemOpen
  \bibfield  {author} {\bibinfo {author} {\bibfnamefont {C.}~\bibnamefont
  {Song}}, \bibinfo {author} {\bibfnamefont {K.}~\bibnamefont {Xu}}, \bibinfo
  {author} {\bibfnamefont {H.}~\bibnamefont {Li}}, \bibinfo {author}
  {\bibfnamefont {Y.-R.}\ \bibnamefont {Zhang}}, \bibinfo {author}
  {\bibfnamefont {X.}~\bibnamefont {Zhang}}, \bibinfo {author} {\bibfnamefont
  {W.}~\bibnamefont {Liu}}, \bibinfo {author} {\bibfnamefont {Q.}~\bibnamefont
  {Guo}}, \bibinfo {author} {\bibfnamefont {Z.}~\bibnamefont {Wang}}, \bibinfo
  {author} {\bibfnamefont {W.}~\bibnamefont {Ren}}, \bibinfo {author}
  {\bibfnamefont {J.}~\bibnamefont {Hao}}, \bibinfo {author} {\bibfnamefont
  {H.}~\bibnamefont {Feng}}, \bibinfo {author} {\bibfnamefont {H.}~\bibnamefont
  {Fan}}, \bibinfo {author} {\bibfnamefont {D.}~\bibnamefont {Zheng}}, \bibinfo
  {author} {\bibfnamefont {D.-W.}\ \bibnamefont {Wang}}, \bibinfo {author}
  {\bibfnamefont {H.}~\bibnamefont {Wang}}, \ and\ \bibinfo {author}
  {\bibfnamefont {S.-Y.}\ \bibnamefont {Zhu}},\ }\bibfield  {title} {\emph
  {\bibinfo {title} {Generation of multicomponent atomic schr\"{o}dinger cat
  states of up to 20 qubits},\ }}\href {\doibase
  https://www.science.org/lookup/doi/10.1126/science.aay0600} {\ \textbf
  {\bibinfo {volume} {365}},\ \bibinfo {pages} {574} (\bibinfo {year}
  {2019})}\BibitemShut {NoStop}%
\bibitem [{\citenamefont {Bouwmeester}\ \emph {et~al.}(1999)\citenamefont
  {Bouwmeester}, \citenamefont {Pan}, \citenamefont {Daniell}, \citenamefont
  {Weinfurter},\ and\ \citenamefont {Zeilinger}}]{GHZ}%
  \BibitemOpen
  \bibfield  {author} {\bibinfo {author} {\bibfnamefont {D.}~\bibnamefont
  {Bouwmeester}}, \bibinfo {author} {\bibfnamefont {J.-W.}\ \bibnamefont
  {Pan}}, \bibinfo {author} {\bibfnamefont {M.}~\bibnamefont {Daniell}},
  \bibinfo {author} {\bibfnamefont {H.}~\bibnamefont {Weinfurter}}, \ and\
  \bibinfo {author} {\bibfnamefont {A.}~\bibnamefont {Zeilinger}},\ }\bibfield
  {title} {\emph {\bibinfo {title} {Observation of three-photon
  greenberger-horne-zeilinger entanglement},\ }}\href {\doibase
  10.1103/PhysRevLett.82.1345} {\bibfield  {journal} {\bibinfo  {journal}
  {Phys. Rev. Lett.}\ }\textbf {\bibinfo {volume} {82}},\ \bibinfo {pages}
  {1345} (\bibinfo {year} {1999})}\BibitemShut {NoStop}%
\bibitem [{\citenamefont {Bishop}\ \emph {et~al.}(2009)\citenamefont {Bishop},
  \citenamefont {Tornberg}, \citenamefont {Price}, \citenamefont {Ginossar},
  \citenamefont {Nunnenkamp}, \citenamefont {Houck}, \citenamefont {Gambetta},
  \citenamefont {Koch}, \citenamefont {Johansson}, \citenamefont {Girvin},\
  and\ \citenamefont {Schoelkopf}}]{GHZstate2}%
  \BibitemOpen
  \bibfield  {author} {\bibinfo {author} {\bibfnamefont {L.~S.}\ \bibnamefont
  {Bishop}}, \bibinfo {author} {\bibfnamefont {L.}~\bibnamefont {Tornberg}},
  \bibinfo {author} {\bibfnamefont {D.}~\bibnamefont {Price}}, \bibinfo
  {author} {\bibfnamefont {E.}~\bibnamefont {Ginossar}}, \bibinfo {author}
  {\bibfnamefont {A.}~\bibnamefont {Nunnenkamp}}, \bibinfo {author}
  {\bibfnamefont {A.~A.}\ \bibnamefont {Houck}}, \bibinfo {author}
  {\bibfnamefont {J.~M.}\ \bibnamefont {Gambetta}}, \bibinfo {author}
  {\bibfnamefont {J.}~\bibnamefont {Koch}}, \bibinfo {author} {\bibfnamefont
  {G.}~\bibnamefont {Johansson}}, \bibinfo {author} {\bibfnamefont {S.~M.}\
  \bibnamefont {Girvin}}, \ and\ \bibinfo {author} {\bibfnamefont {R.~J.}\
  \bibnamefont {Schoelkopf}},\ }\bibfield  {title} {\emph {\bibinfo {title}
  {Proposal for generating and detecting multi-qubit {GHZ} states in circuit
  {QED}},\ }}\href {\doibase 10.1088/1367-2630/11/7/073040} {\bibfield
  {journal} {\bibinfo  {journal} {New J. Phys.}\ }\textbf {\bibinfo {volume}
  {11}},\ \bibinfo {pages} {073040} (\bibinfo {year} {2009})}\BibitemShut
  {NoStop}%
\bibitem [{\citenamefont {Paul}\ and\ \citenamefont
  {Sarma}(2016)}]{bellstate3}%
  \BibitemOpen
  \bibfield  {author} {\bibinfo {author} {\bibfnamefont {K.}~\bibnamefont
  {Paul}}\ and\ \bibinfo {author} {\bibfnamefont {A.~K.}\ \bibnamefont
  {Sarma}},\ }\bibfield  {title} {\emph {\bibinfo {title} {High-fidelity
  entangled bell states via shortcuts to adiabaticity},\ }}\href {\doibase
  10.1103/PhysRevA.94.052303} {\bibfield  {journal} {\bibinfo  {journal} {Phys.
  Rev. A}\ }\textbf {\bibinfo {volume} {94}},\ \bibinfo {pages} {052303}
  (\bibinfo {year} {2016})}\BibitemShut {NoStop}%
\bibitem [{\citenamefont {Pan}\ \emph {et~al.}(2012)\citenamefont {Pan},
  \citenamefont {Chen}, \citenamefont {Lu}, \citenamefont {Weinfurter},
  \citenamefont {Zeilinger},\ and\ \citenamefont {\ifmmode~\dot{Z}\else
  \.{Z}\fi{}ukowski}}]{multiphoton}%
  \BibitemOpen
  \bibfield  {author} {\bibinfo {author} {\bibfnamefont {J.-W.}\ \bibnamefont
  {Pan}}, \bibinfo {author} {\bibfnamefont {Z.-B.}\ \bibnamefont {Chen}},
  \bibinfo {author} {\bibfnamefont {C.-Y.}\ \bibnamefont {Lu}}, \bibinfo
  {author} {\bibfnamefont {H.}~\bibnamefont {Weinfurter}}, \bibinfo {author}
  {\bibfnamefont {A.}~\bibnamefont {Zeilinger}}, \ and\ \bibinfo {author}
  {\bibfnamefont {M.}~\bibnamefont {\ifmmode~\dot{Z}\else \.{Z}\fi{}ukowski}},\
  }\bibfield  {title} {\emph {\bibinfo {title} {Multiphoton entanglement and
  interferometry},\ }}\href {\doibase 10.1103/RevModPhys.84.777} {\bibfield
  {journal} {\bibinfo  {journal} {Rev. Mod. Phys.}\ }\textbf {\bibinfo {volume}
  {84}},\ \bibinfo {pages} {777} (\bibinfo {year} {2012})}\BibitemShut
  {NoStop}%
\bibitem [{\citenamefont {Tashima}\ \emph {et~al.}(2016)\citenamefont
  {Tashima}, \citenamefont {Tame}, \citenamefont {\"Ozdemir}, \citenamefont
  {Nori}, \citenamefont {Koashi},\ and\ \citenamefont {Weinfurter}}]{photonic}%
  \BibitemOpen
  \bibfield  {author} {\bibinfo {author} {\bibfnamefont {T.}~\bibnamefont
  {Tashima}}, \bibinfo {author} {\bibfnamefont {M.~S.}\ \bibnamefont {Tame}},
  \bibinfo {author} {\bibfnamefont {S.~K.}\ \bibnamefont {\"Ozdemir}}, \bibinfo
  {author} {\bibfnamefont {F.}~\bibnamefont {Nori}}, \bibinfo {author}
  {\bibfnamefont {M.}~\bibnamefont {Koashi}}, \ and\ \bibinfo {author}
  {\bibfnamefont {H.}~\bibnamefont {Weinfurter}},\ }\bibfield  {title} {\emph
  {\bibinfo {title} {Photonic multipartite entanglement conversion using
  nonlocal operations},\ }}\href {\doibase 10.1103/PhysRevA.94.052309}
  {\bibfield  {journal} {\bibinfo  {journal} {Phys. Rev. A}\ }\textbf {\bibinfo
  {volume} {94}},\ \bibinfo {pages} {052309} (\bibinfo {year}
  {2016})}\BibitemShut {NoStop}%
\bibitem [{\citenamefont {Strauch}\ \emph {et~al.}(2010)\citenamefont
  {Strauch}, \citenamefont {Jacobs},\ and\ \citenamefont {Simmonds}}]{noon1}%
  \BibitemOpen
  \bibfield  {author} {\bibinfo {author} {\bibfnamefont {F.~W.}\ \bibnamefont
  {Strauch}}, \bibinfo {author} {\bibfnamefont {K.}~\bibnamefont {Jacobs}}, \
  and\ \bibinfo {author} {\bibfnamefont {R.~W.}\ \bibnamefont {Simmonds}},\
  }\bibfield  {title} {\emph {\bibinfo {title} {Arbitrary control of
  entanglement between two superconducting resonators},\ }}\href {\doibase
  10.1103/PhysRevLett.105.050501} {\bibfield  {journal} {\bibinfo  {journal}
  {Phys. Rev. Lett.}\ }\textbf {\bibinfo {volume} {105}},\ \bibinfo {pages}
  {050501} (\bibinfo {year} {2010})}\BibitemShut {NoStop}%
\bibitem [{\citenamefont {Merkel}\ and\ \citenamefont {Wilhelm}(2010)}]{noon2}%
  \BibitemOpen
  \bibfield  {author} {\bibinfo {author} {\bibfnamefont {S.~T.}\ \bibnamefont
  {Merkel}}\ and\ \bibinfo {author} {\bibfnamefont {F.~K.}\ \bibnamefont
  {Wilhelm}},\ }\bibfield  {title} {\emph {\bibinfo {title} {Generation and
  detection of noon states in superconducting circuits},\ }}\href {\doibase
  https://doi.org/10.1088/1367-2630/12/9/093036} {\bibfield  {journal}
  {\bibinfo  {journal} {New J. Phys.}\ }\textbf {\bibinfo {volume} {12}},\
  \bibinfo {pages} {3175} (\bibinfo {year} {2010})}\BibitemShut {NoStop}%
\bibitem [{\citenamefont {Macr\`{\i}}\ \emph {et~al.}(2018)\citenamefont
  {Macr\`{\i}}, \citenamefont {Nori},\ and\ \citenamefont
  {Kockum}}]{bellstate}%
  \BibitemOpen
  \bibfield  {author} {\bibinfo {author} {\bibfnamefont {V.}~\bibnamefont
  {Macr\`{\i}}}, \bibinfo {author} {\bibfnamefont {F.}~\bibnamefont {Nori}}, \
  and\ \bibinfo {author} {\bibfnamefont {A.~F.}\ \bibnamefont {Kockum}},\
  }\bibfield  {title} {\emph {\bibinfo {title} {Simple preparation of bell and
  greenberger-horne-zeilinger states using ultrastrong-coupling circuit qed},\
  }}\href {\doibase 10.1103/PhysRevA.98.062327} {\bibfield  {journal} {\bibinfo
   {journal} {Phys. Rev. A}\ }\textbf {\bibinfo {volume} {98}},\ \bibinfo
  {pages} {062327} (\bibinfo {year} {2018})}\BibitemShut {NoStop}%
\bibitem [{\citenamefont {Tittel}\ \emph {et~al.}(2000)\citenamefont {Tittel},
  \citenamefont {Brendel}, \citenamefont {Zbinden},\ and\ \citenamefont
  {Gisin}}]{cryptography}%
  \BibitemOpen
  \bibfield  {author} {\bibinfo {author} {\bibfnamefont {W.}~\bibnamefont
  {Tittel}}, \bibinfo {author} {\bibfnamefont {J.}~\bibnamefont {Brendel}},
  \bibinfo {author} {\bibfnamefont {H.}~\bibnamefont {Zbinden}}, \ and\
  \bibinfo {author} {\bibfnamefont {N.}~\bibnamefont {Gisin}},\ }\bibfield
  {title} {\emph {\bibinfo {title} {Quantum cryptography using entangled
  photons in energy-time bell states},\ }}\href {\doibase
  10.1103/PhysRevLett.84.4737} {\bibfield  {journal} {\bibinfo  {journal}
  {Phys. Rev. Lett.}\ }\textbf {\bibinfo {volume} {84}},\ \bibinfo {pages}
  {4737} (\bibinfo {year} {2000})}\BibitemShut {NoStop}%
\bibitem [{\citenamefont {Kaszlikowski}\ \emph {et~al.}(2005)\citenamefont
  {Kaszlikowski}, \citenamefont {Lim}, \citenamefont {Oi}, \citenamefont
  {Willeboordse}, \citenamefont {Gopinathan},\ and\ \citenamefont
  {Kwek}}]{cryptography2}%
  \BibitemOpen
  \bibfield  {author} {\bibinfo {author} {\bibfnamefont {D.}~\bibnamefont
  {Kaszlikowski}}, \bibinfo {author} {\bibfnamefont {J.~Y.}\ \bibnamefont
  {Lim}}, \bibinfo {author} {\bibfnamefont {D.~K.~L.}\ \bibnamefont {Oi}},
  \bibinfo {author} {\bibfnamefont {F.~H.}\ \bibnamefont {Willeboordse}},
  \bibinfo {author} {\bibfnamefont {A.}~\bibnamefont {Gopinathan}}, \ and\
  \bibinfo {author} {\bibfnamefont {L.~C.}\ \bibnamefont {Kwek}},\ }\bibfield
  {title} {\emph {\bibinfo {title} {Quantum tomographic cryptography with bell
  diagonal states: Nonequivalence of classical and quantum distillation
  protocols},\ }}\href {\doibase 10.1103/PhysRevA.71.012309} {\bibfield
  {journal} {\bibinfo  {journal} {Phys. Rev. A}\ }\textbf {\bibinfo {volume}
  {71}},\ \bibinfo {pages} {012309} (\bibinfo {year} {2005})}\BibitemShut
  {NoStop}%
\bibitem [{\citenamefont {Kim}\ \emph {et~al.}(2001)\citenamefont {Kim},
  \citenamefont {Kulik},\ and\ \citenamefont {Shih}}]{teleportation}%
  \BibitemOpen
  \bibfield  {author} {\bibinfo {author} {\bibfnamefont {Y.-H.}\ \bibnamefont
  {Kim}}, \bibinfo {author} {\bibfnamefont {S.~P.}\ \bibnamefont {Kulik}}, \
  and\ \bibinfo {author} {\bibfnamefont {Y.}~\bibnamefont {Shih}},\ }\bibfield
  {title} {\emph {\bibinfo {title} {Quantum teleportation of a polarization
  state with a complete bell state measurement},\ }}\href {\doibase
  10.1103/PhysRevLett.86.1370} {\bibfield  {journal} {\bibinfo  {journal}
  {Phys. Rev. Lett.}\ }\textbf {\bibinfo {volume} {86}},\ \bibinfo {pages}
  {1370} (\bibinfo {year} {2001})}\BibitemShut {NoStop}%
\bibitem [{\citenamefont {D\"ur}\ \emph {et~al.}(2000)\citenamefont {D\"ur},
  \citenamefont {Vidal},\ and\ \citenamefont {Cirac}}]{wstate}%
  \BibitemOpen
  \bibfield  {author} {\bibinfo {author} {\bibfnamefont {W.}~\bibnamefont
  {D\"ur}}, \bibinfo {author} {\bibfnamefont {G.}~\bibnamefont {Vidal}}, \ and\
  \bibinfo {author} {\bibfnamefont {J.~I.}\ \bibnamefont {Cirac}},\ }\bibfield
  {title} {\emph {\bibinfo {title} {Three qubits can be entangled in two
  inequivalent ways},\ }}\href {\doibase 10.1103/PhysRevA.62.062314} {\bibfield
   {journal} {\bibinfo  {journal} {Phys. Rev. A}\ }\textbf {\bibinfo {volume}
  {62}},\ \bibinfo {pages} {062314} (\bibinfo {year} {2000})}\BibitemShut
  {NoStop}%
\bibitem [{\citenamefont {Lee}\ and\ \citenamefont {Kim}(2000)}]{wernerstate}%
  \BibitemOpen
  \bibfield  {author} {\bibinfo {author} {\bibfnamefont {J.}~\bibnamefont
  {Lee}}\ and\ \bibinfo {author} {\bibfnamefont {M.~S.}\ \bibnamefont {Kim}},\
  }\bibfield  {title} {\emph {\bibinfo {title} {Entanglement teleportation via
  werner states},\ }}\href {\doibase 10.1103/PhysRevLett.84.4236} {\bibfield
  {journal} {\bibinfo  {journal} {Phys. Rev. Lett.}\ }\textbf {\bibinfo
  {volume} {84}},\ \bibinfo {pages} {4236} (\bibinfo {year}
  {2000})}\BibitemShut {NoStop}%
\bibitem [{\citenamefont {You}\ and\ \citenamefont {Franco}(2011)}]{cqed}%
  \BibitemOpen
  \bibfield  {author} {\bibinfo {author} {\bibfnamefont {J.~Q.}\ \bibnamefont
  {You}}\ and\ \bibinfo {author} {\bibfnamefont {N.}~\bibnamefont {Franco}},\
  }\bibfield  {title} {\emph {\bibinfo {title} {Atomic physics and quantum
  optics using superconducting circuits},\ }}\href
  {https://www.nature.com/articles/nature10122} {\bibfield  {journal} {\bibinfo
   {journal} {Nature (London)}\ }\textbf {\bibinfo {volume} {474}},\ \bibinfo
  {pages} {589} (\bibinfo {year} {2011})}\BibitemShut {NoStop}%
\bibitem [{\citenamefont {Niemczyk}\ \emph {et~al.}(2010)\citenamefont
  {Niemczyk}, \citenamefont {Deppe}, \citenamefont {Huebl}, \citenamefont
  {Menzel}, \citenamefont {Hocke}, \citenamefont {Schwarz}, \citenamefont
  {Garciaripoll}, \citenamefont {Zueco}, \citenamefont {H\"ommer},\ and\
  \citenamefont {Solano}}]{cq}%
  \BibitemOpen
  \bibfield  {author} {\bibinfo {author} {\bibfnamefont {T.}~\bibnamefont
  {Niemczyk}}, \bibinfo {author} {\bibfnamefont {F.}~\bibnamefont {Deppe}},
  \bibinfo {author} {\bibfnamefont {H.}~\bibnamefont {Huebl}}, \bibinfo
  {author} {\bibfnamefont {E.~P.}\ \bibnamefont {Menzel}}, \bibinfo {author}
  {\bibfnamefont {F.}~\bibnamefont {Hocke}}, \bibinfo {author} {\bibfnamefont
  {M.~J.}\ \bibnamefont {Schwarz}}, \bibinfo {author} {\bibfnamefont {J.~J.}\
  \bibnamefont {Garciaripoll}}, \bibinfo {author} {\bibfnamefont
  {D.}~\bibnamefont {Zueco}}, \bibinfo {author} {\bibfnamefont
  {T.}~\bibnamefont {H\"ommer}}, \ and\ \bibinfo {author} {\bibfnamefont
  {E.}~\bibnamefont {Solano}},\ }\bibfield  {title} {\emph {\bibinfo {title}
  {Circuit quantum electrodynamics in the ultrastrong-coupling regime},\
  }}\href {https://www.nature.com/articles/nphys1730} {\bibfield  {journal}
  {\bibinfo  {journal} {Nat. Phys.}\ }\textbf {\bibinfo {volume} {6}},\
  \bibinfo {pages} {772} (\bibinfo {year} {2010})}\BibitemShut {NoStop}%
\bibitem [{\citenamefont {Forn-D\'{\i}az}\ \emph {et~al.}(2019)\citenamefont
  {Forn-D\'{\i}az}, \citenamefont {Lamata}, \citenamefont {Rico}, \citenamefont
  {Kono},\ and\ \citenamefont {Solano}}]{ultrastrong}%
  \BibitemOpen
  \bibfield  {author} {\bibinfo {author} {\bibfnamefont {P.}~\bibnamefont
  {Forn-D\'{\i}az}}, \bibinfo {author} {\bibfnamefont {L.}~\bibnamefont
  {Lamata}}, \bibinfo {author} {\bibfnamefont {E.}~\bibnamefont {Rico}},
  \bibinfo {author} {\bibfnamefont {J.}~\bibnamefont {Kono}}, \ and\ \bibinfo
  {author} {\bibfnamefont {E.}~\bibnamefont {Solano}},\ }\bibfield  {title}
  {\emph {\bibinfo {title} {Ultrastrong coupling regimes of light-matter
  interaction},\ }}\href {\doibase 10.1103/RevModPhys.91.025005} {\bibfield
  {journal} {\bibinfo  {journal} {Rev. Mod. Phys.}\ }\textbf {\bibinfo {volume}
  {91}},\ \bibinfo {pages} {025005} (\bibinfo {year} {2019})}\BibitemShut
  {NoStop}%
\bibitem [{\citenamefont {Kockum}\ \emph {et~al.}(2019)\citenamefont {Kockum},
  \citenamefont {Miranowicz}, \citenamefont {Liberato}, \citenamefont
  {Savasta},\ and\ \citenamefont {Nori}}]{ultrastrong2}%
  \BibitemOpen
  \bibfield  {author} {\bibinfo {author} {\bibfnamefont {A.~F.}\ \bibnamefont
  {Kockum}}, \bibinfo {author} {\bibfnamefont {A.}~\bibnamefont {Miranowicz}},
  \bibinfo {author} {\bibfnamefont {S.~D.}\ \bibnamefont {Liberato}}, \bibinfo
  {author} {\bibfnamefont {S.}~\bibnamefont {Savasta}}, \ and\ \bibinfo
  {author} {\bibfnamefont {F.}~\bibnamefont {Nori}},\ }\bibfield  {title}
  {\emph {\bibinfo {title} {Ultrastrong coupling between light and matter},\
  }}\href {https://www.nature.com/articles/s42254-018-0006-2} {\bibfield
  {journal} {\bibinfo  {journal} {Nat. Rev. Phys.}\ }\textbf {\bibinfo {volume}
  {1}},\ \bibinfo {pages} {19} (\bibinfo {year} {2019})}\BibitemShut {NoStop}%
\bibitem [{\citenamefont {Combescot}(2001)}]{fermigolden}%
  \BibitemOpen
  \bibfield  {author} {\bibinfo {author} {\bibfnamefont {M.}~\bibnamefont
  {Combescot}},\ }\bibfield  {title} {\emph {\bibinfo {title} {On the
  generalized golden rule for transition probabilities},\ }}\href {\doibase
  10.1088/0305-4470/34/31/304} {\bibfield  {journal} {\bibinfo  {journal} {J.
  Phys. A: Math. Gen.}\ }\textbf {\bibinfo {volume} {34}},\ \bibinfo {pages}
  {6087} (\bibinfo {year} {2001})}\BibitemShut {NoStop}%
\bibitem [{\citenamefont {Garziano}\ \emph {et~al.}(2016)\citenamefont
  {Garziano}, \citenamefont {Macr\`{\i}}, \citenamefont {Stassi}, \citenamefont
  {Di~Stefano}, \citenamefont {Nori},\ and\ \citenamefont
  {Savasta}}]{secondorder}%
  \BibitemOpen
  \bibfield  {author} {\bibinfo {author} {\bibfnamefont {L.}~\bibnamefont
  {Garziano}}, \bibinfo {author} {\bibfnamefont {V.}~\bibnamefont
  {Macr\`{\i}}}, \bibinfo {author} {\bibfnamefont {R.}~\bibnamefont {Stassi}},
  \bibinfo {author} {\bibfnamefont {O.}~\bibnamefont {Di~Stefano}}, \bibinfo
  {author} {\bibfnamefont {F.}~\bibnamefont {Nori}}, \ and\ \bibinfo {author}
  {\bibfnamefont {S.}~\bibnamefont {Savasta}},\ }\bibfield  {title} {\emph
  {\bibinfo {title} {One photon can simultaneously excite two or more atoms},\
  }}\href {\doibase 10.1103/PhysRevLett.117.043601} {\bibfield  {journal}
  {\bibinfo  {journal} {Phys. Rev. Lett.}\ }\textbf {\bibinfo {volume} {117}},\
  \bibinfo {pages} {043601} (\bibinfo {year} {2016})}\BibitemShut {NoStop}%
\bibitem [{\citenamefont {Qi}\ and\ \citenamefont {Jing}(2020)}]{noon3}%
  \BibitemOpen
  \bibfield  {author} {\bibinfo {author} {\bibfnamefont {S.-f.}\ \bibnamefont
  {Qi}}\ and\ \bibinfo {author} {\bibfnamefont {J.}~\bibnamefont {Jing}},\
  }\bibfield  {title} {\emph {\bibinfo {title} {Generating noon states in
  circuit qed using a multiphoton resonance in the presence of counter-rotating
  interactions},\ }}\href {\doibase 10.1103/PhysRevA.101.033809} {\bibfield
  {journal} {\bibinfo  {journal} {Phys. Rev. A}\ }\textbf {\bibinfo {volume}
  {101}},\ \bibinfo {pages} {033809} (\bibinfo {year} {2020})}\BibitemShut
  {NoStop}%
\bibitem [{\citenamefont {Ma}\ and\ \citenamefont {Law}(2015)}]{ae}%
  \BibitemOpen
  \bibfield  {author} {\bibinfo {author} {\bibfnamefont {K.~K.~W.}\
  \bibnamefont {Ma}}\ and\ \bibinfo {author} {\bibfnamefont {C.~K.}\
  \bibnamefont {Law}},\ }\bibfield  {title} {\emph {\bibinfo {title}
  {Three-photon resonance and adiabatic passage in the large-detuning rabi
  model},\ }}\href {\doibase 10.1103/PhysRevA.92.023842} {\bibfield  {journal}
  {\bibinfo  {journal} {Phys. Rev. A}\ }\textbf {\bibinfo {volume} {92}},\
  \bibinfo {pages} {023842} (\bibinfo {year} {2015})}\BibitemShut {NoStop}%
\bibitem [{\citenamefont {Kaufman}\ \emph {et~al.}(2020)\citenamefont
  {Kaufman}, \citenamefont {Rozgonyi}, \citenamefont {Marquetand},\ and\
  \citenamefont {Weinacht}}]{ae1}%
  \BibitemOpen
  \bibfield  {author} {\bibinfo {author} {\bibfnamefont {B.}~\bibnamefont
  {Kaufman}}, \bibinfo {author} {\bibfnamefont {T.}~\bibnamefont {Rozgonyi}},
  \bibinfo {author} {\bibfnamefont {P.}~\bibnamefont {Marquetand}}, \ and\
  \bibinfo {author} {\bibfnamefont {T.}~\bibnamefont {Weinacht}},\ }\bibfield
  {title} {\emph {\bibinfo {title} {Adiabatic elimination in strong-field
  light-matter coupling},\ }}\href {\doibase 10.1103/PhysRevA.102.063117}
  {\bibfield  {journal} {\bibinfo  {journal} {Phys. Rev. A}\ }\textbf {\bibinfo
  {volume} {102}},\ \bibinfo {pages} {063117} (\bibinfo {year}
  {2020})}\BibitemShut {NoStop}%
\bibitem [{\citenamefont {Ozeri}\ \emph {et~al.}(2003)\citenamefont {Ozeri},
  \citenamefont {Katz}, \citenamefont {Steinhauer}, \citenamefont {Rowen},\
  and\ \citenamefont {Davidson}}]{threewave1}%
  \BibitemOpen
  \bibfield  {author} {\bibinfo {author} {\bibfnamefont {R.}~\bibnamefont
  {Ozeri}}, \bibinfo {author} {\bibfnamefont {N.}~\bibnamefont {Katz}},
  \bibinfo {author} {\bibfnamefont {J.}~\bibnamefont {Steinhauer}}, \bibinfo
  {author} {\bibfnamefont {E.}~\bibnamefont {Rowen}}, \ and\ \bibinfo {author}
  {\bibfnamefont {N.}~\bibnamefont {Davidson}},\ }\bibfield  {title} {\emph
  {\bibinfo {title} {Three-wave mixing of bogoliubov quasiparticles in a
  bose-einstein condensate},\ }}\href {\doibase 10.1103/PhysRevLett.90.170401}
  {\bibfield  {journal} {\bibinfo  {journal} {Phys. Rev. Lett.}\ }\textbf
  {\bibinfo {volume} {90}},\ \bibinfo {pages} {170401} (\bibinfo {year}
  {2003})}\BibitemShut {NoStop}%
\bibitem [{\citenamefont {Abdo}\ \emph {et~al.}(2013)\citenamefont {Abdo},
  \citenamefont {Kamal},\ and\ \citenamefont {Devoret}}]{threewave2}%
  \BibitemOpen
  \bibfield  {author} {\bibinfo {author} {\bibfnamefont {B.}~\bibnamefont
  {Abdo}}, \bibinfo {author} {\bibfnamefont {A.}~\bibnamefont {Kamal}}, \ and\
  \bibinfo {author} {\bibfnamefont {M.}~\bibnamefont {Devoret}},\ }\bibfield
  {title} {\emph {\bibinfo {title} {Nondegenerate three-wave mixing with the
  josephson ring modulator},\ }}\href {\doibase 10.1103/PhysRevB.87.014508}
  {\bibfield  {journal} {\bibinfo  {journal} {Phys. Rev. B}\ }\textbf {\bibinfo
  {volume} {87}},\ \bibinfo {pages} {014508} (\bibinfo {year}
  {2013})}\BibitemShut {NoStop}%
\bibitem [{\citenamefont {Xu}\ \emph {et~al.}(2018)\citenamefont {Xu},
  \citenamefont {Cai}, \citenamefont {Ma}, \citenamefont {Mu}, \citenamefont
  {Hu}, \citenamefont {Chen}, \citenamefont {Wang}, \citenamefont {Song},
  \citenamefont {Xue}, \citenamefont {Yin},\ and\ \citenamefont
  {Sun}}]{onegate}%
  \BibitemOpen
  \bibfield  {author} {\bibinfo {author} {\bibfnamefont {Y.}~\bibnamefont
  {Xu}}, \bibinfo {author} {\bibfnamefont {W.}~\bibnamefont {Cai}}, \bibinfo
  {author} {\bibfnamefont {Y.}~\bibnamefont {Ma}}, \bibinfo {author}
  {\bibfnamefont {X.}~\bibnamefont {Mu}}, \bibinfo {author} {\bibfnamefont
  {L.}~\bibnamefont {Hu}}, \bibinfo {author} {\bibfnamefont {T.}~\bibnamefont
  {Chen}}, \bibinfo {author} {\bibfnamefont {H.}~\bibnamefont {Wang}}, \bibinfo
  {author} {\bibfnamefont {Y.~P.}\ \bibnamefont {Song}}, \bibinfo {author}
  {\bibfnamefont {Z.-Y.}\ \bibnamefont {Xue}}, \bibinfo {author} {\bibfnamefont
  {Z.-q.}\ \bibnamefont {Yin}}, \ and\ \bibinfo {author} {\bibfnamefont
  {L.}~\bibnamefont {Sun}},\ }\bibfield  {title} {\emph {\bibinfo {title}
  {Single-loop realization of arbitrary nonadiabatic holonomic single-qubit
  quantum gates in a superconducting circuit},\ }}\href {\doibase
  10.1103/PhysRevLett.121.110501} {\bibfield  {journal} {\bibinfo  {journal}
  {Phys. Rev. Lett.}\ }\textbf {\bibinfo {volume} {121}},\ \bibinfo {pages}
  {110501} (\bibinfo {year} {2018})}\BibitemShut {NoStop}%
\end{thebibliography}%

\end{document}